\documentclass[pdflatex,sn-mathphys-num]{sn-jnl}

\usepackage{graphicx}%
\usepackage{multirow}%
\usepackage{amsmath,amssymb,amsfonts}%
\usepackage{amsthm}%
\usepackage{mathrsfs}%
\usepackage[title]{appendix}%
\usepackage{xcolor}%
\usepackage{textcomp}%
\usepackage{manyfoot}%
\usepackage{booktabs}%
\usepackage{algorithm}%
\usepackage{algorithmicx}%
\usepackage{algpseudocode}%
\usepackage{listings}%
\usepackage{braket}
\usepackage{soul}
\usepackage{slashed}

\theoremstyle{thmstyleone}%
\theoremstyle{thmstyletwo}%

\theoremstyle{thmstylethree}%

\begin{document}

\title[Pseudoscalar meson dominance, the pion--nucleon coupling constant and the Goldberger--Treiman discrepancy]{Pseudoscalar meson dominance, the pion--nucleon coupling constant and the Goldberger--Treiman discrepancy}


\author*[1]{\fnm{Enrique} \sur{Ruiz Arriola}}\email{earriola@ugr.es}

\author[1]{\fnm{Pablo} \sur{Sanchez-Puertas}}\email{pablosanchez@ugr.es}
\equalcont{These authors contributed equally to this work.}

\affil*[1]{\orgdiv{Departamento de F{\'i}sica At{\'o}mica, Molecular y Nuclear}, \orgname{Universidad de Granada}, \orgaddress{\street{Av. de la Fuentenueva s/n}, \city{Granada}, \postcode{E-18071}, \country{Spain}}}


\abstract{\unboldmath We analyze the matrix elements of the pseudoscalar density
  with pion-quantum numbers $I^G J^{PC}= 1^- 0^{-+}$ in the nucleon in
  terms of dispersion relations, PCAC and pQCD asymptotic sum rules
  for the pseudoscalar form factor. We show that the corresponding
  spectral density must have at least one zero. A model based on ChPT
  at low energies, resonances at intermediate energies, Regge
  power-like behaviour at high energies and pQCD at asymptotically
  high energies, allows to deduce the pion--nucleon coupling constant and the
  Goldberger--Treiman discrepancy $\Delta_{\rm GT} = 1 -\frac{m_N g_A
  }{F_{\pi}g_{\pi NN}}$, yielding the results for the charged channel
  \[
    g_{\pi^+ pn} = 13.14(^{+6}_{-4})(7)_{\rm IB}, \quad \Delta_{\rm GT} = 1.26(^{+51}_{-34})(50)_{\rm IB}\% ,  
  \]
to be compared with the most precise determinations, $g_{\pi^+ np} =
13.25(5)$ (and hence $\Delta_{\rm GT}=2.1(4) \%$), from $np, pp$ scattering analysis of the Granada-2013
database and $g_{\pi^+pn}=13.11(10)$, $\Delta_{\rm GT}=1.0(7)\%$ from the GMO sum rule. Our work supports the
concept of pseudoscalar dominance in the nucleon structure suggested
by Dominguez long ago. The minimal resonance saturation of the 
pseudoscalar form factor of the nucleon with the lowest isovector-pseudoscalar
mesons compatible with analyticity, pQCD short distance constraints
and chiral symmetry leads to an extended PCAC in the large-$N_c$
limit, and effectively depends on the $\pi(1300)$ excited pion
state. Our results are compatible, though more accurate, than recent
lattice QCD studies and are consistent with almost flat strong
pion-nucleon-nucleon vertices.}




\maketitle

\section{Introduction}\label{sec1}

\begin{figure}[hhh]\center
  \includegraphics[width=8.5cm]{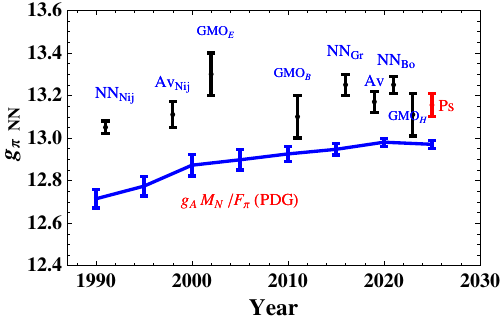}
  \caption{Pion--nucleon coupling constant determinations compared to
    the historical PDG averages of the ratio $g_A m_N/F_\pi$.  Here
    $Av_{\rm Nij}$~\cite{deSwart:1997ep}, GMO$_B$~\cite{Baru:2010xn}, GMO$_E$~\cite{Ericson:2000md}, GMO$_H$ \cite{Hoferichter:2023ptl},
    $NN_{\rm Gr}$~\cite{NavarroPerez:2013mvd},
    Av~\cite{Matsinos:2019kqi}, $NN_{\rm Bo}$~\cite{Reinert:2020mcu},
    Ps represents this work's value, which supersedes our previous
    estimate of Ref.~\cite{RuizArriola:2023xap}. The difference in
    percentage of both determinations corresponds to the GT
    discrepancy $\Delta_{\rm GT}= 1- g_A m_N / (g_{\pi NN} F_\pi) $ and is a direct
    measure of the chiral symmetry breaking in QCD.}
\label{fig:gpiNN}
\end{figure}

The pion--nucleon coupling constant, $g_{\pi NN}$, operates as a
fundamental constant in hadronic and nuclear physics; it determines
the strength of the strong interaction between nucleons at distances
above $(2-3)$~fm. It was first introduced by
Yukawa~\cite{Yukawa:1935xg} as the primary explanation of the finite
range of the nuclear force in terms of pion exchange between
nucleons. Soon thereafter, Bethe realized the tensorial character of
the One Pion Exchange (OPE) interaction, inferring the value $g_{\pi
  NN}=13.2-13.4$ from known deuteron
properties~\cite{Bethe:1940iba,Bethe:1940zz}. Within a field-theoretical 
context, this coupling is rigorously defined as the $\pi NN
$ 3-point vertex function when all three particles are on the mass
shell; an unphysical point which cannot be accessed easily and
directly by a single measurement. This circumstance imposes practical
limitations, particularly when accurate estimates are attempted in
phenomenological terms at the hadronic level. Numerous attempts have been made
 to reproduce the numerical value from the theoretical
side. However, the continuous upgrading of experimental values has
inevitably generated some confusion as to what schemes were
acceptable, see Fig.~\ref{fig:gpiNN}.

In this regard, the ancient
Goldberger--Treiman~\cite{Goldberger:1958tr} (GT) relation was a
spectacular and accurate prediction where weak- and strong-interaction
properties are intertwined, $g_{\pi^+ pn}= (m_p+m_n) g_A/ (2
F_{\pi^+})$, where $g_{\pi^+ pn}$ describes the vertex $p \to n \pi^+$,
$F_{\pi^+}=92.3(1)$~MeV~\cite{ParticleDataGroup:2024cfk}\footnote{We 
employ the experimental result from Ref.~\cite{ParticleDataGroup:2024cfk} 
based on the (charged) pion weak decay, $\pi^+\to\bar \nu_\mu \mu^+$, 
while Ref.~\cite{ParticleDataGroup:2024cfk} also quotes the lattice-driven 
estimate, $92.07(85)$~MeV.} is the charged pion decay 
constant from $\pi^\pm\to\ell\nu$ decay, 
$m_p$ and $m_n$ are the proton and neutron masses, respectively, 
and $g_A=1.2753(13)$~\cite{ParticleDataGroup:2024cfk} is the axial nucleon 
coupling constant for $\beta$-decay $n \to p + e^- + \bar \nu_e $. 
Note that we take $g_A=\lambda$ from PDG~\cite{ParticleDataGroup:2024cfk}. 
Such value is potentially affected by radiative corrections that must be 
subtracted. Former estimates, such as Ref.~\cite{Gorchtein:2021fce},
found negligible $\mathcal{O}(10^{-4})$ corrections. However, missing terms 
affecting the strong vertex found in Ref.~\cite{Cirigliano:2022hob} 
are potentially larger, as later confirmed in Refs.~\cite{Gorchtein:2023srs,Seng:2024ker} 
(see additional references therein, and advances in lattice QCD in Ref.~\cite{Ma:2023kfr}).
In our chosen isospin-breaking scheme, such QED corrections ---necessary to render the physical 
masses--- are implicit in $g_A$ and should not be subtracted.\footnote{In this regard, our chosen $g_A$ differs from its value in the isospin-symmetric limit. Note also that corrections in Ref.~\cite{Gorchtein:2021fce} are specific of neutron decay, by contrast to the strong vertex ones.}
For details on this, the reader is referred to Appendix~\ref{app:ib}.
The GT relation yields the value
$g_{\pi^+ pn}|_{\rm GT}= 12.97(2)_{F_\pi}$ and is a direct consequence of
both Partial Conservation of the Axial Current (PCAC) at the hadronic
level~\cite{Nambu:1960xd} and the fact that in the isospin limit the
vector weak current of the neutron $\beta$-decay corresponds to the
isospin rotated strong conserved current~\cite{Feynman:1958ty}. 
The Goldberger--Treiman
discrepancy is defined as
\begin{eqnarray}\label{eq:GT}
\Delta_{\rm GT}=1-\frac{m_N g_A}{g_{\pi^+ p n} F_{\pi^+}} \, .
\end{eqnarray}
Note that, although this quantity relies on at least four independent 
(direct or indirect) measurements ---$g_{\pi^+ pn}$, $F_{\pi^+}$, $m_N$, and $g_A$--- 
rather than on a single experiment, it has traditionally been interpreted 
as a direct quantitative measure of chiral symmetry departures. 
While we will be dealing specifically with the charged pion case, 
we will often assume isospin conservation and employ 
$g_{\pi NN}$ and $F_\pi$ for ease of notation.

Historically, early attempts to estimate $g_{\pi NN}$
 based on unsubtracted dispersion
relations for the pseudoscalar form factor of the nucleon 
---originally postulated by Nishijima~\cite{nishijima1964unsubtracted} and
further developed by Pagels~\cite{Pagels:1969ne}--- were unable to 
reproduce the relatively large GT discrepancy of around
$8\%$ reported at that time. These analyses included models with intermediate $\rho \pi$ 
and $\sigma \pi$ states dominated by an excited pion resonance, proposed 
by Coleman and Moffat~\cite{Coleman:1969ulw}; current algebra by Pagels and
Zepeda~\cite{Pagels:1972xx}; and the use of a Veneziano-type 
pseudoscalar form factor by Dominguez~\cite{Dominguez:1973jj}. 
Much of the confusion in the old days was actually induced by
the large ``experimental'' GT discrepancy at the time, opposite to the currently
accepted rather small value of about $2\%$, which in fact strongly
reinforces Pagels' original claim that ``chiral symmetry is the best
symmetry in strong interactions after isospin''~\cite{Pagels:1974se}.

Noteworthy, it is the extended PCAC (EPCAC) of
Dominguez~\cite{Dominguez:1976ut,Dominguez:1977nt,Dominguez:1977en}
which provides the field theoretical basis of pseudoscalar meson
dominance (for reviews covering up to the mid 80s see
e.g.~\cite{Pagels:1974se,Dominguez:1984ka} and references therein).

The least accurately known parameter entering the GT discrepancy is
$g_{\pi^+ pn}$, so benchmarking $\Delta_{\rm GT}$ or $g_{\pi^+ pn}$
are somewhat equivalent. The direct and precise determination of the
pion--nucleon coupling constant based on large $\pi N$ and $NN$
scattering databases has a long history (see
\cite{deSwart:1997ep,Sainio:1999ba} for a compilation up to late 90s
and \cite{Matsinos:2019kqi} for a recent historical
perspective). Initially, $\pi^- p$ scattering was used and, over the
years, the combined $np$+$pp$ scattering analysis was gradually
invoked~\cite{deSwart:1997ep,RuizArriola:2016ehc}. Thus, using
pionic-hydrogen and the Goldberger--Miyazawa--Oehme (GMO) 
sum rule, that rigorously isolates $g_{\pi^+pn}$ as 
the residue from the pole, $g_{\pi^+ pn}=13.3(1)$ 
was found in Ref.~\cite{Ericson:2000md} (implying $\Delta_{\rm GT} = 2.5(7)\%$). 
Subsequent studies, including important radiative corrections~\cite{Baru:2011bw}, obtained
$g_{\pi^+ pn}=13.12(10)$~\cite{Baru:2010xn}, corresponding to
$\Delta_{\rm GT}= 1.1(7) \%$. Its 
recent update reads $13.11(10)$~\cite{Hoferichter:2023ptl} and 
therefore $\Delta_{\rm GT}= 1.0(7)\%$ , i.e. a value {\it almost} compatible
with zero.  
Besides, a combined $np$+$pp$ fit to the $3\sigma$
self-consistent Granada-2013 database comprising $6700$ selected
scattering measurements below $350$ MeV ~\cite{NavarroPerez:2013mvd}
yields the value $g_{\pi^+ pn}= 13.25(5)$ with a benchmark accuracy of
$0.3\%$ in a phenomenological approach~\cite{NavarroPerez:2016eli}
and $\Delta_{\rm GT}= 2.1(4) \%$. This result has been confirmed
independently by a fit to a subset of the Granada-2013 database using
Chiral Perturbation Theory in Weinberg's counting up to ${\rm
  N^4LO^+}$~\cite{Reinert:2020mcu}. It must be noted 
that account of radiative corrections is not at the same level as 
in Refs.~\cite{Baru:2011bw,Hoferichter:2023ptl}. This estimate is not far, and in fact
compatible with, Bethe's original estimate. It should be stressed
the considerable complexity and the amount of data required to achieve
these high precisions. The situation is summarized in
Fig.~\ref{fig:gpiNN} where we collect some high precision ($\le 1 \%$)
determinations over time of $g_{\pi^+ p n}$ as well as its GT values
corresponding to the ratio $g_A m_N/F_{\pi^+}$ from PDG yearly averages of
$g_A$, $m_N$ and $F_{\pi^+}$ with their estimated uncertainties. It is
noteworthy that the current overall $2.1(4)\%$ accuracy of the GT
relation is {\it small but significant}.  All these phenomenological
findings are based on data analysis at the purely hadronic level, but
no explicit relation to QCD is invoked.

From the point of view of QCD, the GT relation is regarded as an exact
theorem in the (chiral) limit where the light $u$- and $d$-quarks, and hence the pion,
become massless and the classical QCD Lagrangian becomes invariant
under the chiral symmetry $SU (2)_R \otimes SU (2)_L$ group.  For
finite quark masses this relation is subjected to corrections, so that
the current $2\%$ accuracy of the GT relation not only justifies
the old claim by Pagels, but also suggests computing the discrepancy
separately from an independent source.

While the recent  results from phenomenological approaches~\cite{NavarroPerez:2013mvd,NavarroPerez:2016eli,Reinert:2020mcu} provide rather accurate
results, {\it ab initio} determinations such as lattice QCD do not yet
provide convincingly accurate results, due to the observed deviations from
PCAC, often attributed to excited-states contamination or uncertainties 
on lattice extrapolations. Among them, we find 
$g_{\pi NN}=12.9(8)/14.8(1.8)$ from RQCD~\cite{RQCD:2019jai}, 
depending on extrapolation details; 
$g_{\pi NN} =13.5(3)$ from ETMC~\cite{Alexandrou:2020okk}, 
albeit with the assumption of pion-pole dominance, in 
flagrant violation with PCAC;
$g_{\pi NN} = 12.4(1.2)$ from NME~\cite{Park:2021ypf};
$g_{\pi NN}=14.1(1.2)$ from PNDME~\cite{Jang:2023zts};
$g_{\pi NN} = 13.3(7)/12.6(9)$ from ETMC~\cite{Alexandrou:2023qbg}, 
depending on the treatment of excited states contamination; and, 
more recently, promising results have been obtained by the PACS 
collaboration, obtaining $g_{\pi NN} = 12.8(4)/13.2(3)$ 
for different ensembles.\footnote{Note that lattice 
calculations are almost exclusively performed with an isospin limit
defined by the mass of the neutral pion. As such, these may
be subject to relevant QED corrections outlined in Ref.~\cite{Cirigliano:2022hob}.
}

In this paper we will make a determination of
the Goldberger--Treiman discrepancy, and consequently of $g_{\pi NN}$,
by exploiting its relation to the pseudoscalar form factor and the
concept of pseudoscalar meson dominance, which is achieved by using a
suitable dispersion relation in the $J^ {PC} I^G = 0^{-+} 1^- $
channel. We will establish, based on perturbative QCD (pQCD) valid at
sufficiently high energies, that the corresponding spectral function
{\it must} change sign above the $3\pi$ threshold. This is a
consequence of three sum rules, which require a minimal number of one 
pseudoscalar resonance and some high-energy contribution above. These 
can be naturally associated to the $\pi(1300)$ and a Regge tail from 
heavier resonances, eventually matching the pQCD behaviour. 
We will also show that, although there is 
some arbitrariness in describing resonance profiles, the sum
rules guarantee a robust result when comparing different
lineshapes. Finally, we provide an analysis of the pseudoscalar form
factor, that can be compared with available lattice QCD
results. The present work improves over
Ref.~\cite{RuizArriola:2023xap}, where an incorrect high-energy
behaviour was assumed, and hence our numerical estimates there,
although not fully incompatible with the present ones, should be
considered less reliable.

We emphasize  that the pseudoscalar form factor discussed 
in this work should not be identified with those appearing in 
alternative estimates of $g_{\pi NN}$ based on nucleon--nucleon
interactions derived from microscopic models based on meson 
exchange~\cite{Erkelenz:1974uj,Nagels:1977ze,Machleidt:1987hj}, 
that require phenomenological strong form factors to regularize 
the theory (a feature already recognized by Bethe~\cite{Bethe:1940iba,Bethe:1940zz}; 
for a modern perspective in renormalization, see Ref.~\cite{CalleCordon:2009pit}). 
These form factors introduce a momentum-dependent coupling of the mesons
to the $N\bar{N}$ system, and their residue at the pion pole 
defines the pion--nucleon coupling constant. They 
have been estimated with several models (see \cite{Cohen:1986ux,Melde:2008dg,Coon:1990fh}). 
Other approaches estimating $g_{\pi NN}$ through its connection
to the pseudoscalar form factor include QCD sum
rules~\cite{Meissner:1995ra,Birse:1995zh} as well as studies based on
Dyson--Schwinger equations~\cite{Eichmann:2011pv,Chen:2021guo}, the
latter (and more recent) one obtaining a Goldberger–Treiman discrepancy of
$\Delta_{\rm GT} =3.0(1)\%$. An SU(3) breaking pattern of GT
discrepancy was analysed in Ref.~\cite{Goity:1999by}. A range bound of
$ 1.5\% < \Delta_{\rm GT} < 2.2\% $ has been obtained using QCD sum
rules~\cite{Nasrallah:1999fw}.

While many works adopt the axial and induced pseudoscalar form factors 
---derived from the matrix element of the nucleons with the axial current--- 
as primary quantities, obtaining the pseudoscalar one via the 
chiral Ward identity, in this paper we {\it only} use the pseudoscalar 
form factor.\footnote{Besides, analysing the induced pseudoscalar 
form factor would require computing chiral corrections to the axial 
form factor in pQCD. We leave such a study for future work. 
Unlike the induced pseudoscalar form factor, this 
has well-defined quantum numbers $1^-0^{-+}$ and is better suited 
for a dispersive analysis.
}

The paper is organized as follows: in Sect.~\ref{sec:form} we present
the general properties of the isovector pseudoscalar 
form factor of the nucleon, outlining its connection to the 
GT discrepancy based on analytical properties. The latter are 
analysed in terms of a spectral function, which
corresponds to the $s$-channel discontinuity in the $N \bar N$ for the $1^{-}
0^{-+}$ channel.  This spectral function fulfils a set of sum rules
based on chiral symmetry and pQCD, that allow to show that this function must
have at least one zero. This information allows obtaining a lower
bound for the GT discrepancy. In Sect.~\ref{sec:model} we decompose
the spectral function into three main contributions, with the 
low-energy end ---described in terms of ChPT--- and the asymptotic
contribution ---inferred from pQCD--- being negligible, as we will show. The
intermediate energy region is described in terms of 
the well-established PDG resonance $\pi(1300)$
and a Regge tower of states which can be well approximated by a negative
fractional power, $ \sim s^{-2-2 \epsilon}$, with $\epsilon \sim
0.1-0.2$ required by matching to pQCD. In Sect.~\ref{sec:results} we
present our numerical results for the GT discrepancy 
and the pseudoscalar form factor, that can be compared to the lattice
QCD calculations. Finally, in Sect.~\ref{sec:concl} we come to our main
results and conclusions and present our outlook for further study.
Further information is relegated to the appendices, 
including the pQCD asymptotics, discussions of strong form factors, 
the Regge region, isospin breaking, and finite-width effects.

\section{Formalism}
\label{sec:form}

\subsection{General properties}

In the following, we introduce the axial and pseudoscalar 
currents and their associated nucleon form factors, highlighting their connection 
to the GT discrepancy, which will be exploited in the following sections.
We start by discussing basic and well-known facts in order to clearly state 
our problem and to fix our notation. For quark fields
$q_i$ and $q_f$ with flavors $i$ and $f$ respectively, PCAC in pure QCD
reads, for non-singlet currents,
\begin{eqnarray}\label{eq:WInoEM}
\partial_\mu ( \bar q_f \gamma^\mu \gamma_5 q_i) = (m_f+m_i) \bar q_f i \gamma_5 q_i  \, .
\end{eqnarray}
The right-hand side is called the pseudoscalar density, which has
$J^{PC}=0^{-+}$ quantum numbers. Assuming for the moment isospin
symmetry,\footnote{The modifications due to strong ($m_u \neq m_d$) isospin breaking in essence 
amounts to the $2m_q \to m_u +m_d$ replacement for the charged current here discussed. 
For $\alpha$ corrections, we refer to App.~\ref{app:ib}.} 
i.e.  $m_u=m_d \equiv m_q $ and $\alpha\to 0$, the
pseudoscalar density for light $u,d$ flavours becomes $\bar q
\frac{\vec\tau}{2} i \gamma_5 q $ and it has $I^G=1^- $ quantum numbers.
Therefore, at the hadronic level it has a non-vanishing overlap with
any state with an odd number of pions, $\pi, 3 \pi, 5\pi, \dots$ and
the vacuum. We define 
\begin{equation}
  \vec A^\mu  = \bar q \frac{\vec \tau}2  \gamma^\mu \gamma_5  q , \quad   
  \vec P = \bar q \frac{\vec \tau}{2}  i \gamma_5  q ,  \quad
  \partial_\mu \vec A^\mu = \bar{q} \{ \frac{\vec{\tau}}{2} , \mathcal{M}_q \}i\gamma_5 q =   2 m_q \vec P \, .
\end{equation}
Therefore, the local pseudoscalar density couples to {\it all} $0^{-+}$
states and in particular to any odd and local combination of
(strongly) stable pion fields, so that
\begin{eqnarray}
2 m_q \bar q(x)  \frac{\vec \tau }{2} i \gamma_5 q(x) =2m_q\vec{P}= F_\pi  M_\pi^2 \vec\phi_\pi(x) + {\cal O}(\phi_\pi^3)  \, ,
\end{eqnarray}
with $F_\pi$ the pion weak decay constant and $\vec \phi_\pi(x)$ the
canonical pion field, i.e. the one for which the two point function has 
unit residuum at the pion pole. Neglecting higher order pion fields corresponds
to the pion-pole dominance approximation (PPDA), featuring the pre-QCD PCAC
where $\partial_\mu \vec J_A^\mu(x) = F_\pi M_\pi^2 \vec \phi_\pi(x)$. 
In the pion-dominance approximation, one may deduce some useful but approximate properties,
which will not be assumed here (find further comments in Sect.~\ref{sec:resFF}).

Matrix elements between nucleon states with initial and final momenta
$p$ and $p'$, respectively, correspond to the nucleon form factors. In particular, for the axial-current one has 
\begin{eqnarray}\label{eq:AxMatElIS}
\langle N(p') | \vec A^\mu |N(p)  \rangle = \bar u(p') \frac{\vec \tau}2  \left[ G_A (q^2) \gamma^\mu \gamma_5 + \frac{q^\mu }{2m_N} G_P(q^2) 
   \gamma_5 \right] u(p) \, ,
\end{eqnarray}
where $q_{\mu}=p'_{\mu}-p_{\mu}$ is the momentum transfer and 
$u(p)$are nucleon (proton or neutron), $N=(p,n)$, Dirac spinors 
and $G_A(q^2)$and $G_P(q^2)$ are the axial and induced pseudoscalar 
form factors, respectively. Likewise, for the pseudoscalar density,
  \begin{eqnarray}
    \langle N(p') | \vec P |N(p)  \rangle = 
    \bar u(p') \frac{\vec \tau}{2} i \gamma_5 u(p) F_P (q^2) \, , 
  \end{eqnarray}
where $F_P(q^2)$  is the pseudoscalar isovector nucleon form factor. 
From PCAC, we have the relation 
  \begin{equation}\label{eq:PCAC}
  2 m_N G_A(q^2) + \frac{q^2}{2m_N} G_P(q^2) = 2m_q F_P (q^2) \, ,  
  \end{equation}
where $G_A(0) \equiv g_A$. From the absence of a massless pole 
in $G_P(q^2)$, we get the normalization condition
  \begin{eqnarray}
    2m_N g_A = 2m_q F_P(0) \, .
  \end{eqnarray}
In the chiral limit, conservation of the axial current demands 
($M_\pi\to 0$ understood)
\begin{equation}
  2m_N G_A(q^2) +\frac{q^2}{2m_N}G_P(q^2) = 0 \rightarrow \frac{G_P(q^2)}{2m_N} = \frac{2m_N G_A(q^2)}{M_{\pi}^2 -q^2} \, ,
\end{equation}
also known as pion-pole dominance (PPD) and leading to the GT
prediction upon correct residue identification (see definitions and comments below).

The process in the crossed $s$-channel, $s=(p+\bar p)^2$, reads 
\begin{eqnarray}
  \langle \bar N(\bar p) N(p) | m_q\vec P (0) | 0  \rangle =  \bar u(p) \frac{\vec \tau}{2} i \gamma_5 v(\bar p) m_q F_P (s) ,
\end{eqnarray}
which has $ I^G J^{PC} = 1^- 0^{-+} $ quantum numbers. Inserting a
complete set of states in this channel one gets, schematically,\footnote{Note 
that, for single-particle states, $\int d\Pi_n (2\pi)^4\delta^{4}(p_n -p -\bar{p})\to \delta(M_n^2 - s)$.}
\begin{eqnarray} \frac1{\pi}
 {\rm Im}  \langle \bar N N | P  |0 \rangle= 
 \sum_{n= 1^- 0^{-+}} \int d\Pi_n (2\pi)^4\delta^{4}(p_n -p -\bar{p})
 \langle \bar N N |n \rangle^{\dagger} \langle n | P  |0 \rangle \, ,
\end{eqnarray}
where $n= \pi, 3\pi, 5 \pi, \dots $ are {\it stable} hadronic
states. The first non-trivial contribution corresponds to $3\pi$
states in the $ I^G J^{PC} = 1^- 0^{-+} $ channel, which might
eventually correspond to the excited pion resonances
$\pi(1300), \pi(1800), \dots$.

In the high-energy space-like region, $t \gg 0 $ with $t=-s$, the form factor 
can be studied in the framework of pQCD. 
A great deal of work has been devoted to the case of electromagnetic 
form factors (see the seminal and recent works in 
Refs.~\cite{Lepage:1979za,Korenblit:1979cw,Lepage:1980fj,Chernyak:1984bm,Huang:2024ugd,Chen:2024fhj}) 
and, to a lesser extent, to the axial-vector case~\cite{Brodsky:1980sx,Carlson:1985zu}, predicting the scaling behaviour $G_{A(P)}(-t) \sim \alpha_s^2(t)/t^{2(3)}$. 
Surprisingly, to the best of our knowledge the case of the pseudoscalar (or scalar) form factor has not been studied.
From PCAC, Eq.~\eqref{eq:PCAC}, and the knowledge of $G_{A,P}$ there are two possibilities for the high-energy scaling behaviour: $t^{-2}$ or $t^{-3}$. 
Note in this respect that, for instance, Ref.~\cite{Dominguez:1973jj} assumes a similar scaling to that of $G_A(t)$, 
whereas the recent Ref.~\cite{RQCD:2019jai} (incorrectly) quoted from Ref.~\cite{Alabiso:1974ye} a $t^{-3}$ scaling, emphasizing 
the relevance of clarifying this situation.\footnote{In fact, this is a misquote. In particular, Ref.~\cite{Alabiso:1974ye} 
finds such behaviour for a {\it pseudoscalar} quark-quark interaction, as opposed to the actual {\it vector} quark-quark interaction due to one-gluon exchange.}
Indeed, such scaling will be critical in our analysis and for any kind of dispersive approach aiming to describe this form factor. 
To such endeavour, we have evaluated the leading-order pQCD prediction, obtaining as a result that $F_P \sim \alpha^2_s(t)/t^{2}$ in 
the deep space-like region. The details and comments on such results are relegated to App.~\ref{sec:pQCD}.
For completeness, we note that in the high-energy space-like region 
the general anatomy of the form factor takes the form~\cite{Braun:2006hz}
\begin{eqnarray}
 m_q F_P(t) = A(t) + \frac{\alpha(t)}\pi \frac{B(t)}{t} +  \left(\frac{\alpha(t)}\pi\right)^2 \frac{C}{t^2} + ... \, ,
 \quad A(t)\sim t^{-3}, \quad B(t)\sim t^{-2}  \, , 
\end{eqnarray}
where the terms $A(t)$ and $B(t)$ represent power-suppressed contributions, 
and the $C$-term corresponds to the leading pQCD prediction, as computed 
in App.~\ref{sec:pQCD}.
While the $C$-term dominates at asymptotically large $t$, its numerical 
relevance at accessible energies depends on the onset of the perturbative 
regime ---a topic that remains under active investigation.

Finally, we turn our attention to the quantity of interest in this study, 
$g_{\pi NN}$ and the GT discrepancy. The former can be defined in terms of the 
(amputated) $\pi NN$ vertex function $\Gamma_{\pi NN} (q,p',p)$ which, 
on the mass shell, provides the definition of the $g_{\pi NN}$ coupling constant
\begin{eqnarray}\label{eq:gpinnDEF}
 g_{\pi^i NN} \bar u(p') i \tau^i \gamma_5 u(p) =  \Gamma_{\pi^i NN} (q,p',p)|_{q^2=M_\pi^2,p'^2=p^2= m_N^2} \, ,
\end{eqnarray}  
which corresponds to an unphysical point where all the three particles are 
simultaneously on the mass shell. With such definition, the relation 
\begin{equation}
 M_\pi^2 F_\pi g_{\pi NN} = \lim_{s\to M_{\pi}^2}(M_\pi^2 -s) m_q F_P(s) 
\end{equation}
can be readily obtained. In consequence, the following definition will prove useful 
\begin{equation}
  m_q F_P(s) = \frac{M_\pi^2}{M_\pi^2 -s}D(s) . \label{eq:Dt}
\end{equation}
Such form factor has the following properties: (i) $D(0) = m_N g_A$ ;
(ii) $D(M_\pi^2) = F_\pi g_{\pi NN}$; (iii) asymptotically, in the 
deep space-like region, $D(-Q^2) \sim Q^{-2}$. 
The second property allows to express 
$\Delta_{\rm GT} = 1 - g_A m_N/(F_\pi g_{\pi NN}) = 1 -D(0)/D(M_{\pi}^2)$, 
independent both of $g_A$ and $m_N$ and providing a clean way to access the GT 
discrepancy on the lattice. Note indeed that this relation has been employed in the context of 
Dyson--Schwinger equations~\cite{Chen:2021guo} to study $\Delta_{\rm GT}$. 
Finally, we emphasize once more that the form factor $D(s)$, often
noted as $F_{\pi} G_{\pi NN}(s)$, should not be identified with that 
appearing in $NN$ potentials in meson exchange models, see App.~\ref{sec:EPCAC}.

\subsection{Analytic properties, dispersion relations and sum rules}\label{sec:test}

A dispersive representation for $m_q F_P(s) $ was proposed long
ago~\cite{nishijima1964unsubtracted}, which is ultimately justified by
pQCD~\cite{Alvegard:1979ui,Lepage:1979za,Brodsky:1980sx} (see also App.~\ref{sec:pQCD}).  
Based on quark-hadron duality, the pseudoscalar form factor $ m_qF_P (s) $ satisfies useful analytical
properties in the complex $s-$plane, which we list in what follows:\footnote{In
the time-like region, the form factor corresponds to the process $\nu_e
e^+ \to p \bar n $ which takes place for $\sqrt{s} \ge m_p + m_n$.}
\begin{enumerate}
  
\item $m_q F_P(s)$ is real analytic in the
  space-like region, $s= q^2 = -Q^2 \le 0 $ .
\item  It has a pion pole at $s=M_\pi^2$ , with residue $ F_\pi M_\pi^2 g_{\pi NN}$ .
\item   It has branch cuts along the odd number of pions, $\sqrt{s}= 3M_\pi, 5 M_\pi , \dots ,$ 
  corresponding to the process below the $N \bar N $ threshold. 
\item Its value at the origin is $m_q F_P(0)= m_N g_A$.
\item It falls off as $ \sim m_q  (\alpha_s (Q^2)/Q^2)^2 $ 
      for $Q^2 \to \infty $, see App.~\ref{sec:pQCD}.
\end{enumerate}    
With these properties one can write the following dispersion
relation~\cite{nishijima1964unsubtracted} 
\begin{eqnarray}
m_q  F_P (s) = \frac{F_\pi M_\pi^2 g_{\pi NN}}{M_\pi^2-s}+
  \frac1{\pi} \int_{(3M_\pi)^2}^\infty d\zeta \frac{ m_q \operatorname{Im}F_P(\zeta)}{\zeta-s} \, ,
\end{eqnarray}  
where $2 i \operatorname{Im} m_qF_P(s)= m_qF_P(s+i 0^+)-m_qF_P(s-i 0^+) = \operatorname{Disc} m_q F_P(s)$ 
is the discontinuity across the branch cut, starting at 
$\sqrt{s}= 3M_\pi$ and $g_{\pi NN}$ is the pion--nucleon coupling
constant. With such definition, conditions 4 
(normalization) and 5 (asymptotics) imply the following results, known as sum rules:
\begin{eqnarray}
m_N g_A &=& F_\pi g_{\pi NN}+ \frac1{\pi}\int_{(3
  M_\pi)^2}^{\infty} ds \ \frac{\operatorname{Im} m_q F_P(\zeta)}{\zeta} \, , \label{eq:SR1org}\\
0&=&  F_\pi M_\pi^2 g_{\pi NN} + \frac1{\pi}\int_{(3 M_\pi)^2}^{\infty} d\zeta \  \operatorname{Im} m_q F_P(\zeta) \, , \\ 
0&=&  F_\pi M_\pi^4 g_{\pi NN} + \frac1{\pi} \int_{(3 M_\pi)^2}^{\infty} d\zeta \ \operatorname{Im} m_qF_P(\zeta) \, \zeta \, . 
\end{eqnarray}
Note that the second condition arises from the vanishing of 
the $Q^{-2}$ term, whereas the third condition stems from 
the vanishing of the $1/Q^4$ terms, since pQCD requires 
additional log suppression $\sim 1/[Q^4 \log^2 (Q^2/ \Lambda^2) ]$.
Note however that care needs to be taken to ensure that such a sum 
rule does not imply an (overdamped) $Q^{-6}$ behaviour, 
as it was the case in Ref.~\cite{RuizArriola:2023xap} 
(see also App.~\ref{app:DRasymp}).
Note in addition that the first sum rule implies the relation
\begin{equation}\label{eq:GTSR}
  \Delta_{\rm GT} = \frac{x}{1+x}, \qquad x = \frac{1}{m_N g_A}\frac1{\pi}\int_{(3 M_\pi)^2}^{\infty} d\zeta \ \frac{\operatorname{Im} m_q F_P(\zeta)}{\zeta} \, .
\end{equation}

\subsection{Non-positivity of the spectral function}

In our case, we are interested in the process $N \bar N \to X= \pi,
3\pi, 5\pi, \dots$ in the channel with pion quantum numbers and we 
may define the spectral function
\begin{eqnarray}
  \rho(s) =\frac{1}{\pi}  \operatorname{Im} m_q F_P(s) . 
\end{eqnarray}
As we see, for $\Delta_{\rm GT} > 0$, the three integrals involving
the spectral function suggest individually  that it could be negative 
all the way up to $\infty$. 
However, in assuming so, we will show that one 
runs into $\Delta_{\rm GT}> 1$, i.e., an opposite sign for $g_{\pi NN}$ 
and a flagrant departure from the chiral prediction. Specifically, introducing
\begin{equation}
\rho(s) = \left[ {\rm
  sign} (\rho(s)) \sqrt{|\rho(s)|s } \right] \sqrt{\frac{|\rho(s)|}s} \equiv f(s) g(s) \, , 
\end{equation}
where $f(s)$ and $g(s)$ are obviously defined, for either sign choice (i.e., 
for $\rho(s) = \pm |\rho(s)|$),  we get 
\begin{eqnarray}
  ( F_\pi M_\pi^2 g_{\pi NN})^2 &=& \Big|\int ds \rho(s) \Big|^2  
  = \Big|\int ds f(s) g(s) \Big|^2 
  \le \int ds |f(s)|^2 \int ds |g(s)|^2 \nonumber \\ &=& 
  \int ds |\rho(s)| s \int ds |\rho(s)|/s \nonumber 
  = \int ds \rho(s) s \int ds \rho(s)/s \nonumber \\ &=& 
  F_\pi M_\pi^4 g_{\pi NN} ( F_\pi g_{\pi NN}- m_N g_A ) 
  =  ( F_\pi M_\pi^2 g_{\pi NN})^2 \Delta_{\rm GT} \, ,
\label{eq:schwarz}  
\end{eqnarray}
where Schwartz's inequality has been explicitly used and 
the three sum rules have been inserted. Thus, dividing by the first term, we get
the final inequality $1 \leq \Delta_{\rm GT}$.\footnote{
Note this corrects the misleading statement in our previous 
proceedings~\cite{RuizArriola:2023xap} stating that this 
inequality implies $\Delta_{\rm GT} \leq 0$.} Thus, the assumption 
$\rho(s) = \pm |\rho(s)|$ implies $\Delta_{\rm GT} >1$ and, 
therefore, $g_{\pi NN}<0$, i.e., a flagrant departure from the chiral result. 
Thus, the spectral function {\it must} change sign at least once. 
As we will see below, its sign is well-defined close 
to threshold $(\sqrt{s} \to 3 M_\pi)$, where it is  negative, 
so there must be at least one value $s_0$ where $\rho(s_0)=0$.  
At high-energies, its sign 
is subject to significant uncertainties related to the limited knowledge 
on the nucleon distribution amplitude (DA). 
Nevertheless, most parametrizations predict a positive sign 
(see discussions in App.~\ref{sec:pQCD}), in agreement with the suggestions above.

\subsection{An improved bound on \texorpdfstring{$\Delta_{\rm GT}$}{DeltaGT}}
\label{sec:bound}

As we have seen, the spectral function {\it must} change sign for
$g_{\pi NN}> 0$. If $s_0$ corresponds to the first zero, we can apply
apply Schwarz's inequality in the region above $3\pi$-threshold and $s_0$,
where $\rho(s)$ has a well-defined sign, 
so we have instead of the inequalities of Eq.~\eqref{eq:schwarz}  
\begin{eqnarray}
&&  \left[ F_\pi M_\pi^2 g_{\pi NN}+ \int_{s_0}^\infty ds \rho(s) \right]^2 = \Big|\int_{(3M_\pi)^2}^{s_0} ds \rho(s) \Big|^2  \le 
  \int_{(3M_\pi)^2}^{s_0} ds \rho(s) s \int_{(3M_\pi)^2}^{s_0} ds \frac{\rho(s)}{s}  \nonumber \\ &&= 
  \left[ -F_\pi M_\pi^4 g_{\pi NN}-  \int_{s_0}^\infty  ds \rho(s) s\right]
  \left[ m_N g_A - F_\pi g_{\pi NN}- \int_{s_0}^\infty  ds \frac{\rho(s)}{s} \right] \, .
\label{eq:schwarz2}  
\end{eqnarray}
Defining the dimensionless moments,
\begin{eqnarray}
\langle R x^n \rangle = \frac{1}{F_\pi  M_\pi^2 g_{\pi NN}}\int_{s_0}^\infty ds \rho(s) \left( \frac{s}{M_\pi^2} \right)^n   \, ,
\end{eqnarray}
we get 
\begin{eqnarray}
  (1 + \langle R \rangle)^2 \le (\Delta_{\rm GT} + \langle \frac{R}{x}\rangle) (1 + \langle R x \rangle )  \, ,
\label{eq:schwarz3}  
\end{eqnarray}
which becomes a rigorous upper bound for $\Delta_{\rm GT}$,
\begin{eqnarray}
  \Delta_{\rm GT}  \ge  
  \frac{(1 + \langle R  \rangle )^2}{1+\langle R x\rangle} - \langle \frac{R}{x} \rangle \, .
\label{eq:GTbound}  
\end{eqnarray}
This inequality becomes an equality if and only if the spectral
function for $s>s_0$ is infinitely narrow, $\rho(s) = Z
\delta(s-M^2)$. This fact will be shown to be useful in our numerical
estimates.  In the following, we make use of the properties outlined
above to construct a model for the spectral function and to analyse
the implications for $\Delta_{\rm GT}$ and the pseudoscalar form factor
$m_q F_P(s)$.

\section{Modelling the spectral function}
\label{sec:model}

\begin{figure*}[ttt]\centering
\includegraphics[width=0.6\textwidth]{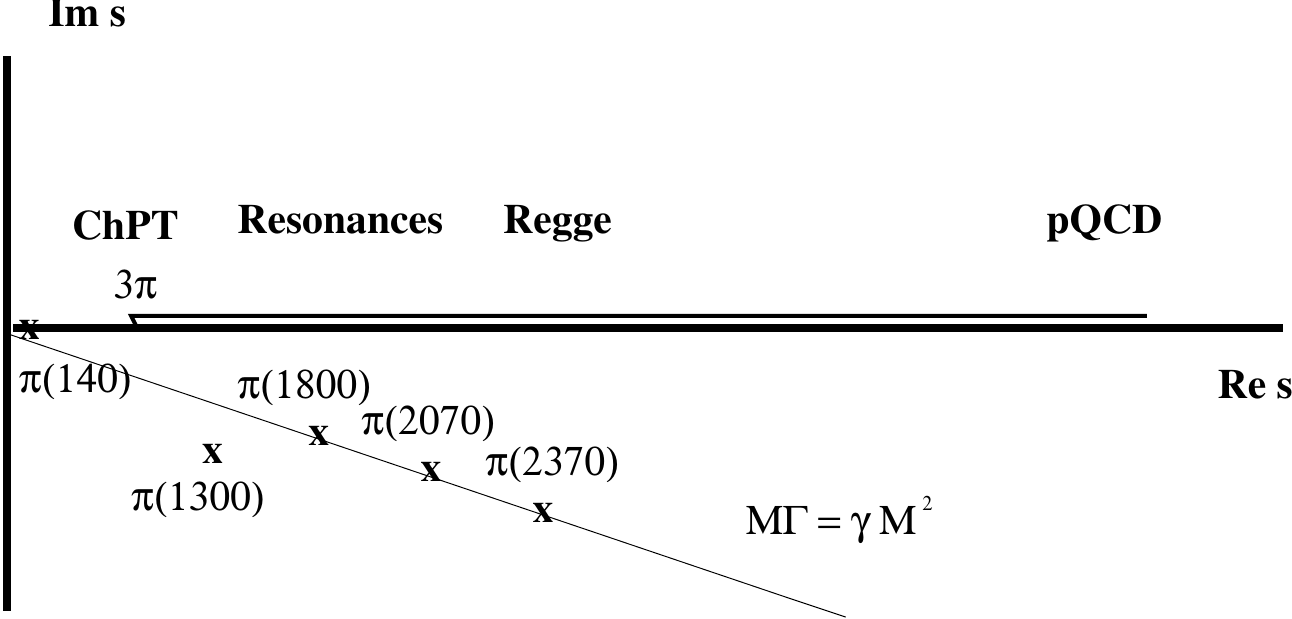}
\caption{ \small Cartoon of the spectral density in the complex $s$-plane  (arbitrary scale) and its different regions, 
  showing the pion pole and the resonance poles (in the second Riemann sheet across the $3\pi$-cut). We also show the assumed
asymptotic Suranyi's radio $\Gamma /M = \gamma$ (see discussion in main text).}
\label{fig:axis}
\end{figure*}

We remind here that, as in any dispersive approach, the issue of
completeness vs predictive power becomes relevant. This is
ameliorated with the imposition of the sum rules. Hence, our analysis 
below is based on the separation of the spectral function into four 
different regions shown in Fig.~\ref{fig:axis}:
\begin{enumerate}
\item The low-energy region, $ 3 M_\pi \le \sqrt{s} \le \Lambda_\chi$, where we expect ChPT to hold. 
\item The intermediate-energy region, $ \Lambda_\chi \le \sqrt{s} \le \Lambda_R $, where we expect non-overlapping resonances to  play the dominant role. 
\item The high-energy Regge region, $ \Lambda_R \le \sqrt{s} \le \Lambda_{\rm pQCD} $, where resonances pileup. 
\item The asymptotically high energy region, $ \Lambda_{\rm pQCD} \le \sqrt{s} $, where pQCD applies.
\end{enumerate}

As we will show first, the extreme regions ---ChPT and pQCD--- have a
negligible impact on our results.  The dominant contribution instead
arises from the intermediate resonances and Regge regions, which are
subsequently discussed in detail.

\subsection{The extreme regions: ChPT and pQCD}

In order to investigate the physics close to the $3\pi$ threshold 
and to assess whether relevant effects are expected in our approach, 
we make use of ChPT. Earlier attempts in this 
regard were carried out by Pagels and Zepeda long ago~\cite{Pagels:1972xx}, 
unable to explain the large $\Delta_{\rm GT} = 8\%$ reported at that time. 
More recently, the relevant spectral functions $G_{A,P}$ have 
been analysed using covariant ChPT in Refs.~\cite{Bernard:1996cc,Kaiser:2003dr,Kaiser:2019irl}, 
that we use to extract $m_q F_P(s)$ via the PCAC relation, Eq.~\eqref{eq:PCAC}.
Importantly, Ref.~\cite{Kaiser:2019irl} found out that the heavy baryon 
approach overestimated the spectral functions by a factor of 2 or 3. 
We therefore employ the following parametrization that describes well 
the heavy baryon limit and account for such correcting factor: 
\begin{equation}\label{eq:SpectralChPT}
\operatorname{Im} m_q F_P(s)|_{\rm Ch PT} = 
\frac{M_{\pi }^2 m_N g_A}{9\pi^3(8 F_{\pi })^4}
\frac{ 1+ g_A^2\left(5+\frac{68 \pi ^2}{35}\right) }{M_{\pi }^2-s}
   (\sqrt{s}-3 M_{\pi })^2 (8 M_{\pi }+\sqrt{s})^2 .
\end{equation}
As usual in ChPT, the result above is ill-behaved at high energies 
and should not be taken seriously above some chiral scale $\Lambda_{\chi}$. 
In order to assess the potential impact of this region and being 
conservative, we evaluate the three sum rules taking a scale as 
large as $\Lambda_\chi=M_\rho$, which is certainly beyond the range 
of applicability of the theory. We obtain 
$\{ \rm SR1, SR2, SR3 \} = -\{ 2.7\times 10^{-4}, 8\times 10^{-5}, 2.5\times 10^{-5} \}$ 
in units of GeV$^{1,3,5}$. 
For comparison, the corresponding $\pi$ contributions adopting 
$g_{\pi NN} \simeq m_N g_A/F_{\pi}$ are
$\{ 1.1974, 0.02, 4.5\times 10^{-4} \}$ in the same units. 
Such a strong difference illustrates that threshold physics and 
chiral behaviour have a negligible impact in this context and can 
be safely ignored. In particular, in the context of the GT 
discrepancy, SR1 translates into $\Delta_{\rm GT} = 0.02\%$. 

At far high energies, one can employ pQCD to describe the behaviour of the form factor. 
Schematically, and along the lines of Ref.~\cite{RuizArriola:2025wyq}, 
the pQCD prediction reads 
\begin{equation}
  m_q F_P(-Q^2) \to \frac{m_q \Lambda_N^4}{Q^4}\alpha_s^2(Q^2), \quad
  \operatorname{Im} m_q F_P(s) \to \frac{m_q \Lambda_N^4}{s^2}
  \operatorname{Im}\left(\frac{4\pi}{\beta_0[L -i\pi]}\right)^2, \label{eq:FFTL}
\end{equation}
with $L=\ln(s/\Lambda_{\rm QCD}^2)$ and $\Lambda_{\rm QCD}=184$~MeV.   
In deriving the equation above, we have made use of the analytic 
continuation of $\alpha_s$ from the 
Euclidean to the time-like region through the relation 
$s = \lim_{\theta\to\pi} e^{-i\theta}Q^2$ at LO (see 
Refs.~\cite{Donoghue:1996bt,Leutwyler:2002hm,RuizArriola:2025wyq} and 
App.~\ref{app:DRasymp})
\begin{equation}
   \alpha_s(Q^2) = \frac{4\pi}{\beta_0 \ln(Q^2/\Lambda_{\rm QCD}^2)}, \qquad 
   \alpha_s(s) = \frac{4\pi}{\beta_0 \ln(s/\Lambda_{\rm QCD}^2) -i \pi} \, ,\label{eq:alphaTL}
\end{equation}
with $\beta_0 = 11 N_c/3 -2n_f/3 $ and $\Lambda_{\rm QCD} =183~\textrm{MeV}$. 
Note in addition that $m_q$, as well as the nucleon distribution amplitude, 
are scale-dependent objects and, at LO,
\begin{equation}
 m_q(\mu) = m_q(\mu_0) \left( \frac{\alpha_S(\mu)}{\alpha_S(\mu_0)} \right)^{4/\beta_0} \, , \qquad
 f_N (\mu) = f_N (\mu_0) \left( \frac{\alpha_S(\mu)}{\alpha_S(\mu_0)} \right)^{2/3\beta_0} .
\label{eq:mq-fN}
\end{equation}  
To account for it in the time-like region, we use the renormalization-group 
equations in two steps. First, one runs 
$m_q(\mu_0^2) \to m_q(Q^2)= m_q(\mu_0)[\alpha_s(Q^2)/\alpha_s(\mu_0^2)]^\gamma$.
Second, one runs from the space-like to the time-like region, corresponding 
to an arc in $Q^2$ plane ($s= e^{-i\pi}Q^2$) and leading to an additional factor 
$m_q(s) = m_q(\mu_0^2)[\alpha_s(s)/\alpha_s(Q^2)]^\gamma$. Overall, the net 
transformation reads $m_q(s) = m_q(\mu_0^2)[\alpha_s(s)/\alpha_s(\mu_0^2)]^\gamma$, 
with $\mu_0$ the reference space-like value and $\alpha_s(s)$ the time-like 
result in Eq.~\eqref{eq:alphaTL}. Such effects can be accounted for upon 
replacing $\alpha_s^2 \to \alpha_s^{2+\gamma}/\alpha_s(\mu_0)^{\gamma}$ 
with $\gamma = (4+4/3)/\beta_0 = 0.59$ in Eq.~\eqref{eq:FFTL}. 
As mentioned earlier, it 
is unclear at which scale pQCD faithfully describes the form factor. 
Taking a value as low as $\Lambda_{\rm pQCD} = 2m_N$, we 
obtain:\footnote{Had we neglected the anomalous dimensions and accounted 
only for the $\alpha_s^2$ factor in Eq.~(\ref{eq:FFTL}), we would obtain
$\{ 5\times 10^{-7}/\Lambda_{\rm QCD}^4, 10^{-4}/\Lambda_{\rm QCD}^2, 6\times 10^{-2} \}m_q \Lambda_N^4$ 
---just a minor modification for the third sum rule. The running of the DA can be easily accounted for 
when expressing it in terms of Appell polynomials. However, its effect is DA-dependent. For instance, 
for the CZ case, modifications are again mild except for the third sum rule, where $25\%$ corrections appear.} 
$\{ 5\times 10^{-7}/\Lambda_{\rm QCD}^4, 10^{-4}/\Lambda_{\rm QCD}^2, 5\times 10^{-2} \}m_q \Lambda_N^4$ 
in terms of $\Lambda_N^4$, which is related to the unknown DA. 
As shown in App.~\ref{sec:pQCD}, current models of DAs display a 
large disparity in sign and magnitude as to deserve a further study, which 
is beyond the scope of the present work (this holds true for vector 
or axial form factors as well). Still, even for the CZ DA~\cite{Chernyak:1984bm}, 
that produces a large value of $\Lambda_N^4 = 13.9~\textrm{GeV}^4$, the contribution 
to the sum rules incorporating the running of the DA reads
$\{ 1.8\times 10^{-5}, 1.2\times 10^{-4}, 1.8\times 10^{-3} \}$, again in 
GeV$^{1,3,5}$ units. 
Except for the third sum rule, these contributions are again small compared to the $\pi$ case, 
pointing to the fact that we are missing the relevant physics ---namely, the 
dominant role of the resonances.
In addition, once resonances are accounted for, the ``natural'' value for 
the third sum rule becomes $m_N g_A M_{\pi}^2 M_R^2\sim 0.023~\textrm{GeV}^5$, 
confirming that pQCD still only plays a minor role, even in this sum rule.

Overall, the previous discussion underlines the fact that all the relevant 
physics occurs in the resonance and, possibly, the high-energy Regge region, 
allowing to dismiss previous contributions and supporting the use of meson 
dominance to be employed below.

\subsection{The resonance region}

At intermediate energies ---around the GeV scale, but still below the
$\sqrt{s} = 2M_N$ threshold--- the physics of $I^G J^{PC} = 1^-0^{+-}$
resonances takes over multipion $(2n+1)\pi$ states, which is
understood on the basis of large-$N_c$ arguments and greatly
simplifies the problem of reconstructing the spectral
function.\footnote{We note that a rigorous reconstruction in terms of
  intermediate multiparticle states has so far been achieved for
  $\pi\pi$ intermediate states alone~\cite{Hoferichter:2016duk,Alarcon:2018irp}.  
  Our case resembles more the
  isoscalar vector form factor of the nucleon which, being coupled to
  the $3\pi$ continuum, is in practice described in terms of $\omega$
  resonances.}  The potential role of pseudoscalar resonances in this
context was already suggested by Pagels~\cite{Pagels:1972xx}, while
the first realization that getting a GT discrepancy with a proper sign
required a resonant behaviour was made in Ref.~\cite{Bhamathi:1977xp}.
Experimentally, the existence of pseudoscalar resonances is well
established, and our current knowledge is summarized in
Table~\ref{tab:ps-states}.  However, the specifics of their masses,
widths, and profiles are not well established even for the lightest
one, the $\pi(1300)$, for which the different mass and width estimates
lie in the $1.128(78)-1.375(40)$~GeV and $218(100)-580(100)$~MeV
range, respectively, see
PDG~\cite{ParticleDataGroup:2024cfk}.\footnote{Actually
    this large width and its range represents a range of values, but
    not necessarily a set of mutually consistent determinations, see App.~\ref{app:pi1300}.}
We refer the interested reader to App.~\ref{app:pi1300} for further details.

It is worth reminding that a large-$N_c$ approach including 
short-distance constraints in the single-resonance approximation 
allows to deduce the result $M_{\pi'} = M_{a_1}= \sqrt{2} M_\rho = 4 \pi F_\pi
\sqrt{3/N_c}$ with nominal ${\cal O} (N_c^{-1})$ accuracy and which
yields $M_{\pi'} = 1100$~MeV and $M_\rho=780 $ MeV for 
$F_\pi= 92$~MeV~\cite{Ledwig:2014cla}. Very recently, a finite volume lattice
calculation has extracted the pole parameters for
$\sqrt{s_{\pi'}}=1169 \pm 46- i (62_{-62}^{+168})$~MeV~\cite{Yan:2025mdm}.

In order to avoid choosing specific models for their profiles and 
the inherent associated model dependency, we opt for a 
narrow-resonance description, and estimate the uncertainties related 
to their unknown profiles via the half-width rule 
(HWR)~\cite{Masjuan:2012sk,RuizArriola:2011nrw,RuizArriola:2012ius} 
(the interested reader is referred to App.~\ref{sec:fin-width} 
for discussions on particular parametrizations).
Specifically, their contribution to the spectral function reads
\begin{equation}\label{eq:FPpsCont}
\operatorname{Im} m_q F_P(s) = \pi \sum_n M_{\pi_n}^2 Z_{\pi_n} \delta(s -M_{\pi_n}^2) \, , \qquad
Z_{\pi_n} \equiv F_{\pi_n} g_{\pi_n NN} \, .
\end{equation}
Note here that $F_{\pi_n} \sim m_q$, approaching zero in the chiral
limit and guaranteeing the conservation of the axial current.  Last,
unless a specific ($n$-dependent) pattern is chosen, a finite number
of resonances must be adopted. Phenomenologically, it has been
experimentally observed that the form factors of ground-state hadrons
are dominated by the lowest-lying resonances required to reproduce the
high-energy power-law behaviour, see Ref.~\cite{Masjuan:2012sk}.  In
our case, this means restricting the sum to a single state, as two of
them would either produce undetermined $Z_{\pi_n}$ coefficients or an
overdamped $Q^{-6}$ behaviour. We choose this state to be the
$\pi(1300)$, while heavier states will be effectively accounted for below
through the high-energy Regge region, that is required to reproduce the
more complicated (non-integer power-law) behaviour of pQCD.\footnote{
  Again, this improves over our previous
  work~\cite{RuizArriola:2023xap}, where two resonances 
  ---the $\pi(1300)$ and $\pi(1800)$--- 
  but no Regge-behaviour was incorporated. This led to 
  an asymptotic $Q^{-6}$ behaviour, which decreases significantly faster 
  than the $Q^{-4} (\ln Q^2)^{-2}$ behaviour of pQCD. 
  }

\subsection{The Regge region}

The contribution of a finite number of resonances is insufficient 
to reproduce the correct high-energy behaviour, which requires 
further $\alpha_s^2(Q^2)$ suppression. 
This limitation can be overcome by considering an infinite 
tower of resonances, which makes it possible to reproduce a non-integer 
power-like behaviour~\cite{Dominguez:2001zu} or even $\alpha_s$-like 
suppression factors~\cite{RuizArriola:2008sq}. 
The presence of such an infinite number of resonances 
can be justified on the grounds of the large-$N_c$ limit of QCD, 
where the study of two-point correlation functions naturally leads 
to this requirement. Such implications will be discussed shortly after.  
Phenomenologically, the QCD resonance spectrum is well described 
by linear Regge trajectories of the form $M_n^2 \sim an+b$ for 
a variety of quantum numbers. 
\begin{table}
\caption{Masses and widths of the isovector pseudoscalar states so far reported. 
         The line marks the separation between states below and above $N \bar N$ threshold.
         Note that PDG~\cite{ParticleDataGroup:2024cfk} reports the first three 
         states only, while the last two come from $p\bar{p}$ annihilation 
         experiments~\cite{Anisovich:2001pn}.}\label{tab:ps-states}    
    \begin{tabular}{cccc|cc}\toprule 
       Name    & $\pi(140)$ & $\pi(1300)$ & $\pi(1800)$ & $\pi(2070)$ & $\pi(2360)$ \\ \midrule
     $M$ (MeV) & $140$ & $1300(100)$ & $1810(10)$ & $2070(35)$ & $2360(25)$ \\
     $\Gamma$ (MeV) & $0$ & $400(200)$ & $215(8)$ & $310^{+100}_{-50}$ & $300^{+100}_{-50}$ \\
     $\Gamma/M$ & $0$ & $0.31(16)$ & $0.12(1)$ & $0.14(4)$ & $0.13(3)$ \\ \bottomrule
    \end{tabular}
\end{table} 
From a theoretical point of view, 
this phenomenological observation can be justified within the 
(relativized) quark model of Isgur and Godfrey~\cite{Godfrey:1985xj}, 
where pseudoscalar resonances are interpreted as $n$-excited 
$^1S_0$ $\bar q q$ states from a two-body Hamiltonian 
of the form $H=2\sqrt{p^2+M_q^2}+ \sigma r$ (here $M_q$ represents the
constituent quark mass). This model suggests a linear Regge-like
excited spectrum (see e.g. \cite{RuizArriola:2006opt,Afonin:2024egd}).
More specifically, the isovector pseudoscalar spectrum ---whose known
states are shown in Table~\ref{tab:ps-states}--- is well reproduced by
choosing $M_{\pi_n}^2=n M_{\pi'}^2 + M_\pi^2$~\cite{Masjuan:2012gc}.\footnote{Non-linear effects in
  Regge trajectories are tiny (see e.g.  \cite{Chen:2022flh} and
  references therein). Note also that the states 
  reported by PDG completely dominate the extraction of the parameters of the Regge trajectory.} Furthermore, the resonances appear to approach
the universal Suranyi's ratio $\Gamma/M =0.12(8)$, which qualitatively
agrees with the large-$N_c$ scaling $\Gamma/M=\mathcal{O}(N_c^{-1})$ (see e.g.
Ref.~\cite{Masjuan:2012gc}).
Adopting such a linear spectrum, one can infer few properties about the
heavy pseudoscalar resonances. Specifically, the two-point
pseudoscalar function at LO in pQCD
reads~\cite{Maltman:2001gc,Kadavy:2020hox,Dominguez:1984eh,Dominguez:1984yx,
  Dominguez:2018azt}
\begin{equation}
i \int d^4 \! x e^{i q\cdot x} \bra{0} T\{ m_q P^i(x) , m_q P^j(0) \}  \ket{0} \equiv \delta^{ij} \Pi_{PP}(q^2) \, , \
  \operatorname{Im} \Pi^{\rm LO}_{PP}(s) = \frac{N_c m_q^2}{16\pi}s \, ,
\end{equation}
whereas the large-$N_c$ hadronic spectral function behaves, for large $s$, as  
\begin{equation}
\frac{1}{\pi}\operatorname{Im} \Pi_{PP}(s) = \sum_n \frac{F_{\pi_n}^2 M_{\pi_n}^4}{4} \delta(s -M_{\pi_n}^2)
     \to \int dn \frac{F_{\pi_n}^2 M_{\pi_n}^4}{4} \delta(s -M_{\pi_n}^2) 
     = \lim_{n\to\infty} \frac{F_{\pi_n}^2 M_{\pi_n}^2 s}{ \frac{dM_{\pi_n}^2}{ds} } \, ,
\end{equation}
demanding that $F_{\pi_n} \sim M_{\pi_n}^{-1} \sim n^{-1/2}$. 
Note therefore that, according to Eq.~\eqref{eq:FPpsCont}, the contributions 
of the pseudoscalar resonances to the spectral function of $m_q F_P(t)$ 
is of the $M_{\pi_n}^2 F_{\pi_n} g_{\pi_n NN} \sim \sqrt{n} g_{\pi_n NN}$ 
kind, demanding $g_{\pi_n NN}$ to decrease faster than $n^{-3/2}$ to 
ensure a convergent sum.

To construct our model for the Regge part in a simple way we note that, at 
high energies, the contribution of the infinite number of resonances 
simplifies considerably (see App.~\ref{sec:regge-asy}). 
In particular, by expressing the individual $\pi_n$ contributions to 
the spectral functions as   
$m_q \operatorname{Im} F_P(s) \simeq \pi\sum_n c_n \delta(s -M_{\pi_n}^2) $, 
where $c_n = M_{\pi_n}^2 Z_{\pi_n}$, cf. Eq.~\eqref{eq:FPpsCont}, one finds 
that, if $c_n \sim n^{-\alpha}$, then  
$\operatorname{Im} F_P(s) \sim s^{-\alpha}$. This motivates a simple 
model, analogous to the one put forward in Ref.~\cite{RuizArriola:2025wyq},
\begin{eqnarray}
  \operatorname{Im} m_q F_P (s)= \operatorname{Im} m_q F_P (\Lambda_R^2)  \left(\frac{\Lambda^2_R}{s} \right)^{2+2 \epsilon} \, , \quad
  \Lambda^2_R \le s \le \Lambda^2_{\rm pQCD} \, . \label{eq:Regge}
\end{eqnarray}
While this model features a fractional power-law behaviour, 
decreasing faster than pQCD, the spectral function could be 
eventually matched to the pQCD prediction, Eq.~\eqref{eq:FFTL}, 
ensuring the correct asymptotic behaviour is recovered. 
In practice, though, we found out that the pQCD region is irrelevant 
for our purposes, and we adopt the model above up to arbitrarily high energies. 
For a more detailed discussion, see Ref.~\cite{RuizArriola:2025wyq}.
To fully specify our model, one parameter in Eq.~(\ref{eq:Regge}) must be 
fixed. Given our ignorance on $\operatorname{Im} F_P (\Lambda_R^2)$, 
we provide an estimate for $\epsilon$ by matching the logarithmic 
derivative of our model to that of pQCD (for simplicity, anomalous 
dimensions are ignored, see App.~\ref{sec:pQCDRegge}), implying
\begin{eqnarray}\label{eq:epsilonMatch}
\epsilon= \frac{2 \log (\Lambda^2_{\rm pQCD}/\Lambda_{\rm QCD}^2)}{ \log^2 (\Lambda^2_{\rm pQCD}/\Lambda_{\rm QCD}^2)+\pi^2} -
 \frac{1}{2 \log (\Lambda^2_{\rm pQCD}/\Lambda_{\rm QCD}^2)} \, .
 \end{eqnarray}
Varying the $\Lambda_{\rm pQCD}$ matching scale across a large 
range of values, one finds $\epsilon=0.1-0.2$ (see App.~\ref{sec:pQCDRegge}).

Having discussed the different regions and justified our model for
the spectral function, we present in the following our results for 
the GT discrepancy and the pseudoscalar form factor.

\section{Results}
\label{sec:results}
 
We justified in the previous sections that the threshold and pQCD regions 
play a marginal role in describing the pseudoscalar form factor, the 
bulk of the contribution being taken by the resonance and Regge regions. 
Including a single resonance, $\pi'\equiv\pi(1300)$, and the Regge model in Eq.~\eqref{eq:Regge}, 
the sum rules read ($Z_{\pi'} = F_{\pi'}g_{\pi'NN}$)
\begin{eqnarray}
  g_A m_N & =& F_\pi g_{\pi NN} + Z_{\pi'} + \frac1{\pi} \frac{ \operatorname{Im} m_q F_P(s_r)}{2 \epsilon+2} \, , \label{eq:resultseq1} \\
  0 & =& F_\pi g_{\pi NN} M_\pi^2 + Z_{\pi'} M_{\pi'}^2 + \frac{s_r}{\pi} \frac{\operatorname{Im} m_q F_P(s_r)}{2 \epsilon+1} \, , \\
  0 & =& F_\pi g_{\pi NN} M_\pi^4 + Z_{\pi'} M_{\pi'}^4+ \frac{s_r^2}{\pi} \frac{\operatorname{Im} m_q F_P(s_r)}{2 \epsilon}  \, ,
\end{eqnarray}  
with $s_r = \Lambda_R^2$, and the form factor
\begin{equation}
   m_q F_P(s) = \frac{M_{\pi}^2 F_{\pi} g_{\pi NN}}{M_{\pi}^2 -s} 
               +\frac{M_{\pi'}^2 Z_{\pi'}}{M_{\pi'}^2 -s}
               +\frac{ \operatorname{Im} m_q F_P (s_r)}{\pi} \int_{s_r}^{\infty} \frac{dt}{\zeta -s -i\epsilon}  \left( \frac{s_r}{\zeta} \right)^{2(1+\epsilon)} \, .
\end{equation}
In the following, we specialize to the charged case, thus
we take $M_{\pi^+} = 139.57039$~MeV, $m_N =(m_p +m_n)/2=938.91875433$~MeV, 
$g_A=1.2753(13)$, $M_{\pi'}=1.3$~GeV~\cite{ParticleDataGroup:2024cfk} and 
adopt the half-width rule~\cite{Masjuan:2012sk} to estimate uncertainties 
on $M_{\pi'}$ as $M_{\pi'}\pm\Gamma_{\pi'}/2$, with 
$\Gamma_{\pi'}=400$~MeV~\cite{ParticleDataGroup:2024cfk},\footnote{We recall again 
that PDG~\cite{ParticleDataGroup:2024cfk} estimates $M_{\pi(1300)}= 1.3(1)$~GeV, 
but quotes different estimates ranging from $1.128(75)$~GeV up to $1.375(4)$~GeV, 
reflecting the large uncertainty on the $\pi(1300)$ mass, which varies 
significantly depending on the chosen process and resonance profile details.}
that provides a successful approach when describing a variety of meson form 
factors~\cite{Masjuan:2012sk}, while avoiding a detailed description of the 
resonance(s) profile (see App.~\ref{sec:fin-width} for 
discussions on particular profiles).
As argued, we take $\epsilon=0.15(5)$, and $\Lambda_R=1.8(2)$ is chosen 
to start at the next resonance, the $\pi(1800)$. 
Our results for the unknown parameters are
\begin{align}
   F_{\pi^+} g_{\pi^+ pn} = 1.2127 (12)_{g_A}  (2)_{\Lambda_R} (4)_{\epsilon} (^{+57}_{-36})_{\rm HWR} [^{+59}_{-39}]  , \\
   Z_{\pi'} = -0.0159 (0)_{g_A}  (5)_{\Lambda_R} (6)_{\epsilon} (^{+34}_{-56})_{\rm HWR} [^{+35}_{-56}]  , \\
   \operatorname{Im} m_q F_P(s_r) = 0.0040 (0)_{g_A} (^{+27}_{-15})_{\Lambda_R} (15)_{\epsilon} (14)_{\rm HWR} [^{+35}_{-24}]  , 
\end{align}
with uncertainties dominated by the half-width rule for $g_{\pi^+ pn}$
and $Z_{\pi'}$. We emphasize that the peak location is more
influential than the detailed lineshape. Note also that, had we
omitted the Regge part, the pseudoscalar form factor would reduce to a
simple dipole model $m_q F_P(s) = g_A m_N (1 -s/M_\pi^2)^{-1}(1
-s/M_{\pi'}^2)^{-1}$.  In addition, one can check that in the chiral
($m_q,M_\pi\to 0$) limit the solutions for $Z_{\pi'},
m_q\operatorname{Im}F_P(s_r)$ vanish, as they should. Concerning
further subtleties on isospin corrections, we refer the reader to
App.~\ref{app:ib} and the error estimate below.

\subsection{Results for \texorpdfstring{$\Delta_{\rm GT}$}{DeltaGT} and \texorpdfstring{$g_{\pi NN}$}{gpiNN}}

Noteworthy, without input for $F_{\pi}$, it is possible to compute 
$\Delta_{\rm GT}$, see Eqs.~(\ref{eq:GT},\ref{eq:GTSR}). 
Moreover, since the sum rules imply a linear dependence in $g_A m_N$ 
for all of the unknowns, and in particular for $F_{\pi} g_{\pi NN}$, 
all dependence in  $m_N g_A$ disappears. 
Our final result for $\Delta_{\rm GT}$, for the charged channel, reads
\begin{align}
   \Delta_{\rm GT} = 1.26 (2)_{\Lambda_R} (3)_{\epsilon} (^{+47}_{-30})_{\rm HWR} (50)_{\rm IB} \% \, .
\end{align}
The result includes an additional uncertainty from isospin-breaking (IB) corrections, as estimated in App.~\ref{app:ib}. 
A full determination of these effects ---and a precise definition $\Delta_{\rm GT}$--- is, however, beyond the scope of the present work. 
The final uncertainty is 
totally dominated by the $\pi(1300)$ uncertainties and IB corrections. This quantity could be obtained in 
lattice studies using $\Delta_{\rm GT} = 1 -D(0)/D(M_{\pi}^2) \simeq 
M_{\pi}^2 D'(0)/D(0)$, see Eq.~\eqref{eq:Dt}. 
Finally, taking $F_{\pi^+}=92.3(1)$~MeV~\cite{ParticleDataGroup:2024cfk}, we find
\begin{align}
   g_{\pi^+ pn} = 13.14 (1)_{g_A}  (^{+6}_{-4})_{\rm HWR}(1)_{F_\pi} (7)_{\rm IB}  \, , 
\end{align} 
totally dominated by the uncertainty of $\pi(1300)$ and IB corrections. 
We note that, had we neglected the Regge region and employed the two sum rules alone, 
the model would reduce to a dipole, and $\Delta_{\rm GT}=M_\pi^2/M_{\pi'}^2 = 1.15\%$, 
pointing to the relevance of the Regge region.\footnote{Equivalently, one may take 
$\Lambda_R\to\infty$. The dipole result is effectively achieved to the digits of accuracy 
here shown if $\Lambda_R > 7$~GeV.} Moreover, our analysis supersedes the simplified 
preliminary result in proceedings~\cite{RuizArriola:2023xap}, where 
the spectral function was modelled using two excited pseudoscalar 
resonances, the $\pi(1300)$ and $\pi(1800)$. In that scheme, the 
$\pi(1800)$ played the role of the Regge contribution, leading to an overdamped 
$Q^{-6}$ (incorrect) behaviour.
Finally, note that our estimate has a spectral function which is 
negative $(Z_{\pi'}<0)$ in the resonance region, changing sign 
somewhere in between the resonance and the Regge region, where 
${\rm Im} F_P(\Lambda_R^2)>0$. As a result, the bound
in Sect.~\ref{sec:bound} applies regardless of specifics of the $\pi(1300)$
lineshape. Provided the Regge contribution is fixed, we expect our 
narrow-width $\pi(1300)$ approximation to represent a lower bound both for 
$\Delta_{\rm GT}$ as well as $g_{\pi NN}$. In App.~\ref{sec:fin-width}
we illustrate finite width effects using a variety of
models, illustrating this and showing that finite-width 
corrections are safely within the half-width rule.

As an aside, our model enables a prediction for $F_{\pi'}g_{\pi'NN}$. 
Adopting the estimate $F_{\pi'}=2.4(6)$~MeV~\cite{Maltman:2001gc} 
based on QCD sum rules, we find 
\begin{align}
   g_{\pi' NN} = -6.6 (2)_{\Lambda_R} (2)_{\epsilon} (^{+1.2}_{-2.3})_{\rm HWR} (^{+1.3}_{-2.2})_{F_{\pi'}}  [^{+2.0}_{-3.2}]  \, .
\end{align} 

An alternative perspective is the following: since $\Delta_{\rm GT}$ mostly depends 
on the peak position of the $\pi(1300)$ (cf. Appendix~\ref{sec:fin-width}), one 
may speculate which mass fulfills the different available extractions for $\Delta_{\rm GT}$.
We present our results for selected values of $g_{\pi NN}$ in Fig.~\ref{fig:gpiNN} in 
Table~\ref{tab:MPIGT}.
\begin{table}
\caption{\label{tab:MPIGT}The $M_{\pi}(1300)$ that would be implied by different $g_{\pi NN}$ extractions based on $NN$ and $\pi N$ GMO analysis.}
\begin{tabular}{ccccccccc} \toprule
                           & GMO$_E$                  & GMO$_B$               & NN$_{\rm{Gr}}$           & NN$_{\rm{Bo}}$           &  GMO$_J$             \\ \midrule
 $g_{\pi NN}$ (input)      &  $13.3(1)$               & $13.12(10)  $         & $13.25(5)$               & $13.23(4)$               &  $13.11(10)$         \\ \midrule
 $M_{\pi'}$ (GeV) (output) & $0.91(^{+0.17}_{-0.11})$ & $1.4(^{+0.9}_{-0.3})$ & $0.99(^{+0.12}_{-0.09})$ & $1.04(^{+0.12}_{-0.09})$ & $1.5(^{+1.0}_{-0.4})$ \\ \bottomrule
\end{tabular}
\end{table}

\subsection{Results for the pseudoscalar form factor\label{sec:resFF}}

\begin{figure}\center
  \includegraphics[width=0.8\textwidth]{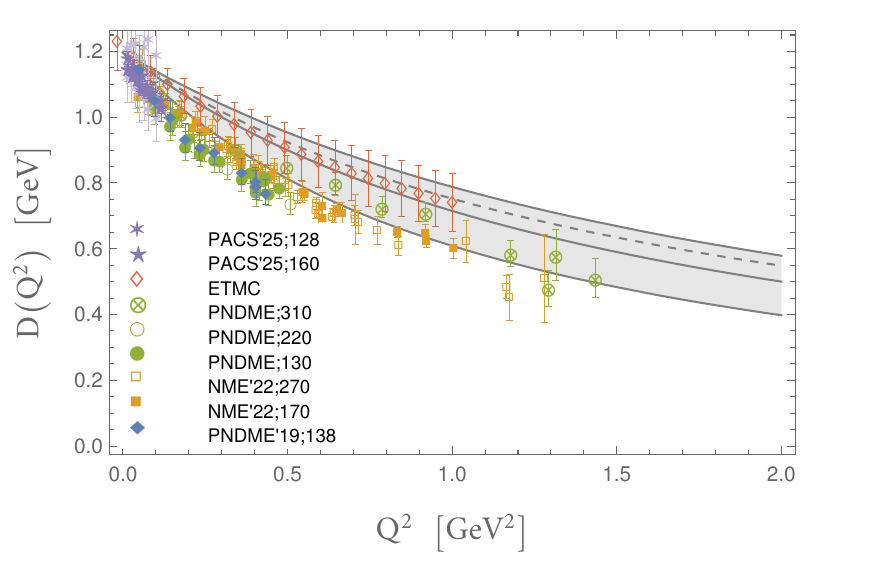}
  \caption{The form factor $D(Q^2)$ (full grey lines and band). In addition, we show as a dashed grey band the pure monopole prediction. 
  Lattice results from PACS'25~\cite{Aoki:2025taf}, ETMC~\cite{Alexandrou:2023qbg}, PNDME~\cite{Jang:2023zts}, NME'22~\cite{Park:2021ypf}, PNDME'19~\cite{Jang:2019vkm}. 
  The final number represents the lattice $\pi$ mass in each ensemble except for ETMC, which results are taken from their table with physical extrapolations.}\label{fig:Dt}
\end{figure}

Finally, the model for the spectral function yields a prediction for 
the space-like pseudoscalar form factor that can be compared to 
lattice QCD results. 
In the following, we present our results for $D(s)$ in Eq.~(\ref{eq:Dt}), 
which removes the $\pi$-pole dependence from $m_q F_P(s)$. 
This choice is convenient, as most collaborations do not provide 
the physical extrapolation for this or $m_q F_P(s)$ itself,  
but rather report their results for a variety of ensembles at unphysical pion masses. 
By factoring out the pion-pole, we remove the dominant dependence on the $\pi$ 
mass. To further account for the physical extrapolation, we rescale 
lattice results by $m_N^{\rm phys}/m_N^{\rm latt}$ along with appropriate 
lattice renormalization factors for the currents, whereas residual effects in 
$g_A^{\textrm{phys}}$ vs. $g_A^{\textrm{latt}}$ as well as continuum 
extrapolation might be present (the exception is ETMC~\cite{Alexandrou:2023qbg}, 
that corresponds to their physical extrapolation). For an 
excellent overview of lattice QCD results, we refer to Ref.~\cite{Gupta:2024krt}. 
The results are shown in Fig.~\ref{fig:Dt} and the errors are dominated by the HWR. 

\begin{figure}\center
  \includegraphics[width=0.8\textwidth]{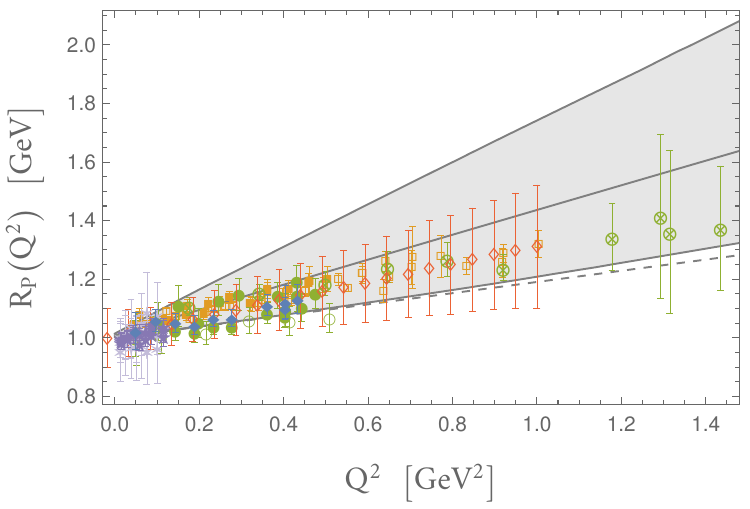}
  \caption{The ratio of pseudoscalar form factors defined in Eq.~\eqref{eq:RP} (full gray lines and band). 
  In addition, we show as a dashed-gray line the pQCD-driven estimate. The colours for the lattice results are analogous to those in Fig.~\ref{fig:Dt}.}\label{fig:RP}
\end{figure}

Our model also helps to illustrate different aspects with regard to 
lattice QCD studies: (i) problems of assuming PPD to extract $g_{\pi NN}$
(ii) pQCD behaviour and $F_{P}/G_P$ with appropriate scaling 
(iii) Expected deviations from PPD.
To do so, we need to briefly comment on $G_A(s)$. To lighten 
the discussion, let us disregard the sum rule obtained from
$\alpha_s$, that would otherwise require a dedicated analysis of 
$G_A$, which is beyond the scope. 
In such limit, the models for $m_q F_P(s)$ and $G_A(s)$ reduce to 
a dipole form. Specifically, 
\begin{equation}
 m_q F_P(s) = m_N g_A \frac{M_\pi^2}{M_\pi^2 -s}\frac{M_{\pi'}^2}{M_{\pi'}^2 -s} \, , \qquad
 G_A(s) = g_A \frac{m_{A_1}^2}{m_{A_1}^2 -s}\frac{m_{A_1'}^2}{m_{A_1'}^2 -s} \, ,
\end{equation}
where $A_{1,2},$ can be identified with the $a_1(1260)$ and 
$a_1(1640)$ axial-vector mesons.
$G_P(s)$ can be readily obtained from the PCAC relation and be 
expressed as 
\begin{align}
  G_P(s) &{}= \frac{4m_N^2 g_A}{M_\pi^2 -s}\left[ 
                   \frac{M_{A_1}^2 M_{A_1'}^2}{(M_{A_1}^2 -s)(M_{A_1'}^2 -s)} 
                 +M_\pi^2 \frac{M_{A_1}^2 M_{A_1'}^2 -M_{\pi'}^2(M_{A_1}^2 +M_{A_1'}^2 -s)}{(M_{A_1}^2 -s)(M_{A_1'}^2 -s)(M_{\pi'}^2 -s)}
                \right] \nonumber\\ \label{eq:GPTexact}
                  &{}\to 4m_N^2 \frac{G_A(s)}{M_\pi^2 -s}\left( 1 - M_\pi^2/M_{A_1'}^2 \right) , 
\end{align}
where the last line follows from the accidental degeneracy 
$M_{\pi(1300)}\sim M_{a_1(1260)}$, which helps to explain the 
phenomenological success of the PPD hypothesis for $G_P$, 
stating that $G_P(s) \simeq 4m_N^2 G_A(s)/(M_\pi^2 -s)$. 
The near-degeneracy and large $M_{\pi}^2/M_{a_1(1640)}^2$ 
suppression provides such an explanation, and helps to provide 
a benchmark for the breaking of PPD 
hypothesis. For related discussions based on a microscopic 
quark-level point of view, rather than the hadronic approach 
adopted here, we refer the reader to Ref.~\cite{Chen:2021guo}. 
We note, however, that the PPD hypothesis for $D(t)$, which assumes 
$D(s) \simeq G_A(s)$, is clearly incorrect for at least 
two reasons: (i) they have different quantum numbers ---specifically, 
$G_A(s)$ features axial-vector meson poles, whereas $D(s)$ 
features pseudoscalar meson poles; (ii) they exhibit  
different asymptotic behaviour. This 
outlines the problems of using PPD 
to extract $g_{\pi NN}$ or $\Delta_{\rm GT}$ based on $G_A$, 
as employed in Ref.~\cite{Alexandrou:2017hac,Alexandrou:2020okk}. 
In particular, such an assumption would imply 
$\Delta_{\rm GT} \sim M_\pi^2/M_{a_1(1260)}^2  +M_\pi^2/M_{a_1(1640)}^2 $
that, besides $M_{a_1(1260)}$ vs $M_{\pi(1300)}$ corrections, would 
imply $+M_\pi^2/M_{a_1(1640)}^2$ non-negligible artifacts, ultimately 
due to the relation in Eq.~\eqref{eq:GPTexact}. 
To illustrate this,  we plot the ratio 
$R_P(Q^2) = (4 m_N / M_\pi^2 ) (m_q F_P(-Q^2) / G_P(-Q^2) )$ 
which, according to pQCD asymptotics, should diverge linearly for large $Q^2$ values. 
Our result is represented in Fig.~\ref{fig:RP} together with lattice calculations.
The uncertainties are again dominated by the HWR that, for the $\pi(1300)$ 
and $a_1(1269)$ mesons, are around $200$~MeV both.
We emphasize that not only ChPT~\cite{Bernard:2001rs,Park:2021ypf} predicts a linear behaviour at 
low $Q^2$, but pQCD guarantees it to hold up to arbitrarily high energies:
\begin{equation}
 R_P(Q^2) = \frac{4 m_N}{M_\pi^2}\frac{m_q F_P(-Q^2)}{G_P(-Q^2)} 
 \sim \frac{4 m_N}{M_\pi^2} \frac{Q^2}{4m_N^2}\frac{m_q F_P(Q^2)}{G_A(-Q^2)} 
  \xrightarrow{Q^2\to\infty} 
  \frac{Q^2}{M_\pi^2} \frac{m_q}{m_N} \, , \label{eq:RP}
\end{equation} 
where use of PCAC in Eq.~(\ref{eq:PCAC}) has been made together 
with the 
fact that most DAs models predict $m_q F_P(s) \sim m_q G_A(s)$.
Such linear rise, normalized at one at the origin, is represented in Fig.~\ref{fig:RP} as a dashed-gray line 
and correctly estimates the order of magnitude. 

\section{Conclusions}
\label{sec:concl}

In this paper, we have analysed the pion--nucleon coupling and the
Goldberger--Treiman discrepancy using dispersion relations for the
pseudoscalar form factor of the nucleon, implementing high-energy
constraints from pQCD which, to the best of our knowledge, are provided
here for the first time.  This framework relies on the pseudoscalar
spectral density, which contains information on the $N \bar N$ system
in the $I^GJ^{PC}=1^- 0^{-+}$ channel. It leads to three sum rules
that are shown to imply the existence of one or more zeros in the
invariant-mass range $ 3M_\pi \le \sqrt{s} < \infty$. To account for
these properties, we model the spectral function by splitting its
behaviour into four different regions: the low, intermediate, Regge,
and pQCD ones.  While the spectral function is expected to be reliably
described by ChPT close to the $3\pi$ (low-energy) threshold region
and by pQCD (high-energy) asymptotically, we estimate their
contributions to actually be tiny. Instead, the most significant
contribution comes from the intermediate resonance region, which can
be approximated by narrow resonances and a Regge tail (eventually
matching pQCD). This result reinforces the use of pseudoscalar meson
dominance dating back to Dominguez~\cite{Dominguez:1984eh}. Our
analysis suggests that, besides IB uncertainties, it is the lowest 
pseudoscalar resonance, the $\pi(1300)$, that  plays the
dominant role for both the signal and the theoretical uncertainties.
As a result, we find for the charged channel,
\begin{equation}
  g_{\pi^+pn} = 13.14(^{+6}_{-4})(7)_{\rm IB}, \quad \Delta_{\rm GT} = 1.26(^{+51}_{-34})(50)_{\rm IB}\% ,  
\end{equation}
compatible and with similar uncertainties to sophisticated analysis based on 
$pp$ and $np$ scattering that find $g_{\pi^+ pn}=13.25(5)$ or 
those from the GMO sum rule, $g_{\pi^+ pn}= 13.11(10)$~\cite{Hoferichter:2023ptl}. Our
findings are efficiently encapsulated, in the isospin-symmetric limit, by the formulae
\begin{eqnarray}
  g_{\pi NN} \simeq \frac{g_A F_\pi}{m_N} \frac{1}{1-M_\pi^2/M_\pi'^2} \, , \qquad
  \Delta_{\rm GT} \simeq \frac{M_\pi^2}{M_{\pi'}^2} \, , 
\end{eqnarray}  
where most of the uncertainty is dominated by the wide and relative
uncertain shape of the $\pi(1300)$ resonance and calls 
for an improved knowledge of the properties of the $\pi(1300)$ meson. Overall, 
our results are more precise than current lattice estimates, which 
suffer from important systematic uncertainties from excited-states 
contamination relevant to fulfil PCAC. Still,  
lattice QCD could potentially reach a competitive prediction for 
$\Delta_{\rm GT}$ by a dedicated analysis of the pseudoscalar 
nucleon form factor and the associated $D(t)$ form factor.  
In this regard, our model also helps in setting a benchmark for the 
breaking of the PPD hypothesis, commonly tested by lattice QCD 
collaborations.

In the future, we expect to apply this framework to the cases of the 
$\eta$ and $\eta'$ mesons by considering the octet and singlet 
pseudoscalar densities, which corresponding GT discrepancies 
are poorly known compared to the pion case. There, chiral 
(as well as $U_A(1)$-anomaly) corrections are expected to be large, 
especially for the heavy $\eta'$ meson. 
A dedicated study is required due to the subtleties of the singlet current, which is 
beyond the scope of this work. 
Such study will be important in estimating the GT discrepancies 
for the neutral channels which, due to 
isospin-breaking effects, have not been discussed here.

\bmhead{Acknowledgments}

We acknowledge R.~Gupta for sharing results from Ref.~\cite{Jang:2019vkm}.
Partially funded by the Spanish Ministerio de Ciencia 
Innovaci{\'o}n y Universidades (MICIU/AEI /10.13039/501100011033
and ERDF/EU) under grants No. PID2020114767GB.I00 and
PID2023.147072NB.I00. PSP is funded by Junta de Andaluc{\'i}a
under the grant POSTDOC 21 00136 and ERA and PSP under
grant FQM225.

\section*{Declarations}

%
%
%
%

\textbf{Funding. }Partially funded by the Spanish Ministerio de Ciencia 
Innovaci{\'o}n y Universidades (MICIU/AEI /10.13039/501100011033
and ERDF/EU) under grants No. PID2020114767GB.I00 and
PID2023.147072NB.I00. PSP is funded by Junta de Andaluc{\'i}a
under the grant POSTDOC 21 00136 and ERA and PSP under
grant FQM225. \textbf{Conflicts of interest/Competing interests. } 
The author declare they have no financial interests.
\textbf{Data availability.} No Data associated in the manuscript.

\begin{appendices}






\section{Details of the pQCD calculation}\label{sec:pQCD}

\begin{figure*}[t]
  \includegraphics[width=\textwidth,height=0.12\textwidth]{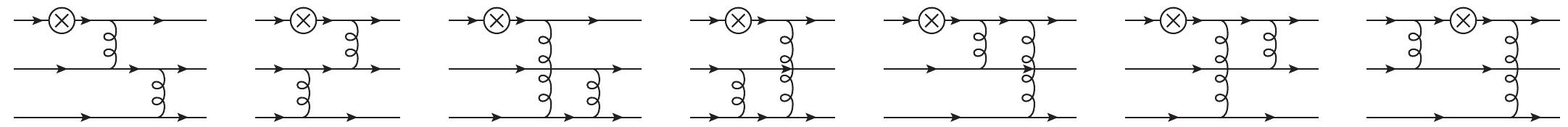}
  \caption{\label{fig:FD}\footnotesize In the diagrams above, the top line 
  is taken as the $u$ quark with momentum fraction $u_1$ or the $d$ quark. 
  The crossed circle stands for the appropriate current insertion.
  All the contributions can be expressed in terms of these.}
\end{figure*}

The asymptotic behaviour of nucleon form factors can be computed in pQCD and can be expressed in terms of convolutions of ---perturbative--- hard-scattering amplitudes and the nucleon distribution amplitude (DA), that encodes the non-perturbative information~\cite{Lepage:1979za,Korenblit:1979cw,Lepage:1980fj,Brodsky:1980sx,Chernyak:1984bm,Carlson:1985zu}. 
For recent advances at next-to-leading order for the vector form factor, see Refs.~\cite{Huang:2024ugd,Chen:2024fhj}. 
At LO, the pQCD prediction for external currents of the form 
$X=\sum_{q} Q_q \bar{q} \Gamma_X q$\footnote{
   $\Gamma_X$ represents the appropriate Dirac matrices for the structures 
   $X=\{V,A,S,P\}$. This is, $\Gamma_X = \{ \gamma^{\mu}, \gamma^{\mu}\gamma_5, 1, i\gamma_5 \}$. 
   Also, for the case of $V,A$, the predictions are for the $F_1(q^2),G_A(q^2)$ form factors.} 
can be expressed as\begin{equation}\label{eq:pQCDgen}
 Q^4 F(Q^2) = \frac{(4\pi\alpha_s f_N)^2}{54} \int du_1 du_2 du_3 dv_1 dv_2 dv_3 \left[ 2 \sum_i T_i^u + \sum_i T_i^d \right],
\end{equation}
where $u_i$ and $v_i$ are the fraction of momenta carried by the quark $i$ in the initial and final nucleon, respectively. 
The constant $f_N$ denotes the nucleon decay constant, that provides the overall normalization of the nucleon DA. 
The hard-scattering amplitudes can be computed from the diagrams in Fig.~(\ref{fig:FD}) and are provided at the end of this appendix for completeness.  
We provide results for the vector, axial, scalar and pseudoscalar form factors. 
Note in particular that results for the scalar case are identical to the pseudoscalar one.  
The DA is typically parametrized in terms of Appell polynomials~\cite{Braun:2014wpa} ($\varphi_{00}=1$) as follows 
\begin{align}
  \phi(x_1,x_2,x_3;\mu) = 120 x_1 x_2 x_3 \sum_{n=0}^{\infty} \sum_{k=0}^n \varphi_{nk}^N(\mu) \mathcal{P}_{nk}(x_i)  , 
\end{align}
a subset of which is given in Ref.~\cite{Braun:2014wpa} (see Eq.~(12) therein). 
These polynomials are the eigenfunctions of the DA evolution equation,  and their 
anomalous dimension are also provided in that reference.
Upon integration of Eq.~\eqref{eq:pQCDgen}, the result can be expressed in a more 
compact form, similar to that in Refs.~\cite{Huang:2024ugd,Chen:2024fhj}, as 
\begin{equation}\label{eq:pQCDint}
 Q^4 F(Q^2) = \frac{(4\pi\alpha_s f_N)^2}{54} 1800 \sum_{j \leq i=0}^{\infty} \sum_{l \leq k}^{ij} (Q_u u_0^{ijkl} +Q_d d_0^{ijkl})\varphi_{ij}^N(\mu) \varphi_{kl}^N(\mu) \, ,
\end{equation}
where $Q_{u,d}$ are the $u$ and $d$ charges of the associated current
and $u_{0}^{ijkl}$ and $d_{0}^{ijkl}$ are numerical coefficients given in Table~\ref{tab:PQCD} 
for each form factor (we remind that scalar and pseudoscalar cases are equivalent). 
These coefficients are in agreement with Refs.~\cite{Huang:2024ugd,Chen:2024fhj} 
for the vector case provided there. 
\begin{table}
\caption{\label{tab:PQCD}The values for $u_0^{ijkl}$ and $d_0^{ijkl}$ in Eq.~(\ref{eq:pQCDint}) 
for the different external currents.}
\begin{tabular}{c|cc|cc|cc} \toprule
         & \multicolumn{2}{|c|}{$V$}              & \multicolumn{2}{|c|}{$A$}               & \multicolumn{2}{c}{$S/P$} \\
  $ijkl$ & $u_0^{ijkl}$       & $d_0^{ijkl}$      & $u_0^{ijkl}$       & $d_0^{ijkl}$       & $u_0^{ijkl}$      & $d_0^{ijkl}$ \\ \midrule
  $0000$ & $1$                & $2$               & $-1$               & $-2$               & $-2$              & $-\frac{3}{2}$ \\ \midrule
  $1000$ & $\frac{49}{3}$     & $\frac{-49}{3}$   & $\frac{49}{3}$     & $-\frac{49}{3}$    & $28$              & $\frac{119}{3}$ \\
  $1010$ & $\frac{1519}{9}$   & $\frac{1568}{9}$  & $\frac{539}{3}$    & $\frac{1568}{3}$   & $0$               & $-\frac{3871}{18}$ \\ \midrule
  $1100$ & $7$                & $-7$              & $\frac{49}{3}$     & $\frac{35}{3}$     & $0$               & $-\frac{35}{3}$ \\
  $1110$ & $\frac{686}{9}$    & $\frac{-686}{9}$  & $\frac{686}{9}$    & -$\frac{686}{9}$   & $\frac{490}{3}$   & $147$ \\
  $1111$ & $\frac{539}{9}$    & $\frac{196}{9}$   & $49$               & $0$                & $\frac{98}{9}$    & $-\frac{343}{18}$ \\ \midrule
  $2000$ & $\frac{427}{10}$   & $\frac{161}{10}$  & $\frac{343}{10}$   & $-\frac{7}{10}$    & $\frac{112}{5}$   & $-\frac{217}{10}$ \\
  $2010$ & $\frac{5
  39}{3}$    & $\frac{-539}{3}$  & $\frac{539}{3}$    & $-\frac{539}{3}$   & $343$             & $\frac{7301}{30}$ \\
  $2011$ & $\frac{931}{5}$    & $\frac{98}{5}$    & $\frac{1029}{5}$   & $\frac{294}{5}$    & $\frac{2254}{15}$ & $-\frac{2401}{30}$ \\
  $2020$ & $\frac{23471}{100}$& $\frac{1372}{25}$ & $\frac{22589}{100}$& $\frac{931}{25}$   & $\frac{8673}{50}$ & $-\frac{13671}{200}$ \\ \midrule
  $2100$ & $\frac{35}{2}$     & $\frac{-35}{2}$   & $\frac{35}{2}$     & $-\frac{35}{2}$    & $35$ & $35$ \\
  $2110$ & $\frac{637}{2}$    & $\frac{637}{2}$   & $\frac{637}{2}$    & $\frac{637}{2}$    & $0$ & $-\frac{1127}{3}$ \\
  $2111$ & $\frac{539}{6}$    & $\frac{-539}{6}$  & $\frac{539}{6}$    & $-\frac{539}{6}$   & $147$ & $\frac{343}{3}$ \\
  $2120$ & $\frac{3381}{20}$  & $-\frac{3381}{20}$& $\frac{3381}{20}$  & $-\frac{3381}{20}$ & $\frac{3479}{10}$ & $\frac{2107}{10}$ \\
  $2121$ & $\frac{637}{4}$    & $\frac{637}{4}$   & $\frac{637}{4}$    & $\frac{637}{4}$    & $0$ & $-147$ \\ \midrule
  $2200$ & $\frac{32}{5}$     & $\frac{1}{5}$     & $\frac{38}{5}$     & $\frac{13}{5}$     & $\frac{34}{5}$    & $-\frac{19}{10}$ \\
  $2210$ & $\frac{49}{2}$     & $\frac{-49}{2}$   & $\frac{49}{2}$     & $-\frac{49}{2}$    & $\frac{112}{3}$ & $\frac{427}{30}$ \\
  $2211$ & $\frac{427}{30}$   & $\frac{161}{30}$  & $\frac{343}{30}$   & $\frac{-7}{30}$    & $\frac{196}{5}$ & $-\frac{287}{30}$ \\
  $2220$ & $\frac{6069}{100}$ & $\frac{987}{100}$ & $\frac{6321}{100}$ & $\frac{1491}{100}$ & $\frac{1106}{25}$ & $-\frac{917}{100}$ \\
  $2221$ & $\frac{287}{20}$   & $\frac{-287}{20}$ & $\frac{287}{20}$   & $-\frac{287}{20}$  & $\frac{483}{10}$ & $\frac{189}{10}$ \\
  $2222$ & $\frac{779}{100}$  & $\frac{43}{25}$   & $\frac{761}{100}$  & $\frac{34}{25}$    & $-\frac{93}{50}$  & $\frac{141}{200}$ \\  \bottomrule
\end{tabular}
\end{table}
The final result notably depends on the input for the DA and $f_N$, where great 
disparities are found across the literature. In Table~\ref{tab:PQCDdas}, we provide 
the results for the different DAs collected in Ref.~\cite{Braun:2014wpa} 
(see Table~IV therein). 
For completeness, we also provide the results for the axial and vector form factors, 
as well as the individual contributions for the $u$ and $d$ quarks. 
\begin{table}
\caption{\label{tab:PQCDdas}The pQCD prediction (in GeV$^4$ units and modulo $\alpha_s^2/Q^4$ overall factor) 
for different DAs in the literature (Latt~\cite{Braun:2014wpa}, KS~\cite{King:1986wi}, 
CZ~\cite{Chernyak:1984bm}, COZ~\cite{Chernyak:1987nu}, SB~\cite{Stefanis:1992nw}, 
BK~\cite{Bolz:1996sw}, BLW~\cite{Braun:2006hz}). 
Results are given at a renormalization scale $\mu=2$~GeV. 
The second row, $P/S$, stands for the isovector pseudoscalar form factor of interest here, 
whereas the third and fourth rows display the result for individual $u,d$ quarks. 
The following rows contain analogous results for axial and vector currents, respectively.}
\begin{tabular}{ccccccccc} \toprule
  & Latt & KS & CZ & COZ & SB & BK & BLW & ABO1/2 \\ \midrule
  $P/S$ & $-0.04(^{+0.40}_{-0.09})$ & $21$ & $14$ & $15$ & $24$ & $-0.04$ & $0.08$ & $1(0.4)/0.07(30)$   \\
  $P/S_u$ & $-0.15(^{+0.11}_{-0.16})$ & $18$ & $13.4$ & $14$ & $11$ & $-0.18$ & $0.05$ & $0.6(5)/-0.08(30)$   \\
  $P/S_d$ & $-0.11(^{+0.00}_{-0.40})$ & $-3$ & $-0.5$ & $-0.7$ & $-13$ & $-0.14$ & $-0.03$ & $-0.4(3)/-0.16(20)$   \\ \midrule
  $A_u$       & $-0.06(^{+0.38}_{-0.00})$ & $21.1$ & $15.2$ & $15.6$ & $22$ & $0.13$ & $0.30$ & $1.1(5)/0.22(22)$   \\
  $A_d$       & $-0.04(^{+0.19}_{-0.04})$ & $-0.4$ & $-0.5$ & $-0.7$ & $-7$ & $-0.5$ & $-0.253$ & $-0.15(10)/-0.28(9)$   \\ \midrule
  $V_u$       & $0.01(^{+0.40}_{-0.01})$ & $22$ & $15.5$ & $16.2$ & $23$ & $0.5$ & $0.5$ & $1.3(5)/0.48(25)$   \\
  $V_d$       & $0.10(^{+0.30}_{-0.05})$ & $2$ & $0.3$ & $0.6$ & $10$ & $0.3$ & $0.12$ & $0.4(2)/0.24(10)$   \\ \bottomrule
\end{tabular}
\end{table}

To close this appendix, we provide the hard-scattering amplitudes obtained from the diagrams in Fig.~\ref{fig:FD}, 
where $Q_q^{X}$ stands for the charge of the quark $q$ associated to each of the currents $X=\{ V,A,S,P \}$, 
\begin{align*}
  T_1^u &{}= \frac{ Q_{u}^{V(A)} [\phi_{123}\bar{\phi}_{123} +(2\bar{T})(2T)]  + Q_{u}^{S/P} [2\bar{T}\phi_{123} +\bar{\phi}_{123}2T]  }{u_3(u_2 +u_3)^2 v_3(v_2 +v_3)^2}  , \quad 
  T_2^{u} = T_4^{u} = 0 ,\\
  T_3^u &{}= \frac{ Q_{u}^{V(A)} [\phi_{123}\bar{\phi}_{123} +(2\bar{T})(2T)]  + Q_{u}^{S/P} [2\bar{T}\phi_{123} +\bar{\phi}_{123}2T]  }{u_2(u_2 +u_3)^2 v_2(v_2 +v_3)^2}  ,\\
  T_5^u &{}= -\frac{ Q_{u}^{V(A)}(2\bar{T})(2T) +Q_u^{S/P}(2\bar{T})\phi_{123}   }{u_2 u_3(u_2 +u_3) v_2 v_3(v_1 +v_3)}  , \quad
  T_6^u = -\frac{ Q_{u}^{V(A)}\bar{\phi}_{123}\phi_{123} +  Q_u^{S/P}\bar{\phi}_{123}(2T)   }{u_2 u_3(u_2 +u_3) v_2 v_3(v_1 +v_2)}  ,\\
  T_7^u &{}= \frac{ \pm Q_{u}^{V(A)}\bar{\phi}_{213}\phi_{213}  -  Q_u^{S/P}(2\bar{T})\phi_{123}   }{u_2 u_3(u_1 +u_2) v_2 v_3(v_1 +v_3)}  ,\\
  T_1^d &{}= \frac{ Q_{u}^{V(A)}[\bar{\phi}_{123}\phi_{123}  +\bar{\phi}_{213}\phi_{213}]   -Q_{u}^{S/P}[\bar{\phi}_{123}\phi_{213} +\bar{\phi}_{213}\phi_{123}]   }{u_1(u_1 +u_2)^2 v_1(v_1 +v_2)^2}  , \quad
  T_2^d = T_4^d = 0  \\
  T_3^d &{}= \frac{ Q_{u}^{V(A)}[\bar{\phi}_{123}\phi_{123}  +\bar{\phi}_{213}\phi_{213}]   -Q_{u}^{S/P}[\bar{\phi}_{123}\phi_{213} +\bar{\phi}_{213}\phi_{123}]   }{u_2(u_1 +u_2)^2 v_2(v_1 +v_2)^2}  ,\\
  T_5^d &{}= -\frac{ Q_{u}^{V(A)} \bar{\phi}_{213}\phi_{213}   + Q_{u}^{S/P}\bar{\phi}_{213}\phi_{123}   }{u_1 u_2(u_1 +u_2) v_1v_2(v_1 +v_3)}  , \quad
  T_6^d = \frac{ -Q_{u}^{V(A)}\bar{\phi}_{123}\phi_{123}   + Q_{u}^{S/P}\bar{\phi}_{123}\phi_{213}   }{u_1 u_2(u_1 +u_2) v_1 v_2 (v_2 +v_3)}  ,\\
  T_7^d &{}= \frac{ \pm Q_{u}^{V(A)}(2\bar{T})(2T)   + Q_{u}^{S/P}\bar{\phi}_{213}\phi_{123}   }{u_1 u_2(u_2 +u_3) v_1 v_2 (v_1 +v_3)}.
\end{align*}
Above, we have introduced the DA
$\phi_{123}\equiv \phi_{123}(u_1,u_2,u_3) = V(u_1,u_2,u_3) -A(u_1,u_2,u_3)$
following Refs.~\cite{Chernyak:1984bm,Braun:2000kw}. 
Permuted versions $\phi_{213} = \phi(u_2,u_1,u_3)$ are also used, and the 
combination 
$2T(u_1,u_2,u_3) = \phi_{132} +\phi_{231}$~\cite{Chernyak:1984bm,Braun:2000kw} 
is employed as well. The corresponding barred expressions refer to the final nucleon 
states and are obtained upon $u_i\to v_i$ exchange.
The result above is in agreement with the classical pQCD prediction for the 
vector form factor in Ref.~\cite{Chernyak:1984bm}.
Finally, we note that the scalar and pseudoscalar currents yield identical 
results, while the vector and axial currents differ only through the sign 
of the $T_7^q$ term.

\section{Remarks on form factors and EPCAC}  
\label{sec:EPCAC}

In this appendix we emphasize the distinction between 
the pseudoscalar form factor, associated to an external pseudoscalar 
current, and strong $\pi NN$ form factors employed in microscopic 
models based on meson-exchange potentials. To start with, note that,
in the narrow-resonance limit, one has EPCAC 
\begin{eqnarray}
\partial^\mu \vec A_\mu = F_\pi M_\pi^2 \vec \phi +   F_{\pi'} M_{\pi'}^2 \vec \phi'  +   \dots \, ,
\end{eqnarray}
where $\vec \phi$ represents the $\pi(140)$ state, $\vec \phi'$ the $\pi(1300)$ state and so on. Note that the pre-QCD PCAC corresponds to take $F_{\pi'}= \dots=0$ and corresponds to the pion pole dominance approximation. From EPCAC we have, for the pseudoscalar form factor, the relation 
\begin{eqnarray}
  m_q F_P(t) = \frac{F_\pi M_\pi^2}{M_\pi^2-t} g_{\pi NN} F_{\pi NN} (t) 
  + \frac{F_{\pi'} M_{\pi'}^2}{M_{\pi'}^2-t} g_{\pi' NN} F_{\pi' NN} (t) + \dots \, ,  
\end{eqnarray}  
where strong form factors (amputated normalized vertex functions) 
have been included ad-hoc (note this is not 
justified from a dispersive point of view). Such form factors
depend only on the momentum transfer $t$, since nucleons in the initial 
and final state are on the mass shell and fulfil the normalization 
condition $F_{\pi NN} (M_\pi^2)= F_{\pi' NN} (M_{\pi'}^2)= \dots = 1$. 
Thus, the $D(s)$ function in Eq.~(\ref{eq:Dt}) becomes
\begin{eqnarray}
D(t) =   F_\pi g_{\pi NN} F_{\pi NN} (t)  
+ F_{\pi'}\frac{ M_{\pi'}^2}{M_\pi^2}
\frac{M_{\pi}^2-t}{M_{\pi'}^2-t} g_{\pi' NN} F_{\pi' NN} (t) + \dots  \, ,
\end{eqnarray}
which clearly illustrates the difference between the strong form factors
and the pseudoscalar vertex with the pion pole removed. 
From the EPCAC condition at $t=0$ we get 
\begin{eqnarray}
  g_A M_N = F_\pi g_{\pi NN} F_{\pi NN} (0) + F_{\pi'} g_{\pi' NN}
  F_{\pi' NN} (0) + \dots\end{eqnarray} With this notation we may make
  contact with $NN$ elastic scattering. Specifically, the on-shell scattering amplitude in 
  Born approximation for $\mathcal{M}(N_1 N_2 \to N_{1'}N_{2'})$ reads
  $\mathcal{M}= (\bar u_1' i \gamma_5 \tau^a u_1)( \bar u_2' i \gamma_5
  \tau^a u_2) V_P(t) $ with
\begin{eqnarray}
V_{P} (t) = g_{\pi NN}^2 \frac{F_{\pi NN}(t)^2}{M_\pi^2-t} +  g_{\pi' NN}^2 \frac{F_{\pi' NN}(t)^2}{M_{\pi'}^2-t}+ \dots \, ,
\end{eqnarray}
where $t=(p_1-p_1')^2$ is the momentum transfer. A common practice is
to retain the $\pi$ piece alone (known as One-Boson-Exchange
approximation) in this channel. The difference with respect to the
pseudoscalar form factor of the nucleon and $F_{\pi NN}(t)$ is clear.
In particular, note that one cannot justify to associate the physics
of $F_{\pi NN}(t)$ to that of heavy pseudoscalar mesons.  As it is
well known, the Born approximation breaks unitarity which is restored
by interpreting $V_P(t)$ as {\it the potential} of the scattering
equation.

The monopole form factor ansatz for the vertex function
reads
\begin{eqnarray}
F_{\pi NN}(t)= \frac{\Lambda_{\pi NN}^2 -M_\pi^2}{\Lambda_{\pi NN}^2 -t } \, ,
\end{eqnarray}
normalized on-shell to unity $F_{\pi NN}(M_\pi^2)=1$. 
For an explicit NN calculation within one-boson-exchange potential 
models, see Ref.~\cite{Holinde:1990fe}. In that work, the inclusion 
of the $\pi(1300)$, introduced to set $\Lambda_{\pi NN}=0.8$ GeV and 
to quantitatively describe NN-scattering data below the pion-production 
threshold, led to
$g_{\pi NN}=13.5$, $g_{\pi'NN}=35.4$ and $\Lambda_{\pi' NN}=2$GeV 
as possible values.

\section{Regge models and asymptotics}
\label{sec:regge-asy}

In this appendix we justify the asymptotic power law of the spectral Regge tail, 
$s^{-2(1+\epsilon)}$  with $\epsilon=0.1-0.2$, based on Breit--Wigner resonance profiles and pQCD matching.

\begin{figure}\center
  \includegraphics[width=0.6\textwidth]{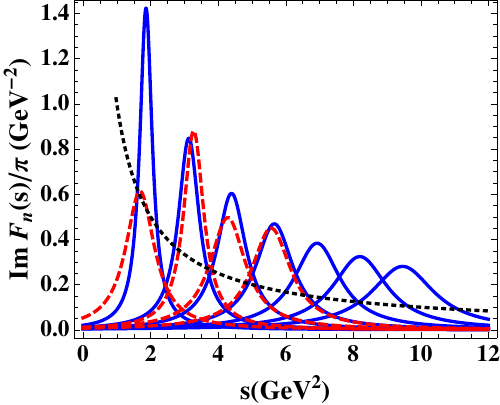}
\caption{ \small Normalised (i.e. equally weighted) Breit--Wigner profiles for the four 
experimental excited pseudoscalar $\pi_n$ states (red, dashed), 
and for the Regge tower of states with a constant Suranyi's ratio
 $\gamma=\Gamma_n/ M_n =0.12$ (blue, solid). For comparison, the $1/s$ behaviour is also shown (black, short-dashed).}
\label{fig:regge-ps}
\end{figure}

\subsection{Breit--Wigner and scaling}

\begin{figure}\center
\includegraphics[width=0.6\textwidth]{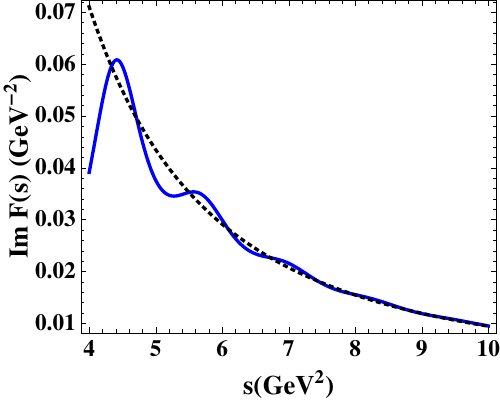}
\caption{ \small Superposition of Breit--Wigner profiles with $c_n \sim n^{-2(1+\epsilon)}$ weights and $\epsilon=0.1$ for the $n \ge 3$ excited pseudoscalar states (dashed-red) and the Regge tower of states with a constant Suranyi's ratio
  $\gamma=\Gamma_n/ M_n =0.12$ (blue, solid). We compare them with $s^{-2(1+\epsilon)}$ (short-dashed black line).}
\label{fig:regge-ps-power}
\end{figure}

To elaborate on the asymptotic behaviour of Regge models we assume,
for simplicity, an energy-independent Breit--Wigner lineshape for the
$\pi_n$-state 
\begin{eqnarray}
{\rm Im} F_{n} (s) =  {\rm Im } \frac{1}{M_n^2-s -i \Gamma_n M_n} 
                 =   \frac{\Gamma_n M_n}{(s-M_n^2)^2+ \Gamma_n^2 M_n^2} \, ,
\end{eqnarray}
which is normalized as $\int ds {\rm Im} F_{n} (s) = \pi$. For illustration purposes, we show in Fig.~\ref{fig:regge-ps}
the comparison between the  four experimental
excited pseudoscalar $\pi_n$ states and the Regge tower of states with
a constant Suranyi's ratio $\gamma=\Gamma_n/ M_n =0.12$ (see
Table~\ref{tab:ps-states}). 
Note that, in the limit $\gamma \to 0$, ${\rm Im} F_n(s) \to \pi \delta
(s-m_n^2)$. More generally, at the peak $s=M_n^2$, 
we have ${\rm Im} F_n(s)|_{s=M_n^2}= 1/(\Gamma_n M_n) = 1/(\gamma M_n^2) = 1/(\gamma s)$. 
Assuming then a Regge spectrum with a constant width-to-mass ratio, and considering a general superposition with couplings $c_n $, we get an expression for a form factor.
In particular, for $c_n = M_n^2 F_{\pi_n} g_{\pi_n NN} $, see Eq.~\eqref{eq:FPpsCont},
\begin{eqnarray}
  \operatorname{Im}m_q F_P(s) = \sum_n c_n \operatorname{Im}m_q F_n(s)  = 
  \sum_n c_n \frac{\gamma M_n^2}{(s-M_n^2)^2+ \gamma^2 M_n^4} \, .
\end{eqnarray}
For large $s$, we may replace the sum in $n$ by an integral, so that 
\begin{eqnarray}
  \operatorname{Im} m_q F_P (s)  & \to & \int dn   c_n \frac{\gamma M_n^2}{(s-M_n^2)^2+ \gamma^2 M_n^4} 
  = \int \frac{dM^2}a c_{M^2/a}  \frac{\gamma M^2}{(s-M^2)^2+ \gamma^2 M^4}
  \nonumber \\
  &=& \int dx \frac1{a} c_{s/a x}  \frac{\gamma x}{(1-x)^2 + \gamma^2 x^2} .
\end{eqnarray}
Clearly, for $c_n \sim n^{-\alpha} $ we have ${\rm Im} F_P(s) \to
s^{-\alpha} $. Of course, this simple behaviour applies to the higher
part of the excited spectrum and above. For a behaviour $c_n \sim
1/n^{2+2\epsilon}$, we get ${\rm Im} F_P(s) \to s^{-2-2\epsilon} $. In
Fig.~\ref{fig:regge-ps-power} we show the sum for $n \ge 3$ and
$\epsilon =0.1$ compared with the corresponding power behaviour. As we
see, the visible oscillations are small around the average value owing
to the overlap of resonances, which leads to an effective coarse
graining of the spectrum. This situation 
resembles the Dominguez model~\cite{Dominguez:1984ka} in its narrow-width limit, which is characterized by a large contribution of the
lowest-lying states and a mild contribution from the heavier excitations, 
that in addition come with opposite sign.

\begin{figure}\center
\includegraphics[width=0.6\textwidth]{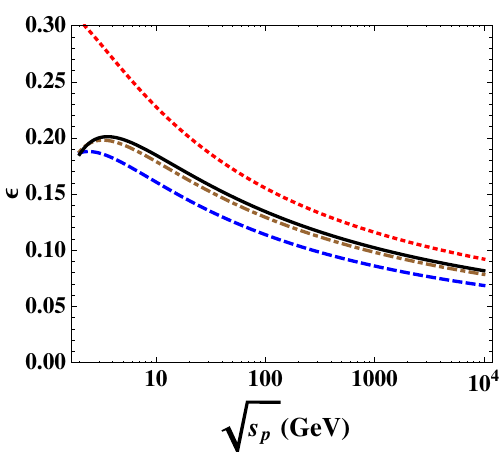}
\caption{ \small Exponent behaviour as a function of the matching scale
  of Regge asymptotics to pQCD. We compare the pion (red, dotted) with
  the nucleon in the cases without anomalous dimension ($\gamma=0$, dashed-blue line), 
  adding the quark mass $m_q$ anomalous dimension 
  ($\gamma=4/\beta_0$, dashed-brown line) and further 
  adding the nucleon wave function anomalous dimension ($\gamma=(4+ 4/3)/\beta_0$, full-black line).}
\label{fig:epsilon}
\end{figure}

\subsection{pQCD matching with Regge\label{sec:pQCDRegge}}

The matching of the logarithmic derivative of the Regge spectral function to
that of pQCD  at the matching point corresponds to take, up to inessential constants,
\begin{eqnarray}
\frac{d}{ds} \log {\rm Im} \alpha_s(s)^{2+ \gamma}  = -2 - 2 \epsilon  \, ,
\end{eqnarray}
where $\gamma = (4 + 4/3)/\beta_0 = 16/27$ (for $n_F=3$) represents the
combined anomalous dimension due to the quark mass, $m_q(\mu)$, and the
nucleon wave function at the origin, $f_N(\mu)^2$ (see Eqs.~\eqref{eq:mq-fN}).  
The strong-coupling constant is analytically continued from the space-like 
to the time-like region $s>0$ as 
$\alpha_s(s) = (4 \pi /\beta_0)/[\log (s/\Lambda_{\rm QCD}^2)- i \pi]$. 
This yields the condition
\begin{eqnarray}
\epsilon=\frac{(\gamma +2) \left(\pi \cot \left((\gamma +2) \tan
  ^{-1}\left(\frac{\pi }{\ln x}\right)\right)+\ln x\right)}{2 \left(\ln ^2 x+\pi ^2\right)}, 
\end{eqnarray}
where $x = \Lambda_{\rm pQCD}^2/\Lambda^2_{\rm QCD}$. For $\gamma\to 0$, 
this result reduces to Eq.~(\ref{eq:epsilonMatch}). 
Finally, to provide an estimate for $\epsilon$, we vary $\Lambda_{\rm pQCD} \in (2,10^4)$~GeV, 
producing the result in Fig.~\ref{fig:epsilon} and leading to our final estimate $\epsilon\in(0.1,0.2)$.

\section{Finite-width effects for isolated resonances}\label{sec:fin-width}

The determination of profiles is very much process dependent, and
hence model dependent, a feature which is in principle reinforced 
for wide resonances, as it happens to be the case for the first
radially excited pion, the $\pi(1300)$ with $\Gamma=400$MeV. In this
appendix, we consider several parameterizations which, despite their
great disparity, lead to moderate uncertainties in our estimate of the
GT discrepancy. These are, in fact, much smaller than those generated 
by the half-width rule consisting on using a narrow resonance profile but located at
$M_{\pi'} \pm \Gamma_{\pi'}/2$.

To assess the impact of the $\pi(1300)$ resonance profile of the 
spectral function in our model, we consider several parametrizations, 
while keeping its mass and width fixed to $M_{\pi'}=1300$~MeV and 
$\Gamma_{\pi'}=400$~MeV. 
To do so, we choose the following generic Breit--Wigner like 
description 
\begin{equation}
  \rho_{\pi(1300)}(s) = Z_{\pi'}\frac{M_R^2}{s - M_R^2 + i M_R \Gamma_R [f(s)/f(M_R^2)] }  \, ,
\end{equation}
incorporating an energy-dependent width through the $f(s)$ function. 
The latter is  varied from model to model as follows:
\begin{align}
  f_{\Pi_3}(s) &{}= \frac{\Pi_3(s)}{s -M_\pi^2} \simeq  2\sqrt{3}\pi M_\pi^2\left( \frac{\sqrt{s} -3M_\pi}{3M_\pi} \right)^2 \frac{1}{s-M_\pi^2} \, ,  \label{eq:prof1} \\
  f_{S}(s) &{}= \lambda^{1/2}(1,M_S^2/s,M_\pi^2/s), \quad \lambda(a,b,c) = a^2 +(b-c)^2 -2a(b+c) \, , \label{eq:prof2} \\
  f_{S}^\pi(s) &{}= \frac{f_S(s)}{s-M_\pi^2} \, , \label{eq:prof3}  \\
  f_{B_0}(s) &{}= B_0(s,M_\pi^2,M_S^2) -B_0(0,M_\pi^2,M_S^2) \, , \label{eq:prof4} \\
  f_{B_0}^\pi(s) &{}= f_{B_0}(s) \frac{s +15M_\pi^2}{s -M_\pi^2} \, .  \label{eq:prof5}
\end{align} 
First of all, since the $\pi(1300)$ resonance decays into 
$3\pi$, we have considered a pure phase-space dependence, 
(see $f_{\Pi_3}(s)$ above) incorporating a pion-pole enhancement
as suggested by ChPT at low-energies, see 
Eq.~\eqref{eq:SpectralChPT}. While the exact phase-space 
is employed in our calculations, we provide the threshold 
behaviour in Eq.~\eqref{eq:prof1}.
\begin{figure}\center
\includegraphics[width=0.5\textwidth]{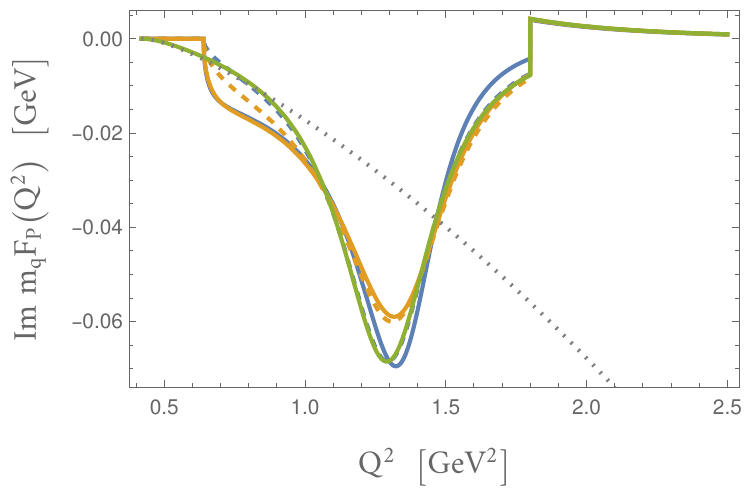}
\includegraphics[width=0.48\textwidth]{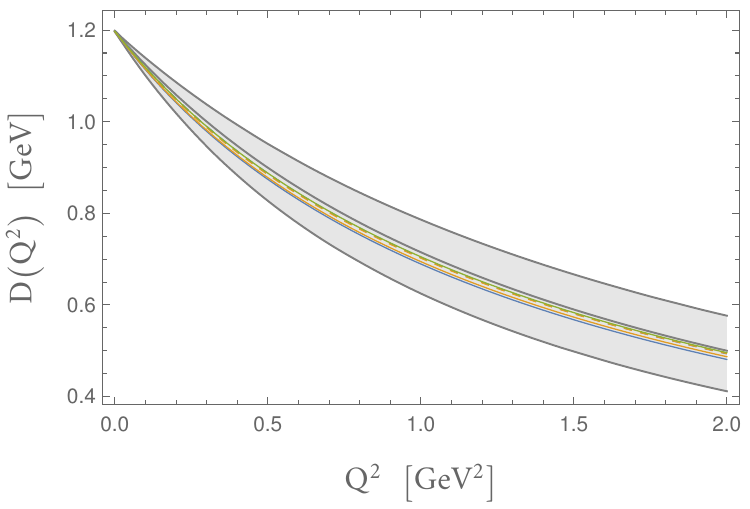}
\caption{Left: different spectral functions in the $3M_{\pi} < \sqrt{s} < 2$~GeV region. 
Variations induced from the chosen profile for the $\pi(1300)$ resonance (see description in the text). 
Right: the corresponding shifts in the space-like region for $D(-Q^2)$.}
\label{fig:1Rprofiles}
\end{figure}
The result is shown as a green 
line in Fig.~\ref{fig:1Rprofiles}. 
Second, the decay is expected to occur via intermediate 
$\pi(1300)\to\rho\pi/\sigma\pi$ states. However, it is up 
to date controversial which is the most favoured option. 
In order to better account for the low-energy behaviour, 
we choose the $\sigma\pi$ channel, which is pure $S$-wave, 
effectively closer to threshold and, therefore, a better 
candidate to match the ChPT result. To do so we choose 
two models: a pure $S$-wave like behaviour, $f_S(s)$, 
and the energy-dependent width motivated by the self-energy 
bubble graph, given in terms of the (subtracted) two-point scalar 
function $B_0$. These are shown as dashed blue and 
dashed orange lines, respectively. In a second step, we 
have incorporated the pion-pole enhancement, which 
is represented for both scenarios with a $\pi$ superscript in Eqs.~(\ref{eq:prof3},\ref{eq:prof5}) 
and are shown as full blue and orange lines in Fig.~\ref{fig:1Rprofiles}. 
Finally, our Eq.~\eqref{eq:SpectralChPT}, that reasonably 
interpolates the ChPT result, appears as a dotted line 
in Fig.~\ref{fig:1Rprofiles} (left). 
We note that,  for $f_{B_0}^\pi(s)$, stronger $\pi$-pole 
enhancements could be achieved. However, these would 
extremely overestimate the ChPT result. 
As shown in Fig.~\ref{fig:1Rprofiles}, variations are not strong in the 
space-like region, and are well within the half-width rule 
estimate. Note however that the profiles above are for 
fixed mass and widths, which vary significantly among 
experiments~\cite{ParticleDataGroup:2024cfk}. In 
particular, it is $M_{\pi(1300)}$ that controls the 
peak position and fully dominates the uncertainty. 

Regarding $\Delta_{\rm GT}$, each of the 
Eqs.~(\ref{eq:prof1}-\ref{eq:prof4}) shift 
our value $\Delta_{\rm GT} = 1.26\%$ to 
$(1.38, 1.38,1.48,1.41,1.46)\%$, within the half-width rule. 
This translates into a shift from $g_{\pi NN}= 13.14$ to 
$13.15, 13.15, 13.17, 13.16,13.17$, in accordance with 
our improved bound discussion in Sect.~\ref{sec:bound}. 

Eventually, once the properties of the 
$\pi(1300)$ ---including its pole parameters and underlying 
dynamics (or decay channels)--- are better understood, the current 
uncertainty which is attributed to the half-width rule, could be reduced. 
This urges for further experimental (or lattice) studies of 
the $\pi(1300)$.

\section{A glimpse on isospin breaking \label{app:ib}}

A pertinent question at this level of precision concerns isospin-breaking effects and the interpretation of our result. 
In order to elucidate this, we shall keep the Ward identity Eq.~\eqref{eq:WInoEM} as our central  
object, in line with Ref.~\cite{Sirlin:1972cs}. 
While the inclusion of strong-IB ($m_u\neq m_d$) effects is obvious, electromagnetic corrections modify 
Eq.~\eqref{eq:WInoEM} as 
\begin{equation}\label{eq:WIem}
  \partial_\mu\left(\bar{q}\gamma^{\mu}\gamma_5 t^i q \right) = \bar{q} i\gamma_5 \{ \mathcal{M}_q, t^i \}  q
       +ie A_{\mu} \bar{q}\gamma^{\mu}\gamma_5 [t^i, \mathcal{Q}_q ] q ,
\end{equation} 
where $t^i = \tau^i/2$ with $\tau^i$ the Pauli matrices in isospin space, 
$\mathcal{Q}_q$ stands for the quark charge matrix,  
and $A_\mu$ is the photon field. Clearly, the last term plays a role for charged currents, which is 
the case that we have discussed in our work.
 
The implications of the former equality for the form factors of interest, 
to leading order in $\alpha$, are
\begin{multline}\label{eq:WIemNP}
  i(p_f -p_i)_{\mu} \bra{p(p_f)}  \bar{q}\gamma^{\mu}\gamma_5 t^+ q \ket{n(p_i)} = 
  (m_u +m_d)  \bra{p(p_f)} \bar{q} i\gamma^5 t^+ q \ket{n(p_i)} \\ 
       -ie^2 \int \frac{d^4 k}{(2\pi)^4}\left( g^{\mu\nu} -\frac{k^\mu k^\nu}{k^2}\xi \right)
             \int d^4x e^{-i kx}   \bra{p(p_f)} T \{ A_{\mu}^+(0) \, V_{\nu}(x) \} \ket{n(p_i)}  ,
\end{multline} 
where $A_{\mu}^+ = \bar{q}\gamma^\mu\gamma^5 t^+ q$ is the axial current and 
$V_{\nu}= \bar{q}\gamma_\nu \mathcal{Q}_q q$ the electromagnetic current.
Of course, consistency demands estimates for the first and second terms at 
$\mathcal{O}(\alpha)$, whereas such effects can be neglected in the last term.
 
Concerning the first term, this can be expressed in terms of the standard form 
factors in Eq.~\eqref{eq:AxMatElIS}.\footnote{IB allows for 
a second-class contribution with tensor structure $i\sigma^{\mu\nu}q_{\nu}\gamma_5$, 
but it does not contribute to the divergence.} 
Here the form factors implicitly include electromagnetic corrections (as extracted 
from experiment; see subtleties below) and physical masses shall be employed, 
i.e. $2m_N \to (m_p +m_n)$. 

The terms remaining on the right-hand side require more care. 
Concerning the first one (the pseudoscalar form factor), we note that 
electromagnetic corrections will provide contributions of the 
$\mathcal{O}(m_q \alpha)$ kind, and therefore provide a small correction to the 
already suppressed quantity $\Delta_{\rm GT}$. 
Thus, the last term, that shall be added to our model
for the pseudoscalar piece with corresponding spectral functions, 
is expected to be the dominant electromagnetic correction.
For the pseudoscalar-pole contributions to the spectral function, 
this justifies keeping their residues in the form $F_{\pi^+} M_{\pi^+}^2 g_{\pi^+pn}$, 
where $M_{\pi^+}$ is the mass (including electromagnetic corrections).
Besides, the last term suggests the 
relevance of photonic intermediate states to the spectral function that 
should be incorporated. In the following, we focus on their contribution 
at $q^2=0$, which is the most relevant for the GT relation.

The central object is the electroweak tensor
\begin{equation}
  \int d^4x e^{-i kx}   \bra{p(p_f)} T \{ A_{\mu}^+(0) \, V_{\nu}(x) \} \ket{n(p_i)} \equiv
  \bar{u}_p T_{\mu\nu} u_n .
\end{equation}
We analyse its impact by modelling this tensor in the chiral limit and 
ignoring additional electromagnetic corrections, that would represent subleading 
$\mathcal{O}(M_\pi^2\alpha,\alpha^2)$ corrections. 
To do so, we split the problem into a low- and a high-energy part.

An important property at this point concerns the Ward 
identities obeyed by this tensor. Neglecting further $m_q$ and $\alpha$ 
corrections, and denoting $p_{V(A)}$ incoming (outgoing) momentum,
\begin{equation}\label{eq:WIEWTensor}
   ip_V^{\nu} \bar{u}_p T_{\mu\nu} u_n  
         =  \bra{p} \bar{q}\gamma_{\mu}\gamma_5 [\mathcal{Q}_q, t^+ ] q  \ket{n} ,  
  -ip_A^{\mu} \bar{u}_p T_{\mu\nu} u_n
         =  \bra{p} \bar{q}\gamma_{\nu}\gamma_5 [ t^+, \mathcal{Q}_q ] q  \ket{n} ,
\end{equation}
which, by contrast to the neutral current, is not conserved due to the non-vanishing 
commutator. Since $p_V^{\nu} \bar{u}_p T_{\mu\nu} u_n $ is independent 
of the photon momentum, the gauge-dependent term in Eq.~\eqref{eq:WIemNP} integrates 
to zero.

\subsection{Low energy elastic part}

\begin{figure}[t]
  \includegraphics[width=\textwidth]{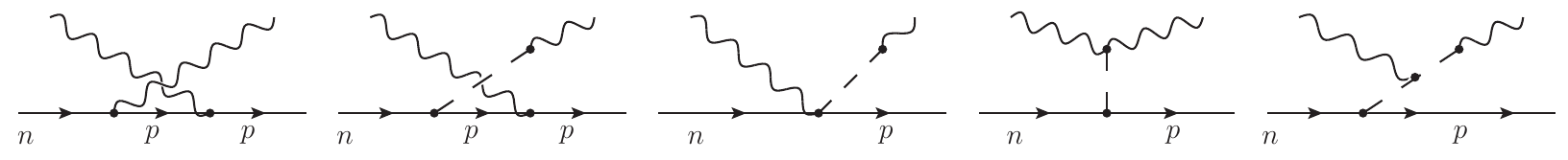}
  \caption{Feynman diagrams at LO in chiral perturbation theory contributing to the electroweak tensor $T^{\mu\nu}$\label{fig:FeynDiagVA}}
\end{figure}

The Ward identities in Eq.~\eqref{eq:WIEWTensor} prevent us from constructing a simple model 
with hadronic form factors. We therefore resort to a LO chiral perturbation theory calculation 
in the chiral limit. Potential pion-pole terms would need to be removed at the end to avoid 
double counting, but no such contributions arise. 
The relevant diagrams are shown in Fig.~\ref{fig:FeynDiagVA} and the result reads (in our 
approximation, $m_p=m_n\equiv m_N$ and $M_\pi =0$) 
\begin{multline}\label{eq:ChPTEWtensor}
  T^{\mu\nu} = i g_A^0\Bigg[\gamma^{\nu}\frac{(\slashed{p}_f -\slashed{p}_V) +m_N}{(p_f -p_V)^2 -m_N^2}\left( \gamma^\mu -\frac{\slashed{p}_A p_A^{\mu}}{p_A^2} \right)\gamma^5
     +\gamma^{\nu}\gamma_5 [t^+,\mathcal{Q}_q]\frac{p_A^{\mu}}{p_A^2} \\
     +\frac{2m_N}{(p_f -p_i)^2} \gamma_5 [t^+,\mathcal{Q}_q] \left( -g^{\mu\nu} +\frac{p_A^\mu (2p_A -p_V)^{\nu}}{p_A^2} \right)  \Bigg] .
\end{multline}

The result must be integrated over the loop momenta, see Eq.~\eqref{eq:WIemNP}.
The three terms without the nucleon propagator 
(corresponding to the last three diagrams) cancel, 
leaving only the terms with the nucleon pole and no pion-pole 
terms. The surviving piece is UV-finite, making the calculation 
predictive without the need of introducing form factors.  
The final result for Eq.~\eqref{eq:WIemNP} reads
\begin{multline}\label{eq:WIDx}
  i(p_f -p_i)_{\mu} \bra{p(p_f)}  \bar{q}\gamma^{\mu}\gamma_5 t^+ q \ket{n(p_i)} = 
  (m_u +m_d)  \bra{p(p_f)} \bar{q} i\gamma^5 t^+ q \ket{n(p_i)} \\ 
       - 2m_N g_A \bar{u}_p i\gamma_5 t^+ u_n \frac{\alpha}{4\pi}\left[ 1 +\frac{1}{2}\ln\left( \frac{m_N^2}{m_{\gamma}^2 -q^2} \right)  \right], 
\end{multline} 
where $m_{\gamma}$ is an IR photon mass, and that 
we shall adopt as an estimate for additional contributions at $q^2\to 0$. 
Once we incorporate the last term to 
Eqs.~(\ref{eq:PCAC},\ref{eq:SR1org},\ref{eq:resultseq1}), the following shift 
is found:
\begin{equation}\label{eq:gAmod}
  \Delta_{\rm GT} \to \Delta_{\rm GT} 
  + \frac{\alpha}{4\pi}\left[ 1 +\ln\left( \frac{m_N}{m_{\gamma}} \right) \right] +\mathcal{O}(\alpha, m_q/\Lambda_{\rm QCD}) .
\end{equation}
One should notice that the result contains IR divergences. 
This is to be expected as outlined in Ref.~\cite{Sirlin:1972cs} and 
forbids an unambiguous definition of the GT discrepancy once EM 
corrections are accounted for. 
Of course, when embedded in a full process, all such divergences cancel 
out~\cite{Sirlin:1972cs}.
In this sense, one could drop such terms in Eq.~\eqref{eq:gAmod} 
(which is analogous to the treatment in \cite{Sirlin:1972cs}), implying a negligible 
$\alpha/(4\pi)$ correction. Another alternative, in the context of neutron decay, 
would be to set $m_{\gamma}$ to be a typical scale in such process, such as 
$m_n -m_p, m_e$, which amounts to a $0.5\%$ correction, again mild. 
A more robust and quantitative assessment would require a dedicated 
and more thorough study of radiative corrections, also for all the processes which inputs enter 
the GT relation, which is beyond the scope of this work.\footnote{Note 
in this context Ref.~\cite{Hoferichter:2016duk}.} 
Note once more here that radiative corrections 
shifting the extraction of $\lambda\to g_A$ from experiment have 
been estimated in Ref.~\cite{Gorchtein:2021fce} to be $\mathcal{O}(10^{-4})$.

Our result might appear to contrast with the recent findings of 
Ref.~\cite{Cirigliano:2022hob}, which, unlike earlier studies such as 
\cite{Gorchtein:2021fce}, report substantial corrections of order $(1-2)\%$. 
This shift is largely driven by electromagnetic corrections 
to the charged-pion mass, and allow the authors to estimate the shift 
of $g_A$ from the pure QCD scenario to the real world, 
including electromagnetic corrections.   
In our approach, $g_A, F_{\pi^+},M_{\pi^+}$ are the experimental ones, 
that are meant to incorporate electromagnetic 
corrections modulo IR ambiguities. 

In summary, isospin-breaking corrections justify the use of 
physical masses in our model for the spectral function. 
Besides, we adopt our simple model estimate for the additional 
contribution at $q^2=0$ as a suggestive estimate of further IB uncertainties 
to $\Delta_{\rm GT}$ (cf. Eq.~\eqref{eq:gAmod}) of order $0.5\%$. 
The high-energy contribution, discussed below, is 
found to be negligible.

\subsection{High-energy part (off-forward)}

In order to evaluate the high-energy contribution, we employ 
the operator product expansion (OPE), obtaining for large $k$
\begin{equation}\label{eq:OPE1}
  \int d^4 x e^{-i kx} \ T\{ A_{\mu}^+(0) V^{\mu}(x) \} = 
  -i \bar{q}\big( 2\slashed{k} +i \overset{\leftrightarrow}{\slashed{\partial}}_+  -4m  \big)\gamma_5\Gamma_-q
  +\bar{q} \overset{\leftrightarrow}{\slashed{\partial}}_+ \gamma_5 \Gamma_+ q \, ,
\end{equation} 
with
\begin{equation}
 \Gamma_\pm = \frac{t^+Q}{(k +i\overset{\rightarrow}{\partial})^2 -m^2} \pm
              \frac{Qt^+}{(k -i\overset{\leftarrow}{\partial})^2 -m^2} \, , \quad 
              \overset{\leftrightarrow}{\partial^{\mu}}_{\pm} =  \overset{\rightarrow}{\partial^{\mu}}  \pm  \overset{\leftarrow}{\partial^{\mu}} \, .  
\end{equation}
Note that the last term in Eq.~\eqref{eq:OPE1} vanishes to leading order in $\alpha_s$ 
upon the use of the equations of motion in the isospin limit.\footnote{Of course, 
the picture holds when evaluated between nucleon states, as one can easily 
explicitly verify.}
Since we are interested in the $q^2\to 0$ limit of Eq.~\eqref{eq:WIemNP}, 
we keep up to linear terms in $q$ that may appear from $k^\mu$ 
integration. Expanding denominators as 
$1/(k +i\overset{\rightarrow}{\partial})^{2} = (1/k^2)\sum_n (-ik\cdot \overset{\leftrightarrow}{\partial}_{+} -ik\cdot \overset{\leftrightarrow}{\partial}_-)^n / k^{2n}$
and anticipating the angular averaging in the loop integration, 
one finds the leading term
\begin{equation}
  \int d^4 x e^{-i kx} \ T\{ A_{\mu}^+(0) V^{\mu}(x) \} \to  \frac{1}{2 k^2}\partial_\mu \left( \bar{q} \gamma^\mu\gamma_5 [t^+,Q] q \right) -\frac{4m}{k^2} \left(\bar{q} i\gamma_5 [t^+,Q]q \right) \, ,
\end{equation}
which vanishes in the chiral limit, as speculated back in 
Ref.~\cite{Sirlin:1972cs}. This demonstrates that the high-energy 
contribution scales as $(\alpha/\pi)(m_q/\Lambda_{\rm QCD})$ and is, 
therefore, highly suppressed.

\section{Dispersion relations and asymptotic logarithmic behaviour\label{app:DRasymp}} 

The asymptotic $\alpha_s$-induced logarithmic behaviour of 
form factors has been traditionally ignored in the literature, 
both because of its slowly logarithmic variation and 
the uncertainty regarding its onset.
Some relevant cases in this study where this has been discussed, 
in more or less detail, are Refs.~\cite{Donoghue:1996bt,Leutwyler:2002hm,RuizArriola:2025wyq}.
In the following, we provide some details calrifying our rationale 
behind our treatment of this effect. 

To such end, consider an analytic function $f(s)$ with singularities 
given by isolated poles and a right-hand cut above at $s=s_0$. 
Assume moreover that $f(-t)$ behaves asymptotically as $1/[t^n\ln^m (t/\Lambda^2)]$ with $n,m >0$. 
It is easy to prove that, for $l \leq n$,
\begin{equation}
 \lim_{t\to\infty} t^l f(-t) = 0 = \frac{1}{\pi}\int_{s_0}^{\infty} dx \ x^{l-1} \operatorname{Im} f(x)  \, ,
\end{equation} 
which are the corresponding sum rules outlined in the main text. 

\begin{figure}\centering
   \includegraphics[width=0.47\textwidth]{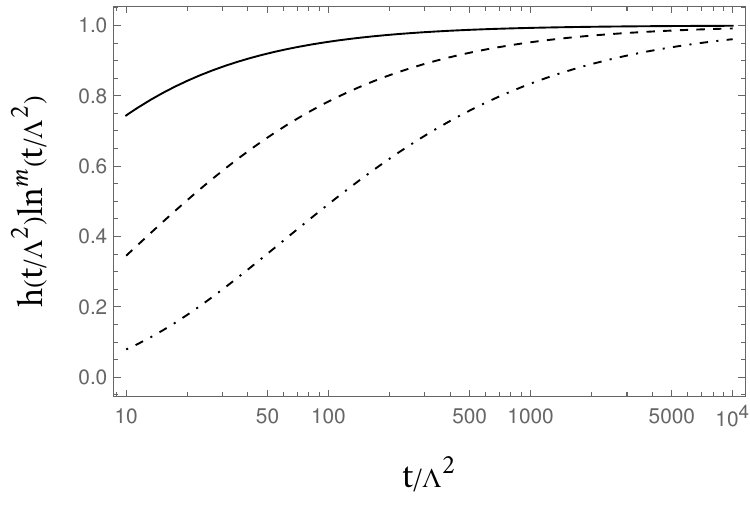}
   \caption{\label{fig:DRexamp}We compare the DR in Eq.~\eqref{eq:DRasympt} to the corresponding asymptotic 
   $\ln^m(t/\Lambda^2)$ behaviour for $m=1$ (full line), $m=2$ (dashed line), $m=3$ (dot-dashed line).
   }
\end{figure}

While these are necessary conditions, the vanishing sum rule for 
$l=n$ applies not only to the desired $1/[t^n\ln^m (t/\Lambda^2)]$ 
suppression, but also to a stronger $1/t^{n+1}$ suppression. 
To guarantee the correct asymptotic behaviour, an appropriate 
asymptotic spectral function must be chosen. 
To motivate and justify our choice, it is convenient to start with 
the standard Cauchy's representation 
\begin{align}
  f(-t) = \frac{1}{2\pi} \oint\limits_{C} dx \ \frac{\operatorname{Im} f(x)}{x+t} 
        &{}=  \frac{1}{2\pi}\oint\limits_{C} dx  \operatorname{Im} f(x) \left(  \frac{(-x)^{n}}{t^n(x+t)} +\frac{1}{t} -\frac{x}{t^2} + ... +\frac{(-x)^{n-1}}{t^{n}} \right) \nonumber \\
        &{}=  \frac{1}{2\pi}\oint\limits_{C} dx  \operatorname{Im} f(x)  \frac{(-x)^{n}}{t^n(x+t)}  , \label{eq:CRepr}
\end{align}
where the usual contour is understood, and the last equality follows 
from the sum rules.
This representation shows that any finite set of pole contributions 
decays as $1/t^{n+1}$; hence the logarithmic suppression 
for large but finite $t$ arises from the branch cut above. 
To reproduce the desired asymptotics and inspired by the  
Phragm{\'e}n--Lindelh{\"o}f theorem  ---that guarantees a similar 
asymptotic behaviour on the cutted complex plane--- we adopt the substitution 
$\ln (t/\Lambda^2) \to \ln (s/\Lambda^2)-i\pi$ discussed in the text. 
This leads to  
\begin{equation}\label{eq:DRasympt}
  \lim_{t\to\infty} t^n f(-t) =  \frac{1}{\pi}\int_{s_0}^{\infty} dx \ \operatorname{Im} \left(\frac{1}{\ln (x/\Lambda^2) -i\pi}\right)^m \frac{1}{x+t} \equiv h_{m}(t/\Lambda^2) \, ,
\end{equation}   
which reproduces the leading asymptotic behavior. To show this, 
note that Cauchy's theorem applied to $1/\ln^m(-t)$ leads to 
\begin{equation}
  \int_0^{\infty} \operatorname{Im} \left(\frac{1}{\ln x -i\pi}\right)^m \frac{dx}{x+t}
     = \frac{1}{\ln^m t} - \sum_{k=1}^m  \frac{c_k^{(m)}}{(t-1)^k},
\end{equation}
with $c_{k}^{(m)}$ given by the Laurent expansion of $1/\ln^m(t)$ at $t=1$ 
(e.g., $c_2^{(2)} = c_1^{(2)} = 1$).
In consequence, 
\begin{align}
  \int_{x_0}^{\infty} \operatorname{Im} \left(\frac{1}{\ln x -i\pi}\right)^m \frac{dx}{x+t}
     &{}= \frac{1}{\ln^m t} - \sum_{k=1}^m \frac{c_k^{(m)}}{(t-1)^k}
        +\int_0^{x_0} \operatorname{Im} \left(\frac{1}{\ln x -i\pi}\right)^m \frac{dx}{x+t} \nonumber\\
     &{}= \frac{1}{\ln^m t} + \mathcal{O}(1/t)   \, .
\end{align}
To further illustrate this, we show 
$h_{m}(t/\Lambda^2) \ln^m(t/\Lambda^2)$ for different values of $m$
in Fig.~\ref{fig:DRexamp}.
It is worth emphasizing that, imposing the 
asymptotic spectral function introduced above for some $s>s_0$ is 
useless unless the sum rules are satisfied. Since this is not the 
case in practice, one must appeal to the pertinent physics before 
the effective onset of pQCD to ensure the 
validity of the sum rules, which in our case is parametrized in 
terms of Regge physics, see Ref.~\cite{RuizArriola:2025wyq}.

\section{Statistical analysis of the  $\pi(1300)$ \label{app:pi1300}}

For our error analysis, the large width of the $\pi(1300)$ plays a
relevant, though not crucial, role. While the PDG recommended values lie
in the range $\Gamma_{\pi'}=(0.2\text{--}0.6)\,\textrm{GeV}$ (so that one naturally would take
$\Gamma_{\pi'}=0.4 \pm 0.2 \,\textrm{GeV}$), we will show that this is in fact a
very conservative estimate which assumes that all quoted values are
eligible. Nonetheless, one should also keep in mind that these values
are not obtained typically as poles of a scattering or decay amplitude
in the second Riemann sheet.

In the latest 2024 edition, twelve measurements are
quoted~\cite{dArgent:2017gzv,Shchegelsky:2006es,OBELIX:2004oio,Chung:2002pu,CrystalBarrel:2001xud,OBELIX:1997zla,CrystalBarrel:1996wfh,Zielinski:1984mt,Bellini:1982ec,Aaron:1980zk,Bonesini:1981sx,ACCMOR:1980llh}. For our analysis, we will only take into account the 10
measurements for which uncertainties are quoted, thus 
excluding~\cite{CrystalBarrel:1996wfh,ACCMOR:1980llh}. 
These are provided in Table~\ref{tab:Gammapi1300} combining the statistical and systematic uncertainties in quadrature when available from the selected references.

\begin{figure}\center
\includegraphics[width=0.5\textwidth]{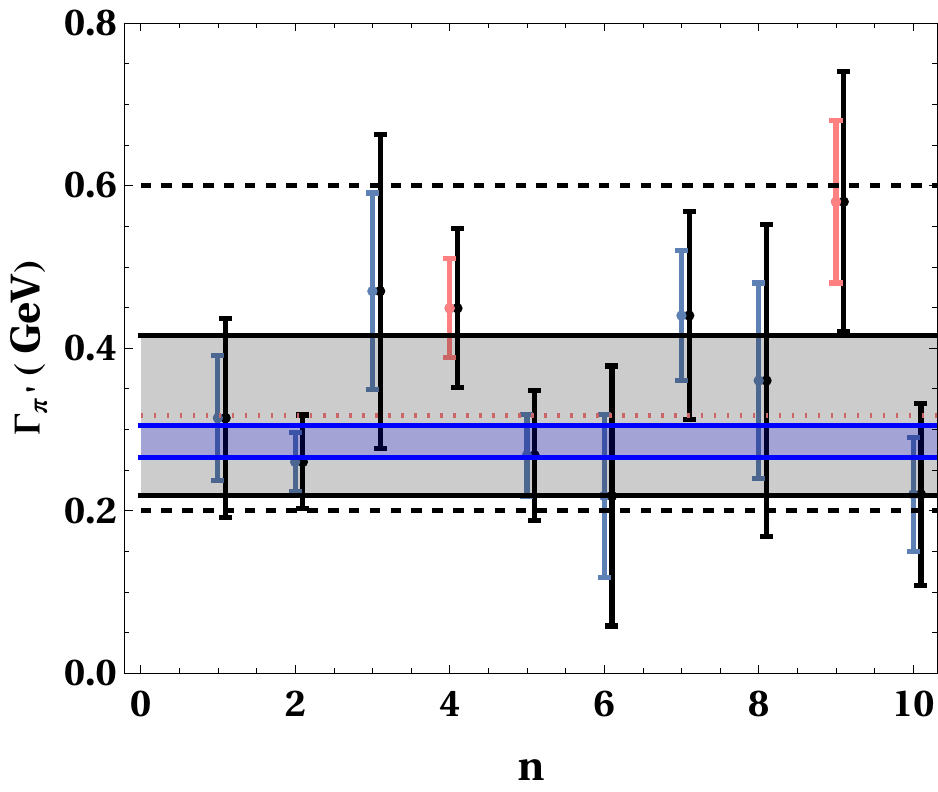}
\caption{ \small Listed PDG widths with uncertainties for the
  $\pi(1300)$ meson shown as blue error bars. The thin black dashed lines represent the
  interval of recommended values by the PDG. The pink dotted line
  represents the initial fit to {\it all} data. The narrow blue band
  represents the statistically meaningful fit to the eight mutually
  compatible measurements, obtained after discarding two outliers (shown in pink).
  The wide grey band represents the fit obtained after applying the Birge-factor enlarged
  uncertainties, corresponding to the enlarged black error bars, 
  slightly displaced to the original estimates.}
\label{fig:gammapi1300}
\end{figure}

\begin{table}
    \caption{Widths with errors reported in the PDG~\cite{ParticleDataGroup:2024cfk}
    for the $\pi(1300)$ state from $n=10$ experiments.}\label{tab:Gammapi1300}    
\begin{tabular}{ccccccccccc}\toprule
  $n$ & 1 
  & 2 
  & 3 
  & 4  & 5  & 6  & 7 & 8  & 9  & 10  \\
  Ref. & \cite{dArgent:2017gzv}
  &  \cite{Shchegelsky:2006es}
  &  \cite{OBELIX:2004oio}
  &  \cite{Chung:2002pu} &  \cite{CrystalBarrel:2001xud} &  \cite{OBELIX:1997zla} &  \cite{Zielinski:1984mt} &  \cite{Bellini:1982ec} &  \cite{Aaron:1980zk} &  \cite{Bonesini:1981sx} \\
\midrule
 $\Gamma_{\pi'}$ & 0.314 & 0.26 & 0.47 & 0.449 & 0.268 & 0.218 & 0.44 & 0.36 & 0.58 & 0.22 \\
 $\Delta\Gamma_{\pi'}$ & 0.076 & 0.03 & 0.12 & 0.061 & 0.050 & 0.1 & 0.08 & 0.12 & 0.1 & 0.07 \\ 
 \bottomrule 
\end{tabular}
\end{table}
A simple $\chi^2$ fit to these ten measurements yields
$\Gamma_{\pi'}=0.317$GeV with $\chi^2/\nu= 22/(10-1)=2.44$, which is $3
\sigma$ away from the expected $\chi^2= 1\pm \sqrt{2/\nu} = 1 \pm
0.44$ for an acceptable confidence level. Consequently, no error propagation
can be performed. There are two possible strategies to address this
problem: data selection or enlarging the uncertainties by means of a Birge factor.
Both approaches are discussed in what follows.
\begin{itemize}
\item \textbf{Data selection.} In order to select mutually compatible data, we  
attribute this incompatibility to the presence of outliers. 
We proceed sequentially by eliminating the data point with the largest contribution to the
$\chi^2$ and refitting the remaining data until the condition $\chi^2= 1\pm
\sqrt{2/\nu}$ is satisfied. The data $n=4$ and $n=9$ are then nominated
as outliers and the resulting fit yields
\begin{eqnarray}
\Gamma_{\pi'}=0.28(2)\,{\rm GeV} \, , \qquad \chi^2/\nu = 8.5/(8-1)=1.22\,, 
\end{eqnarray}
which is acceptable since $\chi^2/\nu$ falls within the interval $ 1
\pm \sqrt{2/(8-1)}= 1 \pm 0.5 $ The situation is illustrated in
Fig.~\ref{fig:gammapi1300}.
\item \textbf{Uncertainty enlargement.} If we assume that all data are compatible but uncertainties are too conservative, 
we may rescale the errors by a Birge factor $B=\sqrt{\chi_{\rm min}^2/\nu}=1.6$ so that 
$\overline{\Delta \Gamma}_{\pi'} = 1.6 \,\Delta \Gamma_{\pi'} $ we get
\begin{eqnarray}
\Gamma_{\pi'}=0.32(10) {\rm GeV} \, , \qquad \overline{\chi}^2/\nu \equiv 1 \,.
\end{eqnarray}
The results are also shown in Fig.~\ref{fig:gammapi1300}.
We remind
that Birge factors of 1.2 are compatible with the error of the
error. In this particular case the large 1.6 factor may be interpreted
as a systematic uncertainty of the width determination compared to the
rigurous determination as a pole in the second Riemann sheet.
\end{itemize}  
In summary, including 2 outliers or not yields a compatible width but
about five times larger uncertainty.

\end{appendices}


\bibliography{sn-bibliography}


\begin{thebibliography}{115}
\ifx \bisbn   \undefined \def \bisbn  #1{ISBN #1}\fi
\ifx \binits  \undefined \def \binits#1{#1}\fi
\ifx \bauthor  \undefined \def \bauthor#1{#1}\fi
\ifx \batitle  \undefined \def \batitle#1{#1}\fi
\ifx \bjtitle  \undefined \def \bjtitle#1{#1}\fi
\ifx \bvolume  \undefined \def \bvolume#1{\textbf{#1}}\fi
\ifx \byear  \undefined \def \byear#1{#1}\fi
\ifx \bissue  \undefined \def \bissue#1{#1}\fi
\ifx \bfpage  \undefined \def \bfpage#1{#1}\fi
\ifx \blpage  \undefined \def \blpage #1{#1}\fi
\ifx \burl  \undefined \def \burl#1{\textsf{#1}}\fi
\ifx \doiurl  \undefined \def \doiurl#1{\url{https://doi.org/#1}}\fi
\ifx \betal  \undefined \def \betal{\textit{et al.}}\fi
\ifx \binstitute  \undefined \def \binstitute#1{#1}\fi
\ifx \binstitutionaled  \undefined \def \binstitutionaled#1{#1}\fi
\ifx \bctitle  \undefined \def \bctitle#1{#1}\fi
\ifx \beditor  \undefined \def \beditor#1{#1}\fi
\ifx \bpublisher  \undefined \def \bpublisher#1{#1}\fi
\ifx \bbtitle  \undefined \def \bbtitle#1{#1}\fi
\ifx \bedition  \undefined \def \bedition#1{#1}\fi
\ifx \bseriesno  \undefined \def \bseriesno#1{#1}\fi
\ifx \blocation  \undefined \def \blocation#1{#1}\fi
\ifx \bsertitle  \undefined \def \bsertitle#1{#1}\fi
\ifx \bsnm \undefined \def \bsnm#1{#1}\fi
\ifx \bsuffix \undefined \def \bsuffix#1{#1}\fi
\ifx \bparticle \undefined \def \bparticle#1{#1}\fi
\ifx \barticle \undefined \def \barticle#1{#1}\fi
\bibcommenthead
\ifx \bconfdate \undefined \def \bconfdate #1{#1}\fi
\ifx \botherref \undefined \def \botherref #1{#1}\fi
\ifx \url \undefined \def \url#1{\textsf{#1}}\fi
\ifx \bchapter \undefined \def \bchapter#1{#1}\fi
\ifx \bbook \undefined \def \bbook#1{#1}\fi
\ifx \bcomment \undefined \def \bcomment#1{#1}\fi
\ifx \oauthor \undefined \def \oauthor#1{#1}\fi
\ifx \citeauthoryear \undefined \def \citeauthoryear#1{#1}\fi
\ifx \endbibitem  \undefined \def \endbibitem {}\fi
\ifx \bconflocation  \undefined \def \bconflocation#1{#1}\fi
\ifx \arxivurl  \undefined \def \arxivurl#1{\textsf{#1}}\fi
\csname PreBibitemsHook\endcsname

\bibitem[\protect\citeauthoryear{Yukawa}{1935}]{Yukawa:1935xg}
\begin{barticle}
\bauthor{\bsnm{Yukawa}, \binits{H.}}:
\batitle{{On the Interaction of Elementary Particles I}}.
\bjtitle{Proc. Phys. Math. Soc. Jap.}
\bvolume{17},
\bfpage{48}--\blpage{57}
(\byear{1935})
\doiurl{10.1143/PTPS.1.1}
\end{barticle}
\endbibitem

\bibitem[\protect\citeauthoryear{Bethe}{1940a}]{Bethe:1940iba}
\begin{barticle}
\bauthor{\bsnm{Bethe}, \binits{H.A.}}:
\batitle{{The Meson Theory of Nuclear Forces I. General Theory}}.
\bjtitle{Phys. Rev.}
\bvolume{57},
\bfpage{260}--\blpage{272}
(\byear{1940})
\doiurl{10.1103/PhysRev.57.260}
\end{barticle}
\endbibitem

\bibitem[\protect\citeauthoryear{Bethe}{1940b}]{Bethe:1940zz}
\begin{barticle}
\bauthor{\bsnm{Bethe}, \binits{H.A.}}:
\batitle{{The Meson Theory of Nuclear Forces. Part 2. Theory of the Deuteron}}.
\bjtitle{Phys. Rev.}
\bvolume{57},
\bfpage{390}--\blpage{413}
(\byear{1940})
\doiurl{10.1103/PhysRev.57.390}
\end{barticle}
\endbibitem

\bibitem[\protect\citeauthoryear{Goldberger and
  Treiman}{1958}]{Goldberger:1958tr}
\begin{barticle}
\bauthor{\bsnm{Goldberger}, \binits{M.L.}},
\bauthor{\bsnm{Treiman}, \binits{S.B.}}:
\batitle{{Decay of the pi meson}}.
\bjtitle{Phys. Rev.}
\bvolume{110},
\bfpage{1178}--\blpage{1184}
(\byear{1958})
\doiurl{10.1103/PhysRev.110.1178}
\end{barticle}
\endbibitem

\bibitem[\protect\citeauthoryear{Navas
  et~al.}{2024}]{ParticleDataGroup:2024cfk}
\begin{barticle}
\bauthor{\bsnm{Navas}, \binits{S.}}, \betal:
\batitle{{Review of particle physics}}.
\bjtitle{Phys. Rev. D}
\bvolume{110}(\bissue{3}),
\bfpage{030001}
(\byear{2024})
\doiurl{10.1103/PhysRevD.110.030001}
\end{barticle}
\endbibitem

\bibitem[\protect\citeauthoryear{Gorchtein and Seng}{2021}]{Gorchtein:2021fce}
\begin{barticle}
\bauthor{\bsnm{Gorchtein}, \binits{M.}},
\bauthor{\bsnm{Seng}, \binits{C.-Y.}}:
\batitle{{Dispersion relation analysis of the radiative corrections to g$_{A}$
  in the neutron {\ensuremath{\beta}}-decay}}.
\bjtitle{JHEP}
\bvolume{10},
\bfpage{053}
(\byear{2021})
\doiurl{10.1007/JHEP10(2021)053}
{\href{https://arxiv.org/abs/2106.09185}{{arXiv:2106.09185}}}
{[hep-ph]}
\end{barticle}
\endbibitem

\bibitem[\protect\citeauthoryear{Cirigliano et~al.}{2022}]{Cirigliano:2022hob}
\begin{barticle}
\bauthor{\bsnm{Cirigliano}, \binits{V.}},
\bauthor{\bsnm{Vries}, \binits{J.}},
\bauthor{\bsnm{Hayen}, \binits{L.}},
\bauthor{\bsnm{Mereghetti}, \binits{E.}},
\bauthor{\bsnm{Walker-Loud}, \binits{A.}}:
\batitle{{Pion-Induced Radiative Corrections to Neutron {\ensuremath{\beta}}
  Decay}}.
\bjtitle{Phys. Rev. Lett.}
\bvolume{129}(\bissue{12}),
\bfpage{121801}
(\byear{2022})
\doiurl{10.1103/PhysRevLett.129.121801}
{\href{https://arxiv.org/abs/2202.10439}{{arXiv:2202.10439}}}
{[nucl-th]}
\end{barticle}
\endbibitem

\bibitem[\protect\citeauthoryear{Gorchtein and Seng}{2023}]{Gorchtein:2023srs}
\begin{barticle}
\bauthor{\bsnm{Gorchtein}, \binits{M.}},
\bauthor{\bsnm{Seng}, \binits{C.-Y.}}:
\batitle{{The Standard Model Theory of Neutron Beta Decay}}.
\bjtitle{Universe}
\bvolume{9}(\bissue{9}),
\bfpage{422}
(\byear{2023})
\doiurl{10.3390/universe9090422}
{\href{https://arxiv.org/abs/2307.01145}{{arXiv:2307.01145}}}
{[hep-ph]}
\end{barticle}
\endbibitem

\bibitem[\protect\citeauthoryear{Seng}{2024}]{Seng:2024ker}
\begin{barticle}
\bauthor{\bsnm{Seng}, \binits{C.-Y.}}:
\batitle{{Hybrid analysis of radiative corrections to neutron decay with
  current algebra and effective field theory}}.
\bjtitle{JHEP}
\bvolume{07},
\bfpage{175}
(\byear{2024})
\doiurl{10.1007/JHEP07(2024)175}
{\href{https://arxiv.org/abs/2403.08976}{{arXiv:2403.08976}}}
{[hep-ph]}
\end{barticle}
\endbibitem

\bibitem[\protect\citeauthoryear{Ma et~al.}{2024}]{Ma:2023kfr}
\begin{barticle}
\bauthor{\bsnm{Ma}, \binits{P.-X.}},
\bauthor{\bsnm{Feng}, \binits{X.}},
\bauthor{\bsnm{Gorchtein}, \binits{M.}},
\bauthor{\bsnm{Jin}, \binits{L.-C.}},
\bauthor{\bsnm{Liu}, \binits{K.-F.}},
\bauthor{\bsnm{Seng}, \binits{C.-Y.}},
\bauthor{\bsnm{Wang}, \binits{B.-G.}},
\bauthor{\bsnm{Zhang}, \binits{Z.-L.}}:
\batitle{{Lattice QCD Calculation of Electroweak Box Contributions to
  Superallowed Nuclear and Neutron Beta Decays}}.
\bjtitle{Phys. Rev. Lett.}
\bvolume{132}(\bissue{19}),
\bfpage{191901}
(\byear{2024})
\doiurl{10.1103/PhysRevLett.132.191901}
{\href{https://arxiv.org/abs/2308.16755}{{arXiv:2308.16755}}}
{[hep-lat]}
\end{barticle}
\endbibitem

\bibitem[\protect\citeauthoryear{Nambu}{1960}]{Nambu:1960xd}
\begin{barticle}
\bauthor{\bsnm{Nambu}, \binits{Y.}}:
\batitle{{Axial vector current conservation in weak interactions}}.
\bjtitle{Phys. Rev. Lett.}
\bvolume{4},
\bfpage{380}--\blpage{382}
(\byear{1960})
\doiurl{10.1103/PhysRevLett.4.380}
\end{barticle}
\endbibitem

\bibitem[\protect\citeauthoryear{Feynman and Gell-Mann}{1958}]{Feynman:1958ty}
\begin{barticle}
\bauthor{\bsnm{Feynman}, \binits{R.P.}},
\bauthor{\bsnm{Gell-Mann}, \binits{M.}}:
\batitle{{Theory of Fermi interaction}}.
\bjtitle{Phys. Rev.}
\bvolume{109},
\bfpage{193}--\blpage{198}
(\byear{1958})
\doiurl{10.1103/PhysRev.109.193}
\end{barticle}
\endbibitem

\bibitem[\protect\citeauthoryear{Nishijima}{1964}]{nishijima1964unsubtracted}
\begin{barticle}
\bauthor{\bsnm{Nishijima}, \binits{K.}}:
\batitle{Unsubtracted dispersion relations in weak interactions and the
  goldberger-treiman relation}.
\bjtitle{Physical Review}
\bvolume{133}(\bissue{4B}),
\bfpage{1092}
(\byear{1964})
\end{barticle}
\endbibitem

\bibitem[\protect\citeauthoryear{Pagels}{1969}]{Pagels:1969ne}
\begin{barticle}
\bauthor{\bsnm{Pagels}, \binits{H.}}:
\batitle{{Hadronic corrections to the Goldberger-Treiman relation}}.
\bjtitle{Phys. Rev.}
\bvolume{179},
\bfpage{1337}--\blpage{1344}
(\byear{1969})
\doiurl{10.1103/PhysRev.179.1337}
\end{barticle}
\endbibitem

\bibitem[\protect\citeauthoryear{Coleman and Moffat}{1969}]{Coleman:1969ulw}
\begin{barticle}
\bauthor{\bsnm{Coleman}, \binits{R.A.}},
\bauthor{\bsnm{Moffat}, \binits{J.W.}}:
\batitle{{Pion lifetime, rho pi and sigma pi intermediate states, and sum
  rules}}.
\bjtitle{Phys. Rev.}
\bvolume{186},
\bfpage{1635}--\blpage{1642}
(\byear{1969})
\doiurl{10.1103/PhysRev.186.1635}
\end{barticle}
\endbibitem

\bibitem[\protect\citeauthoryear{Pagels and Zepeda}{1972}]{Pagels:1972xx}
\begin{barticle}
\bauthor{\bsnm{Pagels}, \binits{H.}},
\bauthor{\bsnm{Zepeda}, \binits{A.}}:
\batitle{{Where are the corrections to the Goldberger-Treiman relation?}}
\bjtitle{Phys. Rev. D}
\bvolume{5},
\bfpage{3262}--\blpage{3268}
(\byear{1972})
\doiurl{10.1103/PhysRevD.5.3262}
\end{barticle}
\endbibitem

\bibitem[\protect\citeauthoryear{Dominguez}{1973}]{Dominguez:1973jj}
\begin{barticle}
\bauthor{\bsnm{Dominguez}, \binits{C.A.}}:
\batitle{{Veneziano-type n n pi form-factor and the Goldberger-Treiman
  relation}}.
\bjtitle{Phys. Rev. D}
\bvolume{7},
\bfpage{1252}--\blpage{1253}
(\byear{1973})
\doiurl{10.1103/PhysRevD.7.1252}
\end{barticle}
\endbibitem

\bibitem[\protect\citeauthoryear{Pagels}{1975}]{Pagels:1974se}
\begin{barticle}
\bauthor{\bsnm{Pagels}, \binits{H.}}:
\batitle{{Departures from Chiral Symmetry: A Review}}.
\bjtitle{Phys. Rept.}
\bvolume{16},
\bfpage{219}
(\byear{1975})
\doiurl{10.1016/0370-1573(75)90039-3}
\end{barticle}
\endbibitem

\bibitem[\protect\citeauthoryear{Dominguez}{1977a}]{Dominguez:1976ut}
\begin{barticle}
\bauthor{\bsnm{Dominguez}, \binits{C.A.}}:
\batitle{{Extended Partially Conserved Axial-Vector Current Hypothesis and
  Chiral Symmetry Breaking}}.
\bjtitle{Phys. Rev. D}
\bvolume{15},
\bfpage{1350}--\blpage{1360}
(\byear{1977})
\doiurl{10.1103/PhysRevD.15.1350}
\end{barticle}
\endbibitem

\bibitem[\protect\citeauthoryear{Dominguez}{1977b}]{Dominguez:1977nt}
\begin{barticle}
\bauthor{\bsnm{Dominguez}, \binits{C.A.}}:
\batitle{{Extended Partially Conserved Axial-Vector Current Hypothesis. 2. Soft
  Meson Theorems}}.
\bjtitle{Phys. Rev. D}
\bvolume{16},
\bfpage{2313}
(\byear{1977})
\doiurl{10.1103/PhysRevD.16.2313}
\end{barticle}
\endbibitem

\bibitem[\protect\citeauthoryear{Dominguez}{1977c}]{Dominguez:1977en}
\begin{barticle}
\bauthor{\bsnm{Dominguez}, \binits{C.A.}}:
\batitle{{Extended Partially Conserved Axial Vector Current Hypothesis and
  Model Dependent Results}}.
\bjtitle{Phys. Rev. D}
\bvolume{16},
\bfpage{2320}
(\byear{1977})
\doiurl{10.1103/PhysRevD.16.2320}
\end{barticle}
\endbibitem

\bibitem[\protect\citeauthoryear{Dominguez}{1985}]{Dominguez:1984ka}
\begin{barticle}
\bauthor{\bsnm{Dominguez}, \binits{C.A.}}:
\batitle{{THE GOLDBERGER-TREIMAN RELATION: A PROBE OF THE CHIRAL SYMMETRIES OF
  QUANTUM CHROMODYNAMICS}}.
\bjtitle{Riv. Nuovo Cim.}
\bvolume{8N6},
\bfpage{1}--\blpage{27}
(\byear{1985})
\doiurl{10.1007/BF02724344}
\end{barticle}
\endbibitem

\bibitem[\protect\citeauthoryear{de~Swart et~al.}{1997}]{deSwart:1997ep}
\begin{barticle}
\bauthor{\bsnm{Swart}, \binits{J.J.}},
\bauthor{\bsnm{Rentmeester}, \binits{M.C.M.}},
\bauthor{\bsnm{Timmermans}, \binits{R.G.E.}}:
\batitle{{The Status of the pion - nucleon coupling constant}}.
\bjtitle{PiN Newslett.}
\bvolume{13},
\bfpage{96}--\blpage{107}
(\byear{1997})
{\href{https://arxiv.org/abs/nucl-th/9802084}{{arXiv:nucl-th/9802084}}}
\end{barticle}
\endbibitem

\bibitem[\protect\citeauthoryear{Baru et~al.}{2011}]{Baru:2010xn}
\begin{barticle}
\bauthor{\bsnm{Baru}, \binits{V.}},
\bauthor{\bsnm{Hanhart}, \binits{C.}},
\bauthor{\bsnm{Hoferichter}, \binits{M.}},
\bauthor{\bsnm{Kubis}, \binits{B.}},
\bauthor{\bsnm{Nogga}, \binits{A.}},
\bauthor{\bsnm{Phillips}, \binits{D.R.}}:
\batitle{{Precision calculation of the $\pi^{-}$ deuteron scattering length and
  its impact on threshold $\pi$ N scattering}}.
\bjtitle{Phys. Lett. B}
\bvolume{694},
\bfpage{473}--\blpage{477}
(\byear{2011})
\doiurl{10.1016/j.physletb.2010.10.028}
{\href{https://arxiv.org/abs/1003.4444}{{arXiv:1003.4444}}}
{[nucl-th]}
\end{barticle}
\endbibitem

\bibitem[\protect\citeauthoryear{Ericson et~al.}{2002}]{Ericson:2000md}
\begin{barticle}
\bauthor{\bsnm{Ericson}, \binits{T.E.O.}},
\bauthor{\bsnm{Loiseau}, \binits{B.}},
\bauthor{\bsnm{Thomas}, \binits{A.W.}}:
\batitle{{Determination of the pion nucleon coupling constant and scattering
  lengths}}.
\bjtitle{Phys. Rev. C}
\bvolume{66},
\bfpage{014005}
(\byear{2002})
\doiurl{10.1103/PhysRevC.66.014005}
{\href{https://arxiv.org/abs/hep-ph/0009312}{{arXiv:hep-ph/0009312}}}
\end{barticle}
\endbibitem

\bibitem[\protect\citeauthoryear{Hoferichter
  et~al.}{2023}]{Hoferichter:2023ptl}
\begin{barticle}
\bauthor{\bsnm{Hoferichter}, \binits{M.}},
\bauthor{\bsnm{Elvira}, \binits{J.R.}},
\bauthor{\bsnm{Kubis}, \binits{B.}},
\bauthor{\bsnm{Mei{\ss}ner}, \binits{U.-G.}}:
\batitle{{On the role of isospin violation in the pion{\textendash}nucleon
  {\ensuremath{\sigma}}-term}}.
\bjtitle{Phys. Lett. B}
\bvolume{843},
\bfpage{138001}
(\byear{2023})
\doiurl{10.1016/j.physletb.2023.138001}
{\href{https://arxiv.org/abs/2305.07045}{{arXiv:2305.07045}}}
{[hep-ph]}
\end{barticle}
\endbibitem

\bibitem[\protect\citeauthoryear{Navarro~P\'erez
  et~al.}{2013}]{NavarroPerez:2013mvd}
\begin{barticle}
\bauthor{\bsnm{Navarro~P\'erez}, \binits{R.}},
\bauthor{\bsnm{Amaro}, \binits{J.E.}},
\bauthor{\bsnm{Ruiz~Arriola}, \binits{E.}}:
\batitle{{Coarse-grained potential analysis of neutron-proton and proton-proton
  scattering below the pion production threshold}}.
\bjtitle{Phys. Rev. C}
\bvolume{88}(\bissue{6}),
\bfpage{064002}
(\byear{2013})
\doiurl{10.1103/PhysRevC.88.064002}
{\href{https://arxiv.org/abs/1310.2536}{{arXiv:1310.2536}}}
{[nucl-th]}.
\bcomment{[Erratum: Phys.Rev.C 91, 029901 (2015)]}
\end{barticle}
\endbibitem

\bibitem[\protect\citeauthoryear{Matsinos}{2019}]{Matsinos:2019kqi}
\begin{botherref}
\oauthor{\bsnm{Matsinos}, \binits{E.}}:
{A brief history of the pion-nucleon coupling constant}
(2019)
{\href{https://arxiv.org/abs/1901.01204}{{arXiv:1901.01204}}}
{[nucl-th]}
\end{botherref}
\endbibitem

\bibitem[\protect\citeauthoryear{Reinert et~al.}{2021}]{Reinert:2020mcu}
\begin{barticle}
\bauthor{\bsnm{Reinert}, \binits{P.}},
\bauthor{\bsnm{Krebs}, \binits{H.}},
\bauthor{\bsnm{Epelbaum}, \binits{E.}}:
\batitle{{Precision determination of pion-nucleon coupling constants using
  effective field theory}}.
\bjtitle{Phys. Rev. Lett.}
\bvolume{126}(\bissue{9}),
\bfpage{092501}
(\byear{2021})
\doiurl{10.1103/PhysRevLett.126.092501}
{\href{https://arxiv.org/abs/2006.15360}{{arXiv:2006.15360}}}
{[nucl-th]}
\end{barticle}
\endbibitem

\bibitem[\protect\citeauthoryear{Ruiz~Arriola and
  Sanchez-Puertas}{2023}]{RuizArriola:2023xap}
\begin{barticle}
\bauthor{\bsnm{Ruiz~Arriola}, \binits{E.}},
\bauthor{\bsnm{Sanchez-Puertas}, \binits{P.}}:
\batitle{{Pseudoscalar Meson Dominance and Nucleon Structure}}.
\bjtitle{Few Body Syst.}
\bvolume{64}(\bissue{3}),
\bfpage{48}
(\byear{2023})
\doiurl{10.1007/s00601-023-01830-z}
{\href{https://arxiv.org/abs/2306.05267}{{arXiv:2306.05267}}}
{[hep-ph]}
\end{barticle}
\endbibitem

\bibitem[\protect\citeauthoryear{Sainio}{1999}]{Sainio:1999ba}
\begin{barticle}
\bauthor{\bsnm{Sainio}, \binits{M.E.}}:
\batitle{{Pion nucleon coupling constant: Working group summary}}.
\bjtitle{PiN Newslett.}
\bvolume{15},
\bfpage{156}--\blpage{161}
(\byear{1999})
{\href{https://arxiv.org/abs/hep-ph/9912337}{{arXiv:hep-ph/9912337}}}
\end{barticle}
\endbibitem

\bibitem[\protect\citeauthoryear{Ruiz~Arriola
  et~al.}{2016}]{RuizArriola:2016ehc}
\begin{barticle}
\bauthor{\bsnm{Ruiz~Arriola}, \binits{E.}},
\bauthor{\bsnm{Amaro}, \binits{J.E.}},
\bauthor{\bsnm{Navarro~P\'erez}, \binits{R.}}:
\batitle{{Three pion nucleon coupling constants}}.
\bjtitle{Mod. Phys. Lett. A}
\bvolume{31}(\bissue{28}),
\bfpage{1630027}
(\byear{2016})
\doiurl{10.1142/S0217732316300275}
{\href{https://arxiv.org/abs/1606.02171}{{arXiv:1606.02171}}}
{[nucl-th]}
\end{barticle}
\endbibitem

\bibitem[\protect\citeauthoryear{Baru et~al.}{2011}]{Baru:2011bw}
\begin{barticle}
\bauthor{\bsnm{Baru}, \binits{V.}},
\bauthor{\bsnm{Hanhart}, \binits{C.}},
\bauthor{\bsnm{Hoferichter}, \binits{M.}},
\bauthor{\bsnm{Kubis}, \binits{B.}},
\bauthor{\bsnm{Nogga}, \binits{A.}},
\bauthor{\bsnm{Phillips}, \binits{D.R.}}:
\batitle{{Precision calculation of threshold $\pi^-d$ scattering, $\pi$N
  scattering lengths, and the GMO sum rule}}.
\bjtitle{Nucl. Phys. A}
\bvolume{872},
\bfpage{69}--\blpage{116}
(\byear{2011})
\doiurl{10.1016/j.nuclphysa.2011.09.015}
{\href{https://arxiv.org/abs/1107.5509}{{arXiv:1107.5509}}}
{[nucl-th]}
\end{barticle}
\endbibitem

\bibitem[\protect\citeauthoryear{Navarro~P\'erez
  et~al.}{2017}]{NavarroPerez:2016eli}
\begin{barticle}
\bauthor{\bsnm{Navarro~P\'erez}, \binits{R.}},
\bauthor{\bsnm{Amaro}, \binits{J.E.}},
\bauthor{\bsnm{Ruiz~Arriola}, \binits{E.}}:
\batitle{{Precise Determination of Charge Dependent Pion-Nucleon-Nucleon
  Coupling Constants}}.
\bjtitle{Phys. Rev. C}
\bvolume{95}(\bissue{6}),
\bfpage{064001}
(\byear{2017})
\doiurl{10.1103/PhysRevC.95.064001}
{\href{https://arxiv.org/abs/1606.00592}{{arXiv:1606.00592}}}
{[nucl-th]}
\end{barticle}
\endbibitem

\bibitem[\protect\citeauthoryear{Bali et~al.}{2020}]{RQCD:2019jai}
\begin{barticle}
\bauthor{\bsnm{Bali}, \binits{G.S.}},
\bauthor{\bsnm{Barca}, \binits{L.}},
\bauthor{\bsnm{Collins}, \binits{S.}},
\bauthor{\bsnm{Gruber}, \binits{M.}},
\bauthor{\bsnm{L{\"o}ffler}, \binits{M.}},
\bauthor{\bsnm{Sch{\"a}fer}, \binits{A.}},
\bauthor{\bsnm{S{\"o}ldner}, \binits{W.}},
\bauthor{\bsnm{Wein}, \binits{P.}},
\bauthor{\bsnm{Weish{\"a}upl}, \binits{S.}},
\bauthor{\bsnm{Wurm}, \binits{T.}}:
\batitle{{Nucleon axial structure from lattice QCD}}.
\bjtitle{JHEP}
\bvolume{05},
\bfpage{126}
(\byear{2020})
\doiurl{10.1007/JHEP05(2020)126}
{\href{https://arxiv.org/abs/1911.13150}{{arXiv:1911.13150}}}
{[hep-lat]}
\end{barticle}
\endbibitem

\bibitem[\protect\citeauthoryear{Alexandrou et~al.}{2021}]{Alexandrou:2020okk}
\begin{barticle}
\bauthor{\bsnm{Alexandrou}, \binits{C.}}, \betal:
\batitle{{Nucleon axial and pseudoscalar form factors from lattice QCD at the
  physical point}}.
\bjtitle{Phys. Rev. D}
\bvolume{103}(\bissue{3}),
\bfpage{034509}
(\byear{2021})
\doiurl{10.1103/PhysRevD.103.034509}
{\href{https://arxiv.org/abs/2011.13342}{{arXiv:2011.13342}}}
{[hep-lat]}
\end{barticle}
\endbibitem

\bibitem[\protect\citeauthoryear{Park et~al.}{2022}]{Park:2021ypf}
\begin{barticle}
\bauthor{\bsnm{Park}, \binits{S.}},
\bauthor{\bsnm{Gupta}, \binits{R.}},
\bauthor{\bsnm{Yoon}, \binits{B.}},
\bauthor{\bsnm{Mondal}, \binits{S.}},
\bauthor{\bsnm{Bhattacharya}, \binits{T.}},
\bauthor{\bsnm{Jang}, \binits{Y.-C.}},
\bauthor{\bsnm{Jo{\'o}}, \binits{B.}},
\bauthor{\bsnm{Winter}, \binits{F.}}:
\batitle{{Precision nucleon charges and form factors using (2+1)-flavor lattice
  QCD}}.
\bjtitle{Phys. Rev. D}
\bvolume{105}(\bissue{5}),
\bfpage{054505}
(\byear{2022})
\doiurl{10.1103/PhysRevD.105.054505}
{\href{https://arxiv.org/abs/2103.05599}{{arXiv:2103.05599}}}
{[hep-lat]}
\end{barticle}
\endbibitem

\bibitem[\protect\citeauthoryear{Jang et~al.}{2024}]{Jang:2023zts}
\begin{barticle}
\bauthor{\bsnm{Jang}, \binits{Y.-C.}},
\bauthor{\bsnm{Gupta}, \binits{R.}},
\bauthor{\bsnm{Bhattacharya}, \binits{T.}},
\bauthor{\bsnm{Yoon}, \binits{B.}},
\bauthor{\bsnm{Lin}, \binits{H.-W.}}:
\batitle{{Nucleon isovector axial form factors}}.
\bjtitle{Phys. Rev. D}
\bvolume{109}(\bissue{1}),
\bfpage{014503}
(\byear{2024})
\doiurl{10.1103/PhysRevD.109.014503}
{\href{https://arxiv.org/abs/2305.11330}{{arXiv:2305.11330}}}
{[hep-lat]}
\end{barticle}
\endbibitem

\bibitem[\protect\citeauthoryear{Alexandrou et~al.}{2024}]{Alexandrou:2023qbg}
\begin{barticle}
\bauthor{\bsnm{Alexandrou}, \binits{C.}},
\bauthor{\bsnm{Bacchio}, \binits{S.}},
\bauthor{\bsnm{Constantinou}, \binits{M.}},
\bauthor{\bsnm{Finkenrath}, \binits{J.}},
\bauthor{\bsnm{Frezzotti}, \binits{R.}},
\bauthor{\bsnm{Kostrzewa}, \binits{B.}},
\bauthor{\bsnm{Koutsou}, \binits{G.}},
\bauthor{\bsnm{Spanoudes}, \binits{G.}},
\bauthor{\bsnm{Urbach}, \binits{C.}}:
\batitle{{Nucleon axial and pseudoscalar form factors using twisted-mass
  fermion ensembles at the physical point}}.
\bjtitle{Phys. Rev. D}
\bvolume{109}(\bissue{3}),
\bfpage{034503}
(\byear{2024})
\doiurl{10.1103/PhysRevD.109.034503}
{\href{https://arxiv.org/abs/2309.05774}{{arXiv:2309.05774}}}
{[hep-lat]}
\end{barticle}
\endbibitem

\bibitem[\protect\citeauthoryear{Erkelenz}{1974}]{Erkelenz:1974uj}
\begin{barticle}
\bauthor{\bsnm{Erkelenz}, \binits{K.}}:
\batitle{{Current status of the relativistic two nucleon one boson exchange
  potential}}.
\bjtitle{Phys. Rept.}
\bvolume{13},
\bfpage{191}--\blpage{258}
(\byear{1974})
\doiurl{10.1016/0370-1573(74)90008-8}
\end{barticle}
\endbibitem

\bibitem[\protect\citeauthoryear{Nagels et~al.}{1978}]{Nagels:1977ze}
\begin{barticle}
\bauthor{\bsnm{Nagels}, \binits{M.M.}},
\bauthor{\bsnm{Rijken}, \binits{T.A.}},
\bauthor{\bsnm{Swart}, \binits{J.J.}}:
\batitle{{A Low-Energy Nucleon-Nucleon Potential from Regge Pole Theory}}.
\bjtitle{Phys. Rev. D}
\bvolume{17},
\bfpage{768}
(\byear{1978})
\doiurl{10.1103/PhysRevD.17.768}
\end{barticle}
\endbibitem

\bibitem[\protect\citeauthoryear{Machleidt et~al.}{1987}]{Machleidt:1987hj}
\begin{barticle}
\bauthor{\bsnm{Machleidt}, \binits{R.}},
\bauthor{\bsnm{Holinde}, \binits{K.}},
\bauthor{\bsnm{Elster}, \binits{C.}}:
\batitle{{The Bonn Meson Exchange Model for the Nucleon Nucleon Interaction}}.
\bjtitle{Phys. Rept.}
\bvolume{149},
\bfpage{1}--\blpage{89}
(\byear{1987})
\doiurl{10.1016/S0370-1573(87)80002-9}
\end{barticle}
\endbibitem

\bibitem[\protect\citeauthoryear{Calle~Cordon and
  Ruiz~Arriola}{2010}]{CalleCordon:2009pit}
\begin{barticle}
\bauthor{\bsnm{Calle~Cordon}, \binits{A.}},
\bauthor{\bsnm{Ruiz~Arriola}, \binits{E.}}:
\batitle{{Renormalization vs Strong Form Factors for One Boson Exchange
  Potentials}}.
\bjtitle{Phys. Rev. C}
\bvolume{81},
\bfpage{044002}
(\byear{2010})
\doiurl{10.1103/PhysRevC.81.044002}
{\href{https://arxiv.org/abs/0905.4933}{{arXiv:0905.4933}}}
{[nucl-th]}
\end{barticle}
\endbibitem

\bibitem[\protect\citeauthoryear{Cohen}{1986}]{Cohen:1986ux}
\begin{barticle}
\bauthor{\bsnm{Cohen}, \binits{T.D.}}:
\batitle{{The $\pi N N$ Form-factor in the Skyrme Model}}.
\bjtitle{Phys. Rev. D}
\bvolume{34},
\bfpage{2187}
(\byear{1986})
\doiurl{10.1103/PhysRevD.34.2187}
\end{barticle}
\endbibitem

\bibitem[\protect\citeauthoryear{Melde et~al.}{2009}]{Melde:2008dg}
\begin{barticle}
\bauthor{\bsnm{Melde}, \binits{T.}},
\bauthor{\bsnm{Canton}, \binits{L.}},
\bauthor{\bsnm{Plessas}, \binits{W.}}:
\batitle{{Structure of meson-baryon interaction vertices}}.
\bjtitle{Phys. Rev. Lett.}
\bvolume{102},
\bfpage{132002}
(\byear{2009})
\doiurl{10.1103/PhysRevLett.102.132002}
{\href{https://arxiv.org/abs/0811.0277}{{arXiv:0811.0277}}}
{[nucl-th]}
\end{barticle}
\endbibitem

\bibitem[\protect\citeauthoryear{Coon and Scadron}{1990}]{Coon:1990fh}
\begin{barticle}
\bauthor{\bsnm{Coon}, \binits{S.A.}},
\bauthor{\bsnm{Scadron}, \binits{M.D.}}:
\batitle{{pi N N couplings, the pi N N form-factor, and the Goldberger-Treiman
  discrepancy}}.
\bjtitle{Phys. Rev. C}
\bvolume{42},
\bfpage{2256}--\blpage{2258}
(\byear{1990})
\doiurl{10.1103/PhysRevC.42.2256}
\end{barticle}
\endbibitem

\bibitem[\protect\citeauthoryear{Meissner}{1995}]{Meissner:1995ra}
\begin{barticle}
\bauthor{\bsnm{Meissner}, \binits{T.}}:
\batitle{{The pi N N form-factor from QCD sum rules}}.
\bjtitle{Phys. Rev. C}
\bvolume{52},
\bfpage{3386}--\blpage{3392}
(\byear{1995})
\doiurl{10.1103/PhysRevC.52.3386}
{\href{https://arxiv.org/abs/nucl-th/9506030}{{arXiv:nucl-th/9506030}}}
\end{barticle}
\endbibitem

\bibitem[\protect\citeauthoryear{Birse and Krippa}{1996}]{Birse:1995zh}
\begin{barticle}
\bauthor{\bsnm{Birse}, \binits{M.C.}},
\bauthor{\bsnm{Krippa}, \binits{B.}}:
\batitle{{Determination of the pion - nucleon coupling constant from QCD sum
  rules}}.
\bjtitle{Phys. Lett. B}
\bvolume{373},
\bfpage{9}--\blpage{15}
(\byear{1996})
\doiurl{10.1016/0370-2693(96)00133-5}
{\href{https://arxiv.org/abs/hep-ph/9512259}{{arXiv:hep-ph/9512259}}}
\end{barticle}
\endbibitem

\bibitem[\protect\citeauthoryear{Eichmann and Fischer}{2012}]{Eichmann:2011pv}
\begin{barticle}
\bauthor{\bsnm{Eichmann}, \binits{G.}},
\bauthor{\bsnm{Fischer}, \binits{C.S.}}:
\batitle{{Nucleon axial and pseudoscalar form factors from the covariant
  Faddeev equation}}.
\bjtitle{Eur. Phys. J. A}
\bvolume{48},
\bfpage{9}
(\byear{2012})
\doiurl{10.1140/epja/i2012-12009-6}
{\href{https://arxiv.org/abs/1111.2614}{{arXiv:1111.2614}}}
{[hep-ph]}
\end{barticle}
\endbibitem

\bibitem[\protect\citeauthoryear{Chen et~al.}{2022}]{Chen:2021guo}
\begin{barticle}
\bauthor{\bsnm{Chen}, \binits{C.}},
\bauthor{\bsnm{Fischer}, \binits{C.S.}},
\bauthor{\bsnm{Roberts}, \binits{C.D.}},
\bauthor{\bsnm{Segovia}, \binits{J.}}:
\batitle{{Nucleon axial-vector and pseudoscalar form factors and PCAC
  relations}}.
\bjtitle{Phys. Rev. D}
\bvolume{105}(\bissue{9}),
\bfpage{094022}
(\byear{2022})
\doiurl{10.1103/PhysRevD.105.094022}
{\href{https://arxiv.org/abs/2103.02054}{{arXiv:2103.02054}}}
{[hep-ph]}
\end{barticle}
\endbibitem

\bibitem[\protect\citeauthoryear{Goity et~al.}{1999}]{Goity:1999by}
\begin{barticle}
\bauthor{\bsnm{Goity}, \binits{J.L.}},
\bauthor{\bsnm{Lewis}, \binits{R.}},
\bauthor{\bsnm{Schvellinger}, \binits{M.}},
\bauthor{\bsnm{Zhang}, \binits{L.-Z.}}:
\batitle{{The Goldberger-Treiman discrepancy in SU(3)}}.
\bjtitle{Phys. Lett. B}
\bvolume{454},
\bfpage{115}--\blpage{122}
(\byear{1999})
\doiurl{10.1016/S0370-2693(99)00217-8}
{\href{https://arxiv.org/abs/hep-ph/9901374}{{arXiv:hep-ph/9901374}}}
\end{barticle}
\endbibitem

\bibitem[\protect\citeauthoryear{Nasrallah}{2000}]{Nasrallah:1999fw}
\begin{barticle}
\bauthor{\bsnm{Nasrallah}, \binits{N.F.}}:
\batitle{{The Goldberger-Treiman discrepancy}}.
\bjtitle{Phys. Rev. D}
\bvolume{62},
\bfpage{036006}
(\byear{2000})
\doiurl{10.1103/PhysRevD.62.036006}
{\href{https://arxiv.org/abs/hep-ph/9904358}{{arXiv:hep-ph/9904358}}}
\end{barticle}
\endbibitem

\bibitem[\protect\citeauthoryear{Lepage and Brodsky}{1979}]{Lepage:1979za}
\begin{barticle}
\bauthor{\bsnm{Lepage}, \binits{G.P.}},
\bauthor{\bsnm{Brodsky}, \binits{S.J.}}:
\batitle{{Exclusive Processes in Quantum Chromodynamics: The Form-Factors of
  Baryons at Large Momentum Transfer}}.
\bjtitle{Phys. Rev. Lett.}
\bvolume{43},
\bfpage{545}--\blpage{549}
(\byear{1979})
\doiurl{10.1103/PhysRevLett.43.545} .
\bcomment{[Erratum: Phys.Rev.Lett. 43, 1625--1626 (1979)]}
\end{barticle}
\endbibitem

\bibitem[\protect\citeauthoryear{Korenblit et~al.}{1979}]{Korenblit:1979cw}
\begin{barticle}
\bauthor{\bsnm{Korenblit}, \binits{G.V.}},
\bauthor{\bsnm{Naumov}, \binits{V.A.}},
\bauthor{\bsnm{Chernyak}, \binits{V.L.}}:
\batitle{{ASYMPTOTICS OF HADRON EXCLUSIVE ELECTROPRODUCTION AMPLITUDES. (IN
  RUSSIAN)}}.
\bjtitle{Sov. J. Nucl. Phys.}
\bvolume{29},
\bfpage{77}
(\byear{1979})
\end{barticle}
\endbibitem

\bibitem[\protect\citeauthoryear{Lepage and Brodsky}{1980}]{Lepage:1980fj}
\begin{barticle}
\bauthor{\bsnm{Lepage}, \binits{G.P.}},
\bauthor{\bsnm{Brodsky}, \binits{S.J.}}:
\batitle{{Exclusive Processes in Perturbative Quantum Chromodynamics}}.
\bjtitle{Phys. Rev.}
\bvolume{D22},
\bfpage{2157}
(\byear{1980})
\doiurl{10.1103/PhysRevD.22.2157}
\end{barticle}
\endbibitem

\bibitem[\protect\citeauthoryear{Chernyak and
  Zhitnitsky}{1984}]{Chernyak:1984bm}
\begin{barticle}
\bauthor{\bsnm{Chernyak}, \binits{V.L.}},
\bauthor{\bsnm{Zhitnitsky}, \binits{I.R.}}:
\batitle{{Nucleon Wave Function and Nucleon Form-Factors in QCD}}.
\bjtitle{Nucl. Phys. B}
\bvolume{246},
\bfpage{52}--\blpage{74}
(\byear{1984})
\doiurl{10.1016/0550-3213(84)90114-7}
\end{barticle}
\endbibitem

\bibitem[\protect\citeauthoryear{Huang et~al.}{2024}]{Huang:2024ugd}
\begin{botherref}
\oauthor{\bsnm{Huang}, \binits{Y.-K.}},
\oauthor{\bsnm{Shi}, \binits{B.-X.}},
\oauthor{\bsnm{Wang}, \binits{Y.-M.}},
\oauthor{\bsnm{Zhao}, \binits{X.-C.}}:
{Next-to-Leading-Order QCD Predictions for the Nucleon Form Factors}
(2024)
{\href{https://arxiv.org/abs/2407.18724}{{arXiv:2407.18724}}}
{[hep-ph]}
\end{botherref}
\endbibitem

\bibitem[\protect\citeauthoryear{Chen et~al.}{2024}]{Chen:2024fhj}
\begin{botherref}
\oauthor{\bsnm{Chen}, \binits{L.-B.}},
\oauthor{\bsnm{Chen}, \binits{W.}},
\oauthor{\bsnm{Feng}, \binits{F.}},
\oauthor{\bsnm{Hu}, \binits{S.}},
\oauthor{\bsnm{Jia}, \binits{Y.}}:
{Next-to-leading-order QCD corrections to nucleon Dirac form factors}
(2024)
{\href{https://arxiv.org/abs/2406.19994}{{arXiv:2406.19994}}}
{[hep-ph]}
\end{botherref}
\endbibitem

\bibitem[\protect\citeauthoryear{Brodsky et~al.}{1981}]{Brodsky:1980sx}
\begin{barticle}
\bauthor{\bsnm{Brodsky}, \binits{S.J.}},
\bauthor{\bsnm{Lepage}, \binits{G.P.}},
\bauthor{\bsnm{Zaidi}, \binits{S.A.A.}}:
\batitle{{Weak and Electromagnetic Form-factors of Baryons at Large Momentum
  Transfer}}.
\bjtitle{Phys. Rev. D}
\bvolume{23},
\bfpage{1152}
(\byear{1981})
\doiurl{10.1103/PhysRevD.23.1152}
\end{barticle}
\endbibitem

\bibitem[\protect\citeauthoryear{Carlson and Poor}{1986}]{Carlson:1985zu}
\begin{barticle}
\bauthor{\bsnm{Carlson}, \binits{C.E.}},
\bauthor{\bsnm{Poor}, \binits{J.L.}}:
\batitle{{The Nucleon Axial Vector Form-factor in Perturbative {QCD}}}.
\bjtitle{Phys. Rev.}
\bvolume{D34},
\bfpage{1478}
(\byear{1986})
\doiurl{10.1103/PhysRevD.34.1478}
\end{barticle}
\endbibitem

\bibitem[\protect\citeauthoryear{Alabiso and Schierholz}{1975}]{Alabiso:1974ye}
\begin{barticle}
\bauthor{\bsnm{Alabiso}, \binits{C.}},
\bauthor{\bsnm{Schierholz}, \binits{G.}}:
\batitle{{Asymptotic Behavior of Form-Factors for Two-Body and Three-Body Bound
  States. 2. Spin 1/2 Constituents}}.
\bjtitle{Phys. Rev. D}
\bvolume{11},
\bfpage{1905}
(\byear{1975})
\doiurl{10.1103/PhysRevD.11.1905}
\end{barticle}
\endbibitem

\bibitem[\protect\citeauthoryear{Braun et~al.}{2006}]{Braun:2006hz}
\begin{barticle}
\bauthor{\bsnm{Braun}, \binits{V.M.}},
\bauthor{\bsnm{Lenz}, \binits{A.}},
\bauthor{\bsnm{Wittmann}, \binits{M.}}:
\batitle{{Nucleon Form Factors in QCD}}.
\bjtitle{Phys. Rev. D}
\bvolume{73},
\bfpage{094019}
(\byear{2006})
\doiurl{10.1103/PhysRevD.73.094019}
{\href{https://arxiv.org/abs/hep-ph/0604050}{{arXiv:hep-ph/0604050}}}
\end{barticle}
\endbibitem

\bibitem[\protect\citeauthoryear{Alvegard and Kogerler}{1979}]{Alvegard:1979ui}
\begin{barticle}
\bauthor{\bsnm{Alvegard}, \binits{C.}},
\bauthor{\bsnm{Kogerler}, \binits{R.}}:
\batitle{{ASYMPTOTIC BEHAVIOR OF WEAK FORM-FACTORS WITHIN QCD}}.
\bjtitle{Z. Phys. C}
\bvolume{2},
\bfpage{173}
(\byear{1979})
\doiurl{10.1007/BF01474131}
\end{barticle}
\endbibitem

\bibitem[\protect\citeauthoryear{Bernard et~al.}{1996}]{Bernard:1996cc}
\begin{barticle}
\bauthor{\bsnm{Bernard}, \binits{V.}},
\bauthor{\bsnm{Kaiser}, \binits{N.}},
\bauthor{\bsnm{Meissner}, \binits{U.-G.}}:
\batitle{{Nucleon electroweak form-factors: Analysis of their spectral
  functions}}.
\bjtitle{Nucl. Phys. A}
\bvolume{611},
\bfpage{429}--\blpage{441}
(\byear{1996})
\doiurl{10.1016/S0375-9474(96)00291-6}
{\href{https://arxiv.org/abs/hep-ph/9607428}{{arXiv:hep-ph/9607428}}}
\end{barticle}
\endbibitem

\bibitem[\protect\citeauthoryear{Kaiser}{2003}]{Kaiser:2003dr}
\begin{barticle}
\bauthor{\bsnm{Kaiser}, \binits{N.}}:
\batitle{{Induced pseudoscalar form-factor of the nucleon at two loop order in
  chiral perturbation theory}}.
\bjtitle{Phys. Rev. C}
\bvolume{67},
\bfpage{027002}
(\byear{2003})
\doiurl{10.1103/PhysRevC.67.027002}
{\href{https://arxiv.org/abs/nucl-th/0301034}{{arXiv:nucl-th/0301034}}}
\end{barticle}
\endbibitem

\bibitem[\protect\citeauthoryear{Kaiser and Passemar}{2019}]{Kaiser:2019irl}
\begin{barticle}
\bauthor{\bsnm{Kaiser}, \binits{N.}},
\bauthor{\bsnm{Passemar}, \binits{E.}}:
\batitle{{Spectral functions of nucleon form factors: Three-pion continua at
  low energies}}.
\bjtitle{Eur. Phys. J. A}
\bvolume{55}(\bissue{2}),
\bfpage{16}
(\byear{2019})
\doiurl{10.1140/epja/i2019-12680-y}
{\href{https://arxiv.org/abs/1901.02865}{{arXiv:1901.02865}}}
{[nucl-th]}
\end{barticle}
\endbibitem

\bibitem[\protect\citeauthoryear{Ruiz~Arriola
  et~al.}{2025}]{RuizArriola:2025wyq}
\begin{barticle}
\bauthor{\bsnm{Ruiz~Arriola}, \binits{E.}},
\bauthor{\bsnm{Sanchez-Puertas}, \binits{P.}},
\bauthor{\bsnm{Weiss}, \binits{C.}}:
\batitle{{Pion transverse charge density from $e^+e^-$ annihilation data and
  logarithmic dispersion relations}}.
\bjtitle{Phys. Lett. B}
\bvolume{866},
\bfpage{139585}
(\byear{2025})
\doiurl{10.1016/j.physletb.2025.139585}
{\href{https://arxiv.org/abs/2503.10465}{{arXiv:2503.10465}}}
{[hep-ph]}
\end{barticle}
\endbibitem

\bibitem[\protect\citeauthoryear{Donoghue and Na}{1997}]{Donoghue:1996bt}
\begin{barticle}
\bauthor{\bsnm{Donoghue}, \binits{J.F.}},
\bauthor{\bsnm{Na}, \binits{E.S.}}:
\batitle{{Asymptotic limits and structure of the pion form factor}}.
\bjtitle{Phys. Rev.}
\bvolume{D56},
\bfpage{7073}--\blpage{7076}
(\byear{1997})
\doiurl{10.1103/PhysRevD.56.7073}
{\href{https://arxiv.org/abs/hep-ph/9611418}{{arXiv:hep-ph/9611418}}}
\end{barticle}
\endbibitem

\bibitem[\protect\citeauthoryear{Leutwyler}{2002}]{Leutwyler:2002hm}
\begin{botherref}
\oauthor{\bsnm{Leutwyler}, \binits{H.}}:
{Electromagnetic form factor of the pion}
(2002)
{\href{https://arxiv.org/abs/hep-ph/0212324}{{arXiv:hep-ph/0212324}}}
\end{botherref}
\endbibitem

\bibitem[\protect\citeauthoryear{Hoferichter
  et~al.}{2016}]{Hoferichter:2016duk}
\begin{barticle}
\bauthor{\bsnm{Hoferichter}, \binits{M.}},
\bauthor{\bsnm{Kubis}, \binits{B.}},
\bauthor{\bsnm{Elvira}, \binits{J.}},
\bauthor{\bsnm{Hammer}, \binits{H.-W.}},
\bauthor{\bsnm{Mei{\ss}ner}, \binits{U.-G.}}:
\batitle{{On the $\pi\pi$ continuum in the nucleon form factors and the proton
  radius puzzle}}.
\bjtitle{Eur. Phys. J. A}
\bvolume{52}(\bissue{11}),
\bfpage{331}
(\byear{2016})
\doiurl{10.1140/epja/i2016-16331-7}
{\href{https://arxiv.org/abs/1609.06722}{{arXiv:1609.06722}}}
{[hep-ph]}
\end{barticle}
\endbibitem

\bibitem[\protect\citeauthoryear{Alarc{\'o}n and Weiss}{2018}]{Alarcon:2018irp}
\begin{barticle}
\bauthor{\bsnm{Alarc{\'o}n}, \binits{J.M.}},
\bauthor{\bsnm{Weiss}, \binits{C.}}:
\batitle{{Accurate nucleon electromagnetic form factors from dispersively
  improved chiral effective field theory}}.
\bjtitle{Phys. Lett. B}
\bvolume{784},
\bfpage{373}--\blpage{377}
(\byear{2018})
\doiurl{10.1016/j.physletb.2018.07.060}
{\href{https://arxiv.org/abs/1803.09748}{{arXiv:1803.09748}}}
{[hep-ph]}
\end{barticle}
\endbibitem

\bibitem[\protect\citeauthoryear{Bhamathi and Raghavan}{1977}]{Bhamathi:1977xp}
\begin{barticle}
\bauthor{\bsnm{Bhamathi}, \binits{G.}},
\bauthor{\bsnm{Raghavan}, \binits{S.}}:
\batitle{{N anti-N Resonance and the Corrections to the Goldberger-Treiman
  Relation}}.
\bjtitle{Pramana}
\bvolume{9},
\bfpage{257}
(\byear{1977})
\doiurl{10.1007/BF02846205}
\end{barticle}
\endbibitem

\bibitem[\protect\citeauthoryear{Ledwig et~al.}{2014}]{Ledwig:2014cla}
\begin{barticle}
\bauthor{\bsnm{Ledwig}, \binits{T.}},
\bauthor{\bsnm{Nieves}, \binits{J.}},
\bauthor{\bsnm{Pich}, \binits{A.}},
\bauthor{\bsnm{Ruiz~Arriola}, \binits{E.}},
\bauthor{\bsnm{Elvira}, \binits{J.}}:
\batitle{{Large-$N_c$ naturalness in coupled-channel meson-meson scattering}}.
\bjtitle{Phys. Rev. D}
\bvolume{90}(\bissue{11}),
\bfpage{114020}
(\byear{2014})
\doiurl{10.1103/PhysRevD.90.114020}
{\href{https://arxiv.org/abs/1407.3750}{{arXiv:1407.3750}}}
{[hep-ph]}
\end{barticle}
\endbibitem

\bibitem[\protect\citeauthoryear{Yan et~al.}{2025}]{Yan:2025mdm}
\begin{botherref}
\oauthor{\bsnm{Yan}, \binits{H.}},
\oauthor{\bsnm{Mai}, \binits{M.}},
\oauthor{\bsnm{Garofalo}, \binits{M.}},
\oauthor{\bsnm{Feng}, \binits{Y.}},
\oauthor{\bsnm{D{\"o}ring}, \binits{M.}},
\oauthor{\bsnm{Liu}, \binits{C.}},
\oauthor{\bsnm{Liu}, \binits{L.}},
\oauthor{\bsnm{Mei{\ss}ner}, \binits{U.-G.}},
\oauthor{\bsnm{Urbach}, \binits{C.}}:
{Emergence of the $\pi(1300)$ Resonance from Lattice QCD}
(2025)
{\href{https://arxiv.org/abs/2510.09476}{{arXiv:2510.09476}}}
{[hep-lat]}
\end{botherref}
\endbibitem

\bibitem[\protect\citeauthoryear{Masjuan et~al.}{2013}]{Masjuan:2012sk}
\begin{barticle}
\bauthor{\bsnm{Masjuan}, \binits{P.}},
\bauthor{\bsnm{Ruiz~Arriola}, \binits{E.}},
\bauthor{\bsnm{Broniowski}, \binits{W.}}:
\batitle{{Meson dominance of hadron form factors and large-$N_c$
  phenomenology}}.
\bjtitle{Phys. Rev.}
\bvolume{D87}(\bissue{1}),
\bfpage{014005}
(\byear{2013})
\doiurl{10.1103/PhysRevD.87.014005}
{\href{https://arxiv.org/abs/1210.0760}{{arXiv:1210.0760}}}
{[hep-ph]}
\end{barticle}
\endbibitem

\bibitem[\protect\citeauthoryear{Ruiz~Arriola and
  Broniowski}{2011}]{RuizArriola:2011nrw}
\begin{bchapter}
\bauthor{\bsnm{Ruiz~Arriola}, \binits{E.}},
\bauthor{\bsnm{Broniowski}, \binits{W.}}:
\bctitle{{$0^{++}$ states in a large-$N_c$ Regge approach}}.
In: \bbtitle{{Mini-Workshop~Bled~2011:~Understanding~Hadronic~Spectra}},
pp. \bfpage{7}--\blpage{17}
(\byear{2011})
\end{bchapter}
\endbibitem

\bibitem[\protect\citeauthoryear{Ruiz~Arriola
  et~al.}{2013}]{RuizArriola:2012ius}
\begin{barticle}
\bauthor{\bsnm{Ruiz~Arriola}, \binits{E.}},
\bauthor{\bsnm{Broniowski}, \binits{W.}},
\bauthor{\bsnm{Masjuan}, \binits{P.}}:
\batitle{{Hadron resonances, large Nc, and the half-width rule}}.
\bjtitle{Acta Phys. Polon. Supp.}
\bvolume{6},
\bfpage{95}--\blpage{102}
(\byear{2013})
\doiurl{10.5506/APhysPolBSupp.6.95}
{\href{https://arxiv.org/abs/1210.7153}{{arXiv:1210.7153}}}
{[hep-ph]}
\end{barticle}
\endbibitem

\bibitem[\protect\citeauthoryear{Dominguez}{2001}]{Dominguez:2001zu}
\begin{barticle}
\bauthor{\bsnm{Dominguez}, \binits{C.A.}}:
\batitle{Pion form factor in large n(c) qcd}.
\bjtitle{Phys. Lett.}
\bvolume{B512},
\bfpage{331}--\blpage{334}
(\byear{2001})
{\href{https://arxiv.org/abs/hep-ph/0102190}{{hep-ph/0102190}}}
\end{barticle}
\endbibitem

\bibitem[\protect\citeauthoryear{Ruiz~Arriola and
  Broniowski}{2008}]{RuizArriola:2008sq}
\begin{barticle}
\bauthor{\bsnm{Ruiz~Arriola}, \binits{E.}},
\bauthor{\bsnm{Broniowski}, \binits{W.}}:
\batitle{{Pion electromagnetic form factor, perturbative QCD, and large-N(c)
  Regge models}}.
\bjtitle{Phys. Rev. D}
\bvolume{78},
\bfpage{034031}
(\byear{2008})
\doiurl{10.1103/PhysRevD.78.034031}
{\href{https://arxiv.org/abs/0807.3488}{{arXiv:0807.3488}}}
{[hep-ph]}
\end{barticle}
\endbibitem

\bibitem[\protect\citeauthoryear{Anisovich et~al.}{2001}]{Anisovich:2001pn}
\begin{barticle}
\bauthor{\bsnm{Anisovich}, \binits{A.V.}},
\bauthor{\bsnm{Baker}, \binits{C.A.}},
\bauthor{\bsnm{Batty}, \binits{C.J.}},
\bauthor{\bsnm{Bugg}, \binits{D.V.}},
\bauthor{\bsnm{Nikonov}, \binits{V.A.}},
\bauthor{\bsnm{Sarantsev}, \binits{A.V.}},
\bauthor{\bsnm{Sarantsev}, \binits{V.V.}},
\bauthor{\bsnm{Zou}, \binits{B.S.}}:
\batitle{{Partial wave analysis of anti-p p annihilation channels in flight
  with I = 1, C = +1}}.
\bjtitle{Phys. Lett. B}
\bvolume{517},
\bfpage{261}--\blpage{272}
(\byear{2001})
\doiurl{10.1016/S0370-2693(01)01017-6}
{\href{https://arxiv.org/abs/1110.0278}{{arXiv:1110.0278}}}
{[hep-ex]}
\end{barticle}
\endbibitem

\bibitem[\protect\citeauthoryear{Godfrey and Isgur}{1985}]{Godfrey:1985xj}
\begin{barticle}
\bauthor{\bsnm{Godfrey}, \binits{S.}},
\bauthor{\bsnm{Isgur}, \binits{N.}}:
\batitle{Mesons in a relativized quark model with chromodynamics}.
\bjtitle{Phys. Rev.}
\bvolume{D32},
\bfpage{189}--\blpage{231}
(\byear{1985})
\end{barticle}
\endbibitem

\bibitem[\protect\citeauthoryear{Ruiz~Arriola and
  Broniowski}{2007}]{RuizArriola:2006opt}
\begin{barticle}
\bauthor{\bsnm{Ruiz~Arriola}, \binits{E.}},
\bauthor{\bsnm{Broniowski}, \binits{W.}}:
\batitle{{Dimension-2 condensates, zeta-regularization and large-N(c) Regge
  Models}}.
\bjtitle{Eur. Phys. J. A}
\bvolume{31},
\bfpage{739}--\blpage{741}
(\byear{2007})
\doiurl{10.1140/epja/i2006-10184-7}
{\href{https://arxiv.org/abs/hep-ph/0609266}{{arXiv:hep-ph/0609266}}}
\end{barticle}
\endbibitem

\bibitem[\protect\citeauthoryear{Afonin and Tsymbal}{2024}]{Afonin:2024egd}
\begin{barticle}
\bauthor{\bsnm{Afonin}, \binits{S.}},
\bauthor{\bsnm{Tsymbal}, \binits{A.}}:
\batitle{{Dynamical ${O(4)}$-Symmetry in the Light Meson Spectrum within the
  Framework of the Regge Approach}}.
\bjtitle{Phys. Atom. Nucl.}
\bvolume{87}(\bissue{Suppl 3}),
\bfpage{477}--\blpage{482}
(\byear{2024})
\doiurl{10.1134/S1063778824701084}
{\href{https://arxiv.org/abs/2502.19562}{{arXiv:2502.19562}}}
{[hep-ph]}
\end{barticle}
\endbibitem

\bibitem[\protect\citeauthoryear{Masjuan et~al.}{2012}]{Masjuan:2012gc}
\begin{barticle}
\bauthor{\bsnm{Masjuan}, \binits{P.}},
\bauthor{\bsnm{Ruiz~Arriola}, \binits{E.}},
\bauthor{\bsnm{Broniowski}, \binits{W.}}:
\batitle{{Systematics of radial and angular-momentum Regge trajectories of
  light non-strange {$\bar{q}q$}-states}}.
\bjtitle{Phys. Rev. D}
\bvolume{85},
\bfpage{094006}
(\byear{2012})
\doiurl{10.1103/PhysRevD.85.094006}
{\href{https://arxiv.org/abs/1203.4782}{{arXiv:1203.4782}}}
{[hep-ph]}
\end{barticle}
\endbibitem

\bibitem[\protect\citeauthoryear{Chen}{2022}]{Chen:2022flh}
\begin{barticle}
\bauthor{\bsnm{Chen}, \binits{J.-K.}}:
\batitle{{Revisiting the pion Regge trajectories}}.
\bjtitle{Nucl. Phys. B}
\bvolume{983},
\bfpage{115911}
(\byear{2022})
\doiurl{10.1016/j.nuclphysb.2022.115911}
{\href{https://arxiv.org/abs/2203.02981}{{arXiv:2203.02981}}}
{[hep-ph]}
\end{barticle}
\endbibitem

\bibitem[\protect\citeauthoryear{Maltman and Kambor}{2002}]{Maltman:2001gc}
\begin{barticle}
\bauthor{\bsnm{Maltman}, \binits{K.}},
\bauthor{\bsnm{Kambor}, \binits{J.}}:
\batitle{{Decay constants, light quark masses and quark mass bounds from light
  quark pseudoscalar sum rules}}.
\bjtitle{Phys. Rev. D}
\bvolume{65},
\bfpage{074013}
(\byear{2002})
\doiurl{10.1103/PhysRevD.65.074013}
{\href{https://arxiv.org/abs/hep-ph/0108227}{{arXiv:hep-ph/0108227}}}
\end{barticle}
\endbibitem

\bibitem[\protect\citeauthoryear{Kadav{\'y} et~al.}{2020}]{Kadavy:2020hox}
\begin{barticle}
\bauthor{\bsnm{Kadav{\'y}}, \binits{T.}},
\bauthor{\bsnm{Kampf}, \binits{K.}},
\bauthor{\bsnm{Novotny}, \binits{J.}}:
\batitle{{OPE of Green functions of chiral currents}}.
\bjtitle{JHEP}
\bvolume{10},
\bfpage{142}
(\byear{2020})
\doiurl{10.1007/JHEP10(2020)142}
{\href{https://arxiv.org/abs/2006.13006}{{arXiv:2006.13006}}}
{[hep-ph]}
\end{barticle}
\endbibitem

\bibitem[\protect\citeauthoryear{Dominguez}{1984}]{Dominguez:1984eh}
\begin{barticle}
\bauthor{\bsnm{Dominguez}, \binits{C.A.}}:
\batitle{{Hadronic Corrections to {QCD} Sum Rules and Light Quark Masses}}.
\bjtitle{Z. Phys. C}
\bvolume{26},
\bfpage{269}
(\byear{1984})
\doiurl{10.1007/BF01421765}
\end{barticle}
\endbibitem

\bibitem[\protect\citeauthoryear{Dominguez et~al.}{1985}]{Dominguez:1984yx}
\begin{barticle}
\bauthor{\bsnm{Dominguez}, \binits{C.A.}},
\bauthor{\bsnm{Kremer}, \binits{M.}},
\bauthor{\bsnm{Papadopoulos}, \binits{N.A.}},
\bauthor{\bsnm{Schilcher}, \binits{K.}}:
\batitle{{Light Quark Condensates From {QCD} Sum Rules}}.
\bjtitle{Z. Phys. C}
\bvolume{27},
\bfpage{481}
(\byear{1985})
\doiurl{10.1007/BF01548656}
\end{barticle}
\endbibitem

\bibitem[\protect\citeauthoryear{Dominguez et~al.}{2019}]{Dominguez:2018azt}
\begin{barticle}
\bauthor{\bsnm{Dominguez}, \binits{C.A.}},
\bauthor{\bsnm{Mes}, \binits{A.}},
\bauthor{\bsnm{Schilcher}, \binits{K.}}:
\batitle{{Up- and down-quark masses from QCD sum rules}}.
\bjtitle{JHEP}
\bvolume{02},
\bfpage{057}
(\byear{2019})
\doiurl{10.1007/JHEP02(2019)057}
{\href{https://arxiv.org/abs/1809.07042}{{arXiv:1809.07042}}}
{[hep-ph]}
\end{barticle}
\endbibitem

\bibitem[\protect\citeauthoryear{Gupta}{2024}]{Gupta:2024krt}
\begin{barticle}
\bauthor{\bsnm{Gupta}, \binits{R.}}:
\batitle{{Isovector Axial Charge and Form Factors of Nucleons from Lattice
  QCD}}.
\bjtitle{PoS}
\bvolume{LATTICE2023},
\bfpage{124}
(\byear{2024})
\doiurl{10.22323/1.453.0124}
\end{barticle}
\endbibitem

\bibitem[\protect\citeauthoryear{Aoki et~al.}{2025}]{Aoki:2025taf}
\begin{botherref}
\oauthor{\bsnm{Aoki}, \binits{Y.}},
\oauthor{\bsnm{Ishikawa}, \binits{K.-I.}},
\oauthor{\bsnm{Kuramashi}, \binits{Y.}},
\oauthor{\bsnm{Sasaki}, \binits{S.}},
\oauthor{\bsnm{Sato}, \binits{K.}},
\oauthor{\bsnm{Shintani}, \binits{E.}},
\oauthor{\bsnm{Tsuji}, \binits{R.}},
\oauthor{\bsnm{Watanabe}, \binits{H.}},
\oauthor{\bsnm{Yamazaki}, \binits{T.}}:
{Method for high-precision determination of the nucleon axial structure using
  lattice QCD: Removing $\pi N$-state contamination}
(2025)
{\href{https://arxiv.org/abs/2505.06854}{{arXiv:2505.06854}}}
{[hep-lat]}
\end{botherref}
\endbibitem

\bibitem[\protect\citeauthoryear{Jang et~al.}{2020}]{Jang:2019vkm}
\begin{barticle}
\bauthor{\bsnm{Jang}, \binits{Y.-C.}},
\bauthor{\bsnm{Gupta}, \binits{R.}},
\bauthor{\bsnm{Yoon}, \binits{B.}},
\bauthor{\bsnm{Bhattacharya}, \binits{T.}}:
\batitle{{Axial Vector Form Factors from Lattice QCD that Satisfy the PCAC
  Relation}}.
\bjtitle{Phys. Rev. Lett.}
\bvolume{124}(\bissue{7}),
\bfpage{072002}
(\byear{2020})
\doiurl{10.1103/PhysRevLett.124.072002}
{\href{https://arxiv.org/abs/1905.06470}{{arXiv:1905.06470}}}
{[hep-lat]}
\end{barticle}
\endbibitem

\bibitem[\protect\citeauthoryear{Alexandrou et~al.}{2017}]{Alexandrou:2017hac}
\begin{barticle}
\bauthor{\bsnm{Alexandrou}, \binits{C.}},
\bauthor{\bsnm{Constantinou}, \binits{M.}},
\bauthor{\bsnm{Hadjiyiannakou}, \binits{K.}},
\bauthor{\bsnm{Jansen}, \binits{K.}},
\bauthor{\bsnm{Kallidonis}, \binits{C.}},
\bauthor{\bsnm{Koutsou}, \binits{G.}},
\bauthor{\bsnm{Vaquero~Aviles-Casco}, \binits{A.}}:
\batitle{{Nucleon axial form factors using $N_f$ = 2 twisted mass fermions with
  a physical value of the pion mass}}.
\bjtitle{Phys. Rev. D}
\bvolume{96}(\bissue{5}),
\bfpage{054507}
(\byear{2017})
\doiurl{10.1103/PhysRevD.96.054507}
{\href{https://arxiv.org/abs/1705.03399}{{arXiv:1705.03399}}}
{[hep-lat]}
\end{barticle}
\endbibitem

\bibitem[\protect\citeauthoryear{Bernard et~al.}{2002}]{Bernard:2001rs}
\begin{barticle}
\bauthor{\bsnm{Bernard}, \binits{V.}},
\bauthor{\bsnm{Elouadrhiri}, \binits{L.}},
\bauthor{\bsnm{Meissner}, \binits{U.-G.}}:
\batitle{{Axial structure of the nucleon: Topical Review}}.
\bjtitle{J. Phys. G}
\bvolume{28},
\bfpage{1}--\blpage{35}
(\byear{2002})
\doiurl{10.1088/0954-3899/28/1/201}
{\href{https://arxiv.org/abs/hep-ph/0107088}{{arXiv:hep-ph/0107088}}}
\end{barticle}
\endbibitem

\bibitem[\protect\citeauthoryear{Braun et~al.}{2014}]{Braun:2014wpa}
\begin{barticle}
\bauthor{\bsnm{Braun}, \binits{V.M.}},
\bauthor{\bsnm{Collins}, \binits{S.}},
\bauthor{\bsnm{Gl\"a\ss{}le}, \binits{B.}},
\bauthor{\bsnm{G\"ockeler}, \binits{M.}},
\bauthor{\bsnm{Sch\"afer}, \binits{A.}},
\bauthor{\bsnm{Schiel}, \binits{R.W.}},
\bauthor{\bsnm{S\"oldner}, \binits{W.}},
\bauthor{\bsnm{Sternbeck}, \binits{A.}},
\bauthor{\bsnm{Wein}, \binits{P.}}:
\batitle{{Light-cone Distribution Amplitudes of the Nucleon and Negative Parity
  Nucleon Resonances from Lattice QCD}}.
\bjtitle{Phys. Rev. D}
\bvolume{89},
\bfpage{094511}
(\byear{2014})
\doiurl{10.1103/PhysRevD.89.094511}
{\href{https://arxiv.org/abs/1403.4189}{{arXiv:1403.4189}}}
{[hep-lat]}
\end{barticle}
\endbibitem

\bibitem[\protect\citeauthoryear{King and Sachrajda}{1987}]{King:1986wi}
\begin{barticle}
\bauthor{\bsnm{King}, \binits{I.D.}},
\bauthor{\bsnm{Sachrajda}, \binits{C.T.}}:
\batitle{{Nucleon Wave Functions and QCD Sum Rules}}.
\bjtitle{Nucl. Phys. B}
\bvolume{279},
\bfpage{785}--\blpage{803}
(\byear{1987})
\doiurl{10.1016/0550-3213(87)90019-8}
\end{barticle}
\endbibitem

\bibitem[\protect\citeauthoryear{Chernyak et~al.}{1988}]{Chernyak:1987nu}
\begin{barticle}
\bauthor{\bsnm{Chernyak}, \binits{V.L.}},
\bauthor{\bsnm{Ogloblin}, \binits{A.A.}},
\bauthor{\bsnm{Zhitnitsky}, \binits{I.R.}}:
\batitle{{Wave Functions of Octet Baryons}}.
\bjtitle{Yad. Fiz.}
\bvolume{48},
\bfpage{1410}--\blpage{1422}
(\byear{1988})
\doiurl{10.1007/BF01557663}
\end{barticle}
\endbibitem

\bibitem[\protect\citeauthoryear{Stefanis and Bergmann}{1993}]{Stefanis:1992nw}
\begin{barticle}
\bauthor{\bsnm{Stefanis}, \binits{N.G.}},
\bauthor{\bsnm{Bergmann}, \binits{M.}}:
\batitle{{On the Nucleon distribution amplitude: the Heterotic solution}}.
\bjtitle{Phys. Rev. D}
\bvolume{47},
\bfpage{3685}--\blpage{3689}
(\byear{1993})
\doiurl{10.1103/PhysRevD.47.R3685}
{\href{https://arxiv.org/abs/hep-ph/9211250}{{arXiv:hep-ph/9211250}}}
\end{barticle}
\endbibitem

\bibitem[\protect\citeauthoryear{Bolz and Kroll}{1996}]{Bolz:1996sw}
\begin{barticle}
\bauthor{\bsnm{Bolz}, \binits{J.}},
\bauthor{\bsnm{Kroll}, \binits{P.}}:
\batitle{{Modeling the nucleon wave function from soft and hard processes}}.
\bjtitle{Z. Phys. A}
\bvolume{356},
\bfpage{327}
(\byear{1996})
\doiurl{10.1007/s002180050186}
{\href{https://arxiv.org/abs/hep-ph/9603289}{{arXiv:hep-ph/9603289}}}
\end{barticle}
\endbibitem

\bibitem[\protect\citeauthoryear{Braun et~al.}{2000}]{Braun:2000kw}
\begin{barticle}
\bauthor{\bsnm{Braun}, \binits{V.}},
\bauthor{\bsnm{Fries}, \binits{R.J.}},
\bauthor{\bsnm{Mahnke}, \binits{N.}},
\bauthor{\bsnm{Stein}, \binits{E.}}:
\batitle{{Higher twist distribution amplitudes of the nucleon in QCD}}.
\bjtitle{Nucl. Phys. B}
\bvolume{589},
\bfpage{381}--\blpage{409}
(\byear{2000})
\doiurl{10.1016/S0550-3213(00)00516-2}
{\href{https://arxiv.org/abs/hep-ph/0007279}{{arXiv:hep-ph/0007279}}}.
\bcomment{[Erratum: Nucl.Phys.B 607, 433--433 (2001)]}
\end{barticle}
\endbibitem

\bibitem[\protect\citeauthoryear{Holinde and Thomas}{1990}]{Holinde:1990fe}
\begin{barticle}
\bauthor{\bsnm{Holinde}, \binits{K.}},
\bauthor{\bsnm{Thomas}, \binits{A.W.}}:
\batitle{{One boson exchange potential based on a soft pion form-factor}}.
\bjtitle{Phys. Rev. C}
\bvolume{42},
\bfpage{1195}--\blpage{1199}
(\byear{1990})
\doiurl{10.1103/PhysRevC.42.R1195}
\end{barticle}
\endbibitem

\bibitem[\protect\citeauthoryear{Sirlin}{1972}]{Sirlin:1972cs}
\begin{barticle}
\bauthor{\bsnm{Sirlin}, \binits{A.}}:
\batitle{{Radiative corrections to the Goldberger-Treiman relation - high-
  frequency contributions}}.
\bjtitle{Phys. Rev. D}
\bvolume{5},
\bfpage{436}--\blpage{444}
(\byear{1972})
\doiurl{10.1103/PhysRevD.5.436}
\end{barticle}
\endbibitem

\bibitem[\protect\citeauthoryear{d'Argent et~al.}{2017}]{dArgent:2017gzv}
\begin{barticle}
\bauthor{\bsnm{d'Argent}, \binits{P.}},
\bauthor{\bsnm{Skidmore}, \binits{N.}},
\bauthor{\bsnm{Benton}, \binits{J.}},
\bauthor{\bsnm{Dalseno}, \binits{J.}},
\bauthor{\bsnm{Gersabeck}, \binits{E.}},
\bauthor{\bsnm{Harnew}, \binits{S.}},
\bauthor{\bsnm{Naik}, \binits{P.}},
\bauthor{\bsnm{Prouve}, \binits{C.}},
\bauthor{\bsnm{Rademacker}, \binits{J.}}:
\batitle{{Amplitude Analyses of $D^0 \to {\pi^+\pi^-\pi^+\pi^-}$ and $D^0 \to
  {K^+K^-\pi^+\pi^-}$ Decays}}.
\bjtitle{JHEP}
\bvolume{05},
\bfpage{143}
(\byear{2017})
\doiurl{10.1007/JHEP05(2017)143}
{\href{https://arxiv.org/abs/1703.08505}{{arXiv:1703.08505}}}
{[hep-ex]}
\end{barticle}
\endbibitem

\bibitem[\protect\citeauthoryear{Shchegelsky et~al.}{2006}]{Shchegelsky:2006es}
\begin{barticle}
\bauthor{\bsnm{Shchegelsky}, \binits{V.A.}},
\bauthor{\bsnm{Sarantsev}, \binits{A.V.}},
\bauthor{\bsnm{Anisovich}, \binits{A.V.}},
\bauthor{\bsnm{Levchenko}, \binits{M.P.}}:
\batitle{{Partial wave analysis of pi+ pi- pi0 production in two-photon
  collisions at LEP}}.
\bjtitle{Eur. Phys. J. A}
\bvolume{27},
\bfpage{199}--\blpage{205}
(\byear{2006})
\doiurl{10.1140/epja/i2005-10266-0}
\end{barticle}
\endbibitem

\bibitem[\protect\citeauthoryear{Salvini et~al.}{2004}]{OBELIX:2004oio}
\begin{barticle}
\bauthor{\bsnm{Salvini}, \binits{P.}}, \betal:
\batitle{{anti-p p annihilation into four charged pions at rest and in
  flight}}.
\bjtitle{Eur. Phys. J. C}
\bvolume{35},
\bfpage{21}--\blpage{33}
(\byear{2004})
\doiurl{10.1140/epjc/s2004-01811-8}
\end{barticle}
\endbibitem

\bibitem[\protect\citeauthoryear{Chung et~al.}{2002}]{Chung:2002pu}
\begin{barticle}
\bauthor{\bsnm{Chung}, \binits{S.U.}}, \betal:
\batitle{{Exotic and q anti-q resonances in the pi+ pi- pi- system produced in
  pi- p collisions at 18-GeV/c/}}.
\bjtitle{Phys. Rev. D}
\bvolume{65},
\bfpage{072001}
(\byear{2002})
\doiurl{10.1103/PhysRevD.65.072001}
\end{barticle}
\endbibitem

\bibitem[\protect\citeauthoryear{Abele et~al.}{2001}]{CrystalBarrel:2001xud}
\begin{barticle}
\bauthor{\bsnm{Abele}, \binits{A.}}, \betal:
\batitle{{Study of f0 decays into four neutral pions}}.
\bjtitle{Eur. Phys. J. C}
\bvolume{19},
\bfpage{667}--\blpage{675}
(\byear{2001})
\doiurl{10.1007/s100520100601}
\end{barticle}
\endbibitem

\bibitem[\protect\citeauthoryear{Bertin et~al.}{1997}]{OBELIX:1997zla}
\begin{barticle}
\bauthor{\bsnm{Bertin}, \binits{A.}}, \betal:
\batitle{{Study of anti-p p --{\ensuremath{>}} 2pi+ 2pi- annihilation from S
  states}}.
\bjtitle{Phys. Lett. B}
\bvolume{414},
\bfpage{220}--\blpage{228}
(\byear{1997})
\doiurl{10.1016/S0370-2693(97)01189-1}
\end{barticle}
\endbibitem

\bibitem[\protect\citeauthoryear{Abele et~al.}{1996}]{CrystalBarrel:1996wfh}
\begin{barticle}
\bauthor{\bsnm{Abele}, \binits{A.}}, \betal:
\batitle{{A Study of f0 (1500) decays into 4 pi0 in anti-p p
  ---{\ensuremath{>}} 5 pi0 at rest}}.
\bjtitle{Phys. Lett. B}
\bvolume{380},
\bfpage{453}--\blpage{460}
(\byear{1996})
\doiurl{10.1016/0370-2693(96)00574-6}
\end{barticle}
\endbibitem

\bibitem[\protect\citeauthoryear{Zielinski et~al.}{1984}]{Zielinski:1984mt}
\begin{barticle}
\bauthor{\bsnm{Zielinski}, \binits{M.}}, \betal:
\batitle{{Partial Wave Analysis of Coherent 3 $\pi$ Production on Nuclei at
  200-{GeV}}}.
\bjtitle{Phys. Rev. D}
\bvolume{30},
\bfpage{1855}
(\byear{1984})
\doiurl{10.1103/PhysRevD.30.1855}
\end{barticle}
\endbibitem

\bibitem[\protect\citeauthoryear{Bellini et~al.}{1982}]{Bellini:1982ec}
\begin{barticle}
\bauthor{\bsnm{Bellini}, \binits{G.}}, \betal:
\batitle{{EVIDENCE FOR NEW 0- S RESONANCES IN THE PI+ PI- PI- SYSTEMS}}.
\bjtitle{Phys. Rev. Lett.}
\bvolume{48},
\bfpage{1697}--\blpage{1700}
(\byear{1982})
\doiurl{10.1103/PhysRevLett.48.1697}
\end{barticle}
\endbibitem

\bibitem[\protect\citeauthoryear{Aaron and Longacre}{1981}]{Aaron:1980zk}
\begin{barticle}
\bauthor{\bsnm{Aaron}, \binits{R.}},
\bauthor{\bsnm{Longacre}, \binits{R.S.}}:
\batitle{{ANALYSIS OF THE J(P) = 1+ AND 0- THREE PION SYSTEMS}}.
\bjtitle{Phys. Rev. D}
\bvolume{24},
\bfpage{1207}--\blpage{1217}
(\byear{1981})
\doiurl{10.1103/PhysRevD.24.1207}
\end{barticle}
\endbibitem

\bibitem[\protect\citeauthoryear{Bonesini et~al.}{1981}]{Bonesini:1981sx}
\begin{barticle}
\bauthor{\bsnm{Bonesini}, \binits{M.}}, \betal:
\batitle{{Evidence for a New Pseudoscalar Meson}}.
\bjtitle{Phys. Lett. B}
\bvolume{103},
\bfpage{75}--\blpage{78}
(\byear{1981})
\doiurl{10.1016/0370-2693(81)90197-0}
\end{barticle}
\endbibitem

\bibitem[\protect\citeauthoryear{Daum et~al.}{1981}]{ACCMOR:1980llh}
\begin{barticle}
\bauthor{\bsnm{Daum}, \binits{C.}}, \betal:
\batitle{{Diffractive Production of 3 $\pi$ States at 63-{GeV} and 94-{GeV}}}.
\bjtitle{Nucl. Phys. B}
\bvolume{182},
\bfpage{269}--\blpage{336}
(\byear{1981})
\doiurl{10.1016/0550-3213(81)90123-1}
\end{barticle}
\endbibitem

\end{thebibliography}

\end{document}